\documentclass[11pt,letterpaper]{JHEP3}
\usepackage{amsmath,epsfig,array}
\usepackage{amssymb}
\usepackage{amsfonts,amstext,latexsym,xspace}
\usepackage{amsfonts}
\usepackage{amsmath}
\usepackage{cite}

\makeatletter
\newif\if@fewtab\@fewtabtrue
\def\moth{\mathsurround=0pt}
\newdimen\zo \zo=0pt

\def\tick{\leaders\hrule height 0.5ex depth 0pt \hskip 0.5pt}
\def\upboxfill{$\moth \setbox\zo\hbox{\tick}%
  \hskip 2pt\hbox to 0pt{$\tick$\hss}\hrulefill \hbox to 2pt{$\tick$\hss}$}

\def\dtick{\leaders\hrule height .34pt depth 0.5ex \hskip 0.5pt}
\def\downboxfill{$\moth \setbox\zo\hbox{\dtick}%
  \hskip 2pt\hbox to 0pt{$\dtick$\hss}\hrulefill%
  \hbox to 2pt{$\dtick$\hss}$}

\usepackage{array}
\usepackage{amsmath}
\usepackage{amssymb}
\usepackage{graphicx}
\usepackage{longtable}
\usepackage[vcentermath]{youngtab}

\newcommand{\commute}[2]{\left[ #1 \, , \, #2 \right]}
\newcommand{\anticommute}[2]{\left\{ #1 \, , \, #2 \right\}}
\newcommand{\ket}[1]{\left\lvert #1 \right\rangle}
\newcommand{\eq}{\begin{equation}}
\newcommand{\en}{\end{equation}}
\newcommand{\eqq}{\begin{eqnarray}}
\newcommand{\enn}{\end{eqnarray}}

\newcommand{\stln}{\setlength{\unitlength}{10pt}}  % Set unit length for YT boxes
\newcommand{\lfr}{\framebox(2,1){}}  % Draws a long empty box
\newcommand{\sfr}{\framebox(1,1)[bl]{\begin{picture}(1,1)(0,0)
                                      \put(0,0){\line(1,1){1}}
                                     \end{picture}
                                     }
                  }  % Draws a `slashed' box
\newcommand{\dotfr}{\stln \lower2.0pt \hbox{\begin{picture}(2,1)(0,0)
                                                 \put(0,0){\lfr}
                                                 \put(0.5,0.5){\dots}
                                                \end{picture}
                                                }
                        } % Draws a long box with dots
\newcommand{\sonebox}{\stln \lower2.0pt \hbox{\begin{picture}(1,1)(0,0)
                                               \put(0,0){\sfr}
                                              \end{picture}
                                              }
                      }
\newcommand{\stwobox}{\stln \lower2.0pt \hbox{\begin{picture}(2,1)(0,0)
                                               \multiput(0,0)(1,0){2}{\sfr}
                                              \end{picture}
                                              }
                      }
\newcommand{\sgenrowbox}{\stln \lower2.0pt \hbox{\begin{picture}(5,1)(0,0)
                                               \multiput(0,0)(1,0){2}{\sfr}
                                               \put(2,0.2){\dotfr}
                                               \put(4,0){\sfr}
                                              \end{picture}
                                              }
                      }

\title{Minimal unitary representation of $SU(2,2)$  and its deformations as massless conformal fields and their supersymmetric extensions}

\author{
Sudarshan Fernando$^{1}$\footnote{fernando@kutztown.edu}  and 
Murat G\"{u}naydin$^{2}$\footnote{murat@phys.psu.edu} 
\\

$^{1}$\emph{Physical Sciences Department\\Kutztown University\\ Kutztown, PA 19530, USA}  \\

$^{2}$\emph{Institute for Gravitation and the Cosmos \\ Physics Department \\
Pennsylvania State University\\
University Park, PA 16802, USA} }

\abstract{ We study the minimal unitary representation (minrep) of 
$SO(4,2)$ over an Hilbert space of functions of three variables, obtained 
by quantizing its quasiconformal action on a  five dimensional space. The 
minrep of $SO(4,2)$, which coincides with the minrep of $SU(2,2)$ similarly 
constructed, corresponds to a massless conformal scalar in four spacetime 
dimensions. There exists  a one-parameter  family of 
deformations of the minrep of $SU(2,2)$. For positive (negative) integer 
values of the deformation 
parameter $\zeta$, one obtains positive energy unitary irreducible 
representations corresponding to massless conformal fields transforming in  
$(0, \zeta/2) (( -\zeta/2 , 0))$ representation of the $SL(2,\mathbb{C})$ 
subgroup. We construct the supersymmetric extensions of the minrep of 
$SU(2,2)$ and its deformations to those of $SU(2,2\,|\,N)$.   The minimal 
unitary supermultiplet of $SU(2,2\,|\,4)$, in the undeformed case, 
simply corresponds to the massless $N=4$ Yang-Mills supermultiplet in four 
dimensions. For each given non-zero integer value of $\zeta$, one obtains a 
unique supermultiplet of massless conformal fields of higher spin. For 
$SU(2,2\,|\,4)$ these supermultiplets are simply the doubleton 
supermultiplets studied in \hepth{9806042}. }

\preprint{arXiv:0908.3624 [hep-th]}

\begin{document}

%%%%%%%%%%%%%%%%%%%%%%%%%%%%%%%%%%%%%%%%%%%%%%%%%%%%%%%%%%%%%%%%%%%%
%%%%%% Introduction %%%%%%%%%%%%%%%%%%%%%%%%%%%%%%%%%%%%%%%%%%%%%%%%
%%%%%%%%%%%%%%%%%%%%%%%%%%%%%%%%%%%%%%%%%%%%%%%%%%%%%%%%%%%%%%%%%%%%

\section{Introduction}
The concept of minimal unitary realizations of Lie algebras was introduced 
by Joseph in \cite{MR0342049} and    was inspired by the work of  physicists on 
spectrum generating symmetry groups.
Minimal unitary representation of a Lie algebra  exponentiates to a unitary
representation of the corresponding noncompact group over a Hilbert space
of functions depending on the smallest (minimal) number of variables 
possible.  Joseph presented  the
minimal realizations of the complex forms of classical Lie algebras
and of $G_2$ in a Cartan-Weil basis.   The
existence of the minimal unitary representation of $E_{8(8)}$ using Langland's classification was first shown by Vogan
\cite{MR644845}.  The minimal unitary representations of 
simply laced groups were studied by Kazhdan and Savin\cite{MR1159103},
and Brylinski and Kostant
\cite{MR1372999,MR1278630}. 
The minimal representations of quaternionic real forms of exceptional Lie 
groups were later studied by Gross and Wallach \cite{MR1327538}. For a review and  more complete list of references 
on the subject in the mathematics literature prior to 2000, we
refer  to the lectures of Jian-Shu Li~\cite{MR1737731}.

Pioline, Kazhdan and Waldron~\cite{Kazhdan:2001nx} 
reformulated  the minimal unitary representations of simply laced
groups given in \cite{MR1159103} and  gave explicit
realizations of the simple root (Chevalley) generators in terms of
pseudo-differential operators for the simply laced exceptional
groups as well as  the spherical vectors necessary for the
construction of modular forms.

 The first known geometric realization of $E_{8(8)}$ 
 as a  quasiconformal group that   leaves invariant
a generalized light-cone with respect to a quartic distance function  in 57
dimensions was given in \cite{Gunaydin:2000xr}. Quasiconformal realizations exist for various  real forms of all noncompact groups as well as for their complex forms  \cite{Gunaydin:2000xr,Gunaydin:2005zz}. 
 
 Remarkably, the quantization of  geometric quasiconformal action of a noncompact group leads directly to its minimal unitary representation. This was first shown  explicitly for the  maximally split 
exceptional group  $E_{8(8)}$ with the maximal compact subgroup $SO(16)$, which is the U-duality group of maximal supergravity in three dimensions 
\cite{Gunaydin:2001bt}.
 The minimal unitary representation  of  three dimensional U-duality group $E_{8(-24)}$  of
the exceptional supergravity \cite{Gunaydin:1983rk} by quantization of its
 quasiconformal realization  was 
given in \cite{Gunaydin:2004md}.  $E_{8(-24)}$ is a quaternionic real form of $E_8$ with the maximal compact subgroup   $E_7 \times SU(2)$. 

The quasiconformal realizations of  noncompact groups correspond to
natural extensions of   generalized conformal realizations of some
of their subgroups and were studied from a generalized spacetime point of view
in \cite{Gunaydin:2005zz}. The class of  generalized spacetimes studied in \cite{Gunaydin:2005zz} are defined by Jordan
algebras of degree three that contain Minkowskian spacetimes as subspacetimes. 
For example,  spacetimes defined by the generic {\it non-simple}  Jordan family of Euclidean
Jordan algebras of
 degree three describe extensions of the Minkowskian spacetimes
 by an extra ``dilatonic'' coordinate. Their quasiconformal groups are $SO(d+2,4)$, which contain the conformal groups $SO(d,2)$ as subgroups. 
  The generalized spacetimes described by {\it simple } Euclidean Jordan
 algebras of degree three extend the  Minkowskian
 spacetimes in the critical dimensions $(d=3,4,6,10)$ by a dilatonic
\emph{and} extra ($2,4,8,16$) commuting \emph{spinorial}
coordinates, respectively.       Their  quasiconformal groups  are  $\mathrm{F}_{4(4)}$, $\mathrm{E}_{6(2)}$,
 $\mathrm{E}_{7(-5)}$ and $\mathrm{E}_{8(-24)}$, which have the generalized conformal
 subgroups  $Sp(6,\mathbb{R}),
 SU^*(6), SO^*(12)$ and $E_{7(-25)}$, respectively.
The minimal unitary representations of these quasiconformal groups obtained by quantization were given in
\cite{Gunaydin:2005zz,Gunaydin:2004md}.
%%%%%%%%%%%%%%%%%%%%%%%%%%%%%%%%%%%%%%

%%%%%%%%%%%%%%%%%%%%%%%%%%%%%%
In \cite{Gunaydin:2006vz}   a unified
formulation of the minimal unitary representations of certain non-compact real forms of 
 groups  of type $A_2$, $G_2$, $D_4$, $F_4$, $E_6$, $E_7$,
$E_8$ and $Sp\left(2n,\mathbb{R}\right)$ was given. The minimal unitary
representations of $Sp\left(2n,\mathbb{R}\right)$ are simply the
singleton representations. The formulation of minimal unitary representations of 
 noncompact groups $SU\left(m,n\right)$, $SO\left(m,n\right)$,
$SO^*(2n)$ and $SL\left(m,\mathbb{R}\right)$ requires slight modification of the unified construction and was also given
explicitly in \cite{Gunaydin:2006vz}. 
Furthermore, this unified  approach was used  to define and construct the corresponding
minimal representations of non-compact supergroups $G$ whose even
subgroups are of the form $H\times SL(2,\mathbb{R})$ with $H$ 
compact.\footnote{ If $H$ is
also noncompact then the supergroup $G $ does not admit any unitary
representations, in general.}  The unified construction with $H$
simple or Abelian leads to the minimal unitary representations of $G(3),
F(4)$ and $OSp\left(n|2,\mathbb{R}\right)$. The minimal unitary representations of
$OSp\left(n|2,\mathbb{R}\right)$ with  even subgroups $SO(n)\times
Sp(2,\mathbb{R})$ are the singleton representations. The minimal realization of the one parameter family of Lie superalgebras $D\left(2,1;\sigma\right)$ with even subgroup $ SU(2)\times SU(2) \times SU(1,1)$  was also presented in \cite{Gunaydin:2006vz}.

%%%%%%%%%%%%%%%%%%%%%%%%%%%%%%%%%%%%%%tbm
%%%%%%%%%%%%%%%%%%%%%%%%%%%%%%%%%%%%%%%%%%%%%%%%tbm

Unitary representations of rank two quaternionic groups 
 $SU(2,1)$ and $G_{2(2)}$ induced by their
geometric quasiconformal actions were studied in great detail in \cite{Gunaydin:2007qq}. 
The set of unitary representations thus obtained 
 include the quaternionic discrete
series representations that were studied in mathematics literature
using other methods \cite{MR1421947}.  In the construction  of  unitary representations via the quasiconformal approach \cite{Gunaydin:2007qq},
spherical vectors of maximal compact subgroups of $SU(2,1)$ and $G_{2(2)}$ play a fundamental  role.  Authors of \cite{Gunaydin:2007qq} studied  the minimal unitary representations of $SU(2,1)$ and $G_{2(2)}$ obtained by quantization  as well.\footnote{The minrep of $SU(2,1)$   was constructed earlier in \cite{Gunaydin:2001bt}.}
Later in
\cite{Gunaydin:2009dq}  a unified quasiconformal realization of
three dimensional U-duality groups $QConf(J)$ of all $N=2$ MESGTs with
symmetric scalar manifolds defined by Euclidean Jordan algebras $J$  of
degree three was given. These three-dimensional U-duality groups are
$F_{4(4)}$, $E_{6(2)}$, $E_{7(-5)}$, $E_{8(-24)}$ and $SO(n_V+2,4)$. Spherical vectors of the quasiconformal 
actions of all these groups with respect to their maximal compact subgroups as well as the eigenvalues of  their quadratic Casimir
operators were also presented  in \cite{Gunaydin:2009dq}.
 These results were then extended to the
split exceptional groups $E_{6(6)}$, $E_{7(7)}$, $E_{8(8)}$ and
$SO(n+3,m+3)$ in \cite{Gunaydin:2009zz}.  

%%%%%%%%%%%%%%%%%%%%%%%%%%%%%%%%%%%%%%%%%%
In this paper we give a detailed study of the minrep of  $SO(4,2)$ obtained by quantizing its realization as a quasiconformal group that leaves invariant a quartic light-cone in five dimensions, its deformations and their 
supersymmetric extensions. The motivations for our work are multifold. 
First we would like to extend the results of \cite{Gunaydin:2001bt,Gunaydin:2006vz} to construct the minimal unitary 
representations of more general noncompact  supergroups such as  $SU(n,m\,|\,p+q)$ in general. Since the group $SU(2,2)$ is the covering group of $SO(4,2)$,   
the family $SU(2,2\,|\,N)$ corresponds to   four dimensional conformal  or five dimensional anti-de Sitter superalgebras and have important applications to $AdS_5/CFT_4$ dualities \cite{Maldacena:1997re}. 
The noncompact groups that are not  of Hermitian symmetric type, in general,  admit a unique or at most finitely many minimal unitary representations \cite{MR1737731}. The unified approach to  minimal unitary representations given in \cite{Gunaydin:2006vz} is applicable to all noncompact groups including those that are of Hermitian symmetric type. We shall  extend  the quasiconformal formalism of \cite{Gunaydin:2006vz} by introducing a deformation parameter $\zeta$ so as to be able to construct all the ``minimal'' unitary representations of $SU(2,2)$, which is of hermitian symmetric type. We shall refer to the representation with $\zeta =0$ as the minimal unitary representation and the representations with nonzero $\zeta$ as deformations of the minimal representation. For each integer value of $\zeta$ one obtains a unique unitary irreducible representation of $SU(2,2)$. 
We then extend these results to the minimal unitary representations of $SU(2,2\,|\,N)$ and their deformations.   Again in the supersymmetric case, each integer value of the deformation parameter
$\zeta$ leads to  a unique unitary supermultiplet of $SU(2,2\,|\,N))$.    The minimal unitary supermultiplet of $SU(2,2\,|\,N)$ and its deformations turn out to be the doubleton supermultiplets that were constructed and studied using the oscillator method \cite{Gunaydin:1984fk,Gunaydin:1998sw,Gunaydin:1998jc} earlier.  Our results extend to the minreps of $SU(m,n)$ and of $SU(m,n\,|\,p+q)$ and their deformations in a straightforward manner.

Now the unitary representations of  $SO(4,2)$ or its covering group $SU(2,2)$ have been studied very extensively over the last half century. The so-called ladder representations 
of $SU(2,2)$ constructed using bosonic annihilation and creation operators appeared in the physics literature as early as 1960s in at least three different contexts. First, in the formulation of $SO(4,2)$ as a spectrum generating symmetry group of the Hydrogen atom \cite{malkin_manko,nambu1,nambu2,barut_kleinert1,Barut:1967zz}. Second, in hadron physics as symmetry of infinite component fields \cite{barut_kleinert3}. Thirdly in studies of massless wave equations in  four dimensional spacetime, for which we refer to \cite{Mack:1969dg} and the references cited therein.\footnote{ We thank Professor Ivan Todorov for bringing reference \cite{Mack:1969dg} to our attention.}
 A full classification of positive energy unitary representations  of $SU(2,2)$ was given in \cite{Mack:1975je}, to which we refer for the earlier literature on  the subject. A complete classification of all unitary representations (unitary dual) of $SU(2,2)$, which include the positive energy representations, was  given in the mathematics literature \cite{MR645645}. A classification of the positive energy unitary representations of $SU(2,2\,|\,N))$ using the formalism of  Kac \cite{Kac:1977em,Kac:1977qb} was given in \cite{Dobrev:1985vh,Dobrev:1985qv}. 
 
 The minimal unitary representations of symplectic groups $Sp(2n, \mathbb{R})$ are the singleton representations which are known as the metaplectic representations in the mathematics literature. Since the singleton representations of $Sp(2n, \mathbb{R})$ can be realized over the Fock space of bosonic oscillators transforming in the fundamental representation of its maximal compact subgroup $U(n)$, they are also sometimes referred to as the ``oscillator representation.''
The entire Fock space created by the action of  $n$ bosonic creation operators transforming in the fundamental representation of $U(n)$  decomposes into a direct sum of the two singleton representations of $Sp(2n, \mathbb{R})$.  Dirac discovered the two singleton representations of the covering group $Sp(4,\mathbb{R})$ of four dimensional anti-de Sitter group $SO(3,2)$ without using oscillators and referred to them as  remarkable representations of anti-de Sitter group \cite{Dirac:1963ta}. 
The wave functions corresponding to the remarkable representations do not depend on the radial coordinate  of  the four dimensional anti-de Sitter space ($AdS_4$), suggesting that they should be interpreted as living on the boundary of $AdS_4$. The term singletons for these remarkable representations of $SO(3,2)$ was coined by Fronsdal and collaborators later \cite{Flato:1980we,Fronsdal:1981gq,Angelopoulos:1980wg}, who showed that the singleton representations do not have a Poincar\'{e} limit. They also showed these representations have the additional remarkable property that by tensoring two copies of the singleton representations one obtains all the massless representations of $SO(3,2)$ which do have a smooth Poincar\'{e} limit. 

Using oscillators to construct representations of symmetry groups is a time honored tradition in physics.  Here we should stress that using the oscillator method one can construct more general representations than what is commonly referred to as the ``oscillator representation(s)'' in the mathematics literature. For  symplectic groups the term ``oscillator representations'' typically refers to the  singleton (metaplectic) representations of $Sp(2n,\mathbb{R})$. A general method for constructing  more general classes of unitary representations of noncompact groups was formulated in \cite{Gunaydin:1981yq}, which unified and generalized the known  constructions in special cases in the physics literature.  The formulation of  \cite{Gunaydin:1981yq} was later extended to give a general method for  constructing unitary representations of noncompact supergroups in \cite{Bars:1982ep} using bosonic as well as fermionic oscillators. In these generalized formulations of the oscillator method the generators of noncompact groups or supergroups are realized as bilinears of an arbitrary number  $P$  (colors) of  sets of oscillators transforming in an irreducible representation of their maximal compact subgroups or supergroups. 
  For symplectic groups $Sp(2n,\mathbb{R})$ the minimum possible value of $P$ is one and the resulting unitary representations are simply the singletons. If the minimum allowed value of $P_{min}$ is two, the resulting unitary representations were later referred to as doubleton representations. For example, the groups $SU(n,m)$ and $SO^*(2n)$, with maximal compact subgroups $SU(m)\times SU(n) \times U(1)$ and $U(n)$, respectively, admit doubleton representations. There exists only two singleton representations of $Sp(2n,\mathbb{R})$, for which the minimum value of $P_{min}$ is one.  When the minimum allowed number $P_{min}$  of colors is two, one finds an infinite number of doubleton irreducible representations of the respective noncompact groups or supergroups. 
  Since the general oscillator method realizes the generators as bilinear of free bosonic and fermionic oscillators, the tensoring of the resulting representations is very straightforward. Even though the singletons or doubletons themselves do not belong to the discrete series, by tensoring them one obtains unitary representations that belong, in general, to the holomorphic discrete series. 
  
 The Kaluza-Klein spectrum of IIB supergravity over the $AdS_5 \times S^5$ space was first obtained via the oscillator method by simple tensoring of the CPT self-conjugate doubleton supermultiplet of  $N=8$, $AdS_5$ superalgebra $SU(2,2\,|\,4)$ with itself repeatedly and restricting to the CPT self-conjugate short supermultiplets of $SU(2,2\,|\,4)$ \cite{Gunaydin:1984fk}. The CPT self-conjugate doubleton supermultiplet itself decouples from the Kaluza-Klein spectrum as gauge modes. Again  in \cite{Gunaydin:1984fk} it was pointed out that the  CPT self-conjugate doubleton supermultiplet $SU(2,2\,|\,4)$ does not have a Poincar\'{e} limit in five dimensions and its field theory  lives on the boundary of $AdS_5$ on which $SO(4,2)$ acts as a conformal group and that the unique candidate for this theory is the  four dimensional $N=4$ super Yang-Mills theory that is  conformally invariant. Analogous results were obtained for the compactifications of 11 dimensional supergravity over $AdS_4\times S^7$ and $AdS_7 \times S^4$ with the  symmetry superalgebras $OSp(8\,|\,4,\mathbb{R})$ and $OSp(8^*\,|\,4)$ in \cite{Gunaydin:1985tc} and \cite{Gunaydin:1984wc}, respectively.  
 These results have become an integral part of the work on AdS/CFT dualities in M/superstring theory which has seen an exponential growth since the famous paper of Maldacena\cite{Maldacena:1997re}.  The connection  between the minimal representations and the more general  representations of symmetry groups or supergroups obtained by tensoring them lie at the heart of AdS/CFT dualities in a true Wignerian sense.   AdS/CFT dualities have also found applications  in different areas of physics over the last decade.  These developments show the fundamental importance of  the minimal unitary representations of symmetry groups and supergroups  in physics. 
 
The plan of our paper is as follows. In section \ref{sec:quasiconf}  we  review 
the geometric quasiconformal realizations of  groups $SO(d+2,4)$  as 
 invariance groups of a light-cone with respect to a quartic distance 
function in $2d+5$ dimensional space. The minimal unitary realization of 
the Lie algebra of $SO(d+2,4)$, obtained by  
quantizing this geometric 
action over an Hilbert space of functions in $d+3$ variables, is reviewed 
in section \ref{sec:minrepSO(4,2)}.  We then specialize and study the case of $SO(4,2)$ in 
detail. In section \ref{sec:changeofbasis}, we review the minimal unitary 
realization of $SU(2,2)$ as a special case of $SU(n,m)$ given in 
\cite{Gunaydin:2006vz} and show that it coincides with the minrep 
of $SO(4,2)$. We give the K-type decomposition of the minrep of $SU(2,2)$ 
in section \ref{sec:SU2SU2U1} and show that it coincides with the K-type 
decomposition of scalar doubleton representation 
corresponding to a massless conformal scalar field in 4 dimensions 
\cite{Gunaydin:1984fk,Gunaydin:1998sw,Gunaydin:1998jc}.

We then show, in section \ref{sec:deformations}, that there exists a 
one-parameter ($\zeta$) family of  
deformations of the minrep of $SU(2,2)$. For every positive  
(negative)  integer value of the deformation parameter $\zeta$  one obtains 
a positive energy    
unitary irreducible representation of  $SU(2,2)$ corresponding to a  
massless conformal field in four dimensions  transforming in  $\left( 0 
\,,\, \frac{\zeta}{2} \right)$ $\left( \left( -\frac{\zeta}{2} \,,\, 0 
\right) \right)$ representation of $SL ( 2 , \mathbb{C} )$ subgroup of 
$SU(2,2)$. These are simply the doubleton representations of $SU(2,2)$. They were referred to as ladder (or most degenerate discrete series) unitary representations by Mack and Todorov who showed that they remain  irreducible under restriction to the Poincar\'{e} subgroup \cite{Mack:1969dg}.    

In sections \ref{sec:SU(2,2|4)} and \ref{sec:SU(2,2|p+q)}  we give the supersymmetric extension of the minrep of  
$SU(2,2)$ to the minrep $SU ( 2 , 2 \,|\, \mathfrak{p} + \mathfrak{q} )$  
which has a unique irreducible unitary supermultiplet.
For $SU ( 2 , 2 \,|\, 4 )$ the minimal unitary supermultiplet is simply the  
unique CPT self-conjugate massless $N=4$ Yang-Mills supermultiplet in  
four dimensions \cite{Gunaydin:1984fk}. For every  
deformed minimal unitary  irreducible representation of $SU(2,2)$  
there exists a unique extension to a unitary irreducible deformed  
unitary supermultiplet  of $SU ( 2 , 2 \,|\, \mathfrak{p} + \mathfrak{q} 
)$. For $SU ( 2 , 2 \,|\, 4 )$ these supermultiplets turn out to be 
precisely the doubleton supermultiplets constructed and studied in 
\cite{Gunaydin:1998sw,Gunaydin:1998jc} and correspond to massless conformal  
supermultiplets involving higher spin fields.

%%%%%%%%%%%%%%%%%%%%%%%%%%%%%%%%%%%%%%%%%%%%%%%%%%%%%%%%%%%%%%%%%%%%
%%%%%% Section: Quasiconformal realizations and minrep %%%%%%%%%%%%%
%%%%%%%%%%%%%%%%%%%%%%%%%%%%%%%%%%%%%%%%%%%%%%%%%%%%%%%%%%%%%%%%%%%%

\section{Quasiconformal Realizations of $SO\left(d+2,4\right)$ and Their Minimal Unitary Representations }
\label{sec:quasiconf}

\renewcommand{\theequation}{\arabic{section}.\arabic{equation}}
\setcounter{equation}{0}

%%%%%%%%%%%%%%%%%%%%%%%%%%%%%%%%%%%%%%%%%%%%%%%%%%%%%%%%%%%%%%%%%%%%
%%%%%% Subsection: Geometric realization %%%%%%%%%%%%%%%%%%%%%%%%%%%
%%%%%%%%%%%%%%%%%%%%%%%%%%%%%%%%%%%%%%%%%%%%%%%%%%%%%%%%%%%%%%%%%%%%

\subsection{Geometric realizations of $SO\left(d+2,4\right)$ as 
quasiconformal groups}

Lie algebra of $SO \left( d+2 , 4 \right)$ can be given a
5-graded decomposition with respect to its subalgebra $\mathfrak{so}(d,2) \oplus \mathfrak{so}(1,1)$ \cite{Gunaydin:2005zz}
\begin{equation}
  \mathfrak{so}\left(d+2,4\right) = \mathbf{1}^{(-2)} \oplus \left(\mathbf{d+2}, \mathbf{2} \right)^{(-1)} \oplus
    \left(\Delta \oplus \mathfrak{sp}\left(2, \mathbb{R}\right) \oplus \mathfrak{so}\left(d,2\right) \right)
   \oplus \left(\mathbf{d+2}, \mathbf{2} \right)^{(+1)} \oplus  \mathbf{1}^{(+2)}
\end{equation}
where $\Delta$ is the $SO(1,1)$ generator that determines the five grading and non-zero exponents $m$ label the grade of a generator
\[
[ \Delta, \mathfrak{g}^{(m)} ] = m \, \mathfrak{g}^{(m)} \,.
\]
Generators are realized as differential operators acting on a $(2d+5)$
dimensional space $\mathcal{T}$  corresponding to  the Heisenberg subalgebra generated by elements of $( \mathfrak{g}^{(-2)} \oplus
\mathfrak{g}^{(-1)} ) $ subspace, whose coordinates we shall denote as 
$\mathcal{X}= ( X^{\mu, a}$, $x$), where $X^{\mu,a}$ transform in the 
$ (d+2,2)$  representation of $SO(d,2)\times Sp(2,\mathbb{R})$,  with $a=1,2$ and 
$\mu =1,2,..., d+2$,  and  $x$ is a singlet coordinate.

Let $\epsilon_{ab}$ be the symplectic metric of $Sp(2,\mathbb{R})$, and
$\eta_{\mu\nu}$ 
the $SO(d,2)$ invariant metric ($\eta_{\mu\nu} = ( -,-,\dots,-,+,+))$. Then the quartic polynomial in $X^{\mu,a}$ 
\begin{equation}
   \mathcal{I}_4 (X) = \eta_{\mu\nu} \eta_{\rho\tau} \epsilon_{ac}  \epsilon_{bd} X^{\mu, a} X^{\nu, b} X^{\rho, c} X^{\tau, d}
\end{equation}
is   invariant under $SO(d,2) \times Sp(2,\mathbb{R})$ subgroup.

 We shall label the generators belonging to various grade subspaces as follows
 \begin{equation}
 \mathfrak{so}(d+2,4) = K_{-} \oplus U_{\mu,a} \oplus ( \Delta + M_{\mu\nu} + J_{ab} ) \oplus \Tilde{U}^{\mu,a} \oplus K_{+} 
 \end{equation} 
 where $M_{\mu\nu} $ and $J_{ab}$ are the generators of $SO(d,2)$ and $Sp(2,\mathbb{R})$ subgroups, respectively. 
 The  infinitesimal generators of the quasiconformal action of $SO(d+2,4)$ is then given by
\begin{equation}
\begin{split}
   K_+ &= \frac{1}{2} \left( 2 x^2 - \mathcal{I}_4 \right)  \frac{\partial}{\partial x} -
  \frac{1}{4} \frac{\partial \mathcal{I}_4}{ \partial X^{\mu, a} } \eta^{\mu\nu} \epsilon^{ab} \frac{\partial}{\partial X^{\nu, b}} +  x \, X^{\mu, a} \frac{\partial}{\partial X^{\mu, a}} \\
  U_{\mu, a} &= \frac{\partial}{\partial X^{\mu, a}} - \eta_{\mu, \nu} \epsilon_{a b} X^{\nu, b} \frac{\partial}{\partial x} \\
  M_{\mu\nu} &= \eta_{\mu\rho} X^{\rho,a} \frac{\partial}{\partial X^{\nu,a}} -
                \eta_{\nu\rho} X^{\rho,a} \frac{\partial}{\partial X^{\mu,a}} \\
  J_{ab} &= \epsilon_{ac} X^{\mu, c}  \frac{\partial}{\partial X^{\mu,b}} +
            \epsilon_{bc} X^{\mu, c}  \frac{\partial}{\partial X^{\mu,a}} \\
   K_- &= \frac{\partial}{\partial x}  \phantom{ and also}
  \Delta = 2 \, x \frac{\partial}{\partial x} + X^{\mu, a} \frac{\partial}{\partial X^{\mu, a}}
   \phantom{ and also}   \Tilde{U}_{\mu, a} = \left[ U_{\mu, a}, K_+ \right]
\end{split}
\end{equation}
where $\epsilon^{ab}$ denotes the inverse symplectic metric, such that
$\epsilon^{ab} \epsilon_{bc} = {\delta^a}_c$. Using
\begin{equation*}
   \frac{\partial \mathcal{I}_4}{ \partial X^{\mu, a} } = - 4 \, \eta_{\mu\nu} \, \eta_{\lambda\rho} \,
     X^{\nu,b} X^{\lambda, c}  X^{\rho, d} \, \epsilon_{bc} \epsilon_{ad}
\end{equation*}
one obtains the explicit form of $\Tilde{U}^{\mu,a}$
\begin{equation}
\begin{split}
\Tilde{U}_{\mu, a} 
&= \eta_{\mu\nu} \epsilon_{ad} 
   \left( \eta_{\lambda\rho} \epsilon_{bc} 
          X^{\nu,b} X^{\lambda, c} X^{\rho, d} 
          - x X^{\nu, d} \right) \frac{\partial}{\partial x}
   +  x \frac{\partial}{\partial X^{\mu, a}} \\
& \quad
   - \eta_{\mu\nu}\epsilon_{ab} X^{\nu,b} X^{\rho, c}\frac{\partial}{\partial X^{\rho, c}}
         -  \epsilon_{ad} \eta_{\lambda \rho} X^{\rho, d} X^{\lambda, c} \frac{\partial}{\partial X^{\mu, c}}
         \\ 
& \quad
  +  \epsilon_{ad} \eta_{\mu\nu} X^{\rho, d} X^{\nu, b} \frac{\partial}{\partial X^{\rho, b}}
          +  \eta_{\mu\nu} \epsilon_{bc} X^{\nu, b} X^{\rho, c} \frac{\partial}{\partial X^{\rho, a}} \,.
\end{split}
\end{equation}

These generators satisfy the following commutation relations:
\begin{subequations}
\label{eq:sod4algebra}
\begin{equation}
\begin{split}
    \left[ M_{\mu\nu}, M_{\rho\tau} \right] &= \eta_{\nu\rho} M_{\mu\tau} - \eta_{\mu\rho} M_{\nu\tau} + \eta_{\mu\tau} M_{\nu\rho} - \eta_{\nu\tau} M_{\mu\rho} \\
    \left[ J_{ab} , J_{cd} \right] &= \epsilon_{cb} J_{ad} + \epsilon_{ca} J_{bd} + \epsilon_{db} J_{ac} + \epsilon_{da} J_{bc} \\
   \left[ \Delta, K_\pm \right] &= \pm 2 \, K_\pm \phantom{also} \left[ K_-, K_+ \right] = \Delta \\
   \left[ \Delta, U_{\mu, a} \right] &= - U_{\mu, a} \phantom{also}
   \left[ \Delta, \Tilde{U}_{\mu, a} \right] =  \Tilde{U}_{\mu, a} \\
   \left[ U_{\mu, a} , K_+ \right] &= \Tilde{U}_{\mu, a} \phantom{also }
   \left[ \Tilde{U}_{\mu, a} , K_- \right] = -U_{\mu, a} \\
    \left[ U_{\mu, a}, U_{\nu, b} \right] &= 2 \, \eta_{\mu\nu} \epsilon_{ab} K_- \phantom{also}
    \left[ \Tilde{U}_{\mu, a}, \Tilde{U}_{\nu, b} \right] = 2 \, \eta_{\mu\nu} \epsilon_{ab} K_+
\end{split}
\end{equation}
\begin{equation}
\begin{split}
   \left[ M_{\mu\nu}, U_{\rho, a} \right] &= \eta_{\nu\rho} U_{\mu, a} - \eta_{\mu\rho} U_{\nu, a}  \phantom{also}
   \left[ M_{\mu\nu}, \Tilde{U}_{\rho, a} \right] = \eta_{\nu\rho} \Tilde{U}_{\mu, a} - \eta_{\mu\rho} \Tilde{U}_{\nu, a} \\
   \left[ J_{ab}, U_{\mu, c} \right] &= \epsilon_{cb} U_{\mu,a} + \epsilon_{ca} U_{\mu, b}  \phantom{also}
   \left[ J_{ab}, \Tilde{U}_{\mu, c} \right] = \epsilon_{cb} \Tilde{U}_{\mu,a} + \epsilon_{ca} \Tilde{U}_{\mu, b} \\
\end{split}
\end{equation}
\begin{equation}
  \left[ U_{\mu, a} , \Tilde{U}_{\nu, b} \right] = \eta_{\mu\nu} \epsilon_{ab} \, \Delta - 2 \, \epsilon_{ab} M_{\mu\nu} -
          \eta_{\mu\nu} J_{ab}
\end{equation}
\end{subequations}

One defines the ``length'' (norm)  of a vector $\mathcal{X}= (X^{\mu, a}, x)$ as
\begin{equation}
   \ell \left(\mathcal{X} \right) = \mathcal{I}_4\left(X\right) + 2 \, x^2
\end{equation}
and the ``symplectic'' difference of
 two vectors $\mathcal{X}$ and $\mathcal{Y}$ in the $(2d+5)$ dimensional space $\mathcal{T}$ as
\begin{equation}
    \delta \left( \mathcal{X}, \mathcal{Y} \right) = \left( X^{\mu, a} - Y^{\mu, a}, x-y - \eta_{\mu\nu}\epsilon_{ab} X^{\mu,a} Y^{\nu,b} \right) \,.
\end{equation}
 
 The ``quartic distance'' between any two points labelled by vectors $\mathcal{X}$ and $\mathcal{Y}$ is defined as
 \begin{equation}
d(\mathcal{X},\mathcal{Y}) := \ell \left( \delta \left( \mathcal{X}, \mathcal{Y} \right) \right) \,.
\end{equation}
Under the quasiconformal action of the generators of $SO(d+2,4)$ the distance function transforms as 
\begin{equation}
\begin{split}
   \Delta  d \left(\mathcal{X}, \mathcal{Y}\right) &=
        4 \, d\left(\mathcal{X}, \mathcal{Y}\right)\\
   \Tilde{U}_{\mu, a} d\left(\mathcal{X}, \mathcal{Y}\right) &=
         - 2 \, \eta_{\mu\nu} \epsilon_{ab} \left( X^{\nu, b} + Y^{\nu,b} \right)
          d\left(\mathcal{X}, \mathcal{Y}\right) \\
    K_+  d\left(\mathcal{X}, \mathcal{Y}\right) &=
         2  \left( x + y \right)
         d\left(\mathcal{X}, \mathcal{Y}\right) \\
   M_{\mu\nu} \, d\left(\mathcal{X}, \mathcal{Y}\right) &=  0 \\
    J_{ab} \, d\left(\mathcal{X}, \mathcal{Y}\right) &=  0  \\ 
    U_{\mu,a} \, d\left(\mathcal{X}, \mathcal{Y}\right) &=  0 \\
    K_{-}\, d\left(\mathcal{X}, \mathcal{Y}\right) &=  0 \,.
\end{split}
\end{equation}
They imply that light-like separations
\[ d \left(\mathcal{X}, \mathcal{Y}\right) =0 \]
are left invariant under the quasiconformal action.
In other words, quasiconformal action of $SO(d+2,4)$ leaves the light-cone with respect to the quartic distance function invariant.   

By replacing the $SO(d,2)$ invariant metric $\eta_{\mu\nu}$ by an $SO(p,q)$ invariant metric in the above construction, one can obtain the quasiconformal realization of 
$SO(p+2,q+2)$ in a straightforward manner. Of these noncompact real forms, only the groups of the form $SO(n,4)$ admit quaternionic discrete series representations and the groups of the form $SO(m,2)$ admit holomorphic discrete series representations. Of course, the group $SO(4,2)$ admits  quaternionic as well as holomorphic discrete series representations.

%%%%%%%%%%%%%%%%%%%%%%%%%%%%%%%%%%%%%%%%%%%%%%%%%%%%%%%%%%%%%%%%%%%%
%%%%%% Subsection: Minrep from quantization %%%%%%%%%%%%%%%%%%%%%%%%
%%%%%%%%%%%%%%%%%%%%%%%%%%%%%%%%%%%%%%%%%%%%%%%%%%%%%%%%%%%%%%%%%%%%

\subsection{Minimal unitary representations of $SO\left(d+2,4\right)$ from 
the quantization of their quasiconformal realizations}

  Minimal unitary representations of noncompact  groups can be obtained by the quantization of their geometric realizations as quasiconformal groups \cite{Gunaydin:2001bt, Gunaydin:2004md, Gunaydin:2005zz, Gunaydin:2006vz,Gunaydin:2007qq}. In this section we shall review the minimal unitary representations of quaternionic orthogonal groups   
    $SO\left(d+2, 4\right)$ obtained by the quantization of their
geometric realizations as  quasiconformal groups given in the previous section following \cite{Gunaydin:2005zz,Gunaydin:2006vz} closely. Let $X^\mu$ and $P_\mu$ be the quantum mechanical 
 coordinate  and momentum  operators on 
$\mathbb{R}^{\left(2,d\right)}$ satisfying the canonical commutation relations
\begin{equation}
    \left[ X^\mu, P_\nu\right] = i \, {\delta^\mu_\nu} \,.
\end{equation}
The grade $-2$  and $-1$ generators of $SO(d+2,4)$ form an Heisenberg algebra 
\begin{equation}\label{heisenberg}
 \left[ U_{\mu, a}, U_{\nu, b} \right] = 2 \, \eta_{\mu\nu} \epsilon_{ab} K_- 
 \end{equation}
with $K_-$ playing the role of the central charge.  We shall relabel the generators and define
\begin{equation}
U_{\mu,1} \equiv U_{\mu}
\qquad \qquad \qquad
U_{\mu,2} \equiv V_{\mu}
\end{equation}
and realize the Heisenberg algebra (\ref{heisenberg}) in terms of coordinate and momentum operators $X^\mu$, $P_\mu$ and an extra
``central charge coordinate'' $x$:
\begin{equation}
\begin{split}
U_\mu = x P_\mu
&\qquad \qquad
V^\mu = x X^\mu
\\
&K_- = \frac{1}{2} x^2
\end{split}
\end{equation}

\begin{equation}
\commute{V^\mu}{U_\nu} = 2 i \, {\delta^\mu}_\nu \, K_-
\end{equation}

By introducing the quantum mechanical momentum operator $p$  conjugate to the central charge coordinate $x$
\begin{equation}
   \left[ x, p \right] = i
\end{equation}
one can realize the grade zero generators of $SO(d+2,4)$ as bilinears of canonically conjugate pairs of coordinates and momenta \cite{Gunaydin:2005zz,Gunaydin:2006vz}:
\begin{equation}
\begin{split}
M_{\mu\nu} &= i \, \eta_{\mu\rho} X^\rho P_\nu - i \, \eta_{\nu\rho} X^\rho P_\mu  \\
J_0 &= \frac{1}{2} \left( X^\mu P_\mu + P_\mu X^\mu \right) \\
J_- &= X^\mu X^\nu \eta_{\mu\nu} \\
J_+ &= P_\mu P_\nu \eta^{\mu\nu} \\
\Delta & = \frac{1}{2} ( x p + p x )
\end{split}
\label{gradezeroSO(d+2,2)}
\end{equation}
They  satisfy the commutation relations 
\begin{equation}
\begin{split}
       \left[ M_{\mu\nu}, M_{\rho\tau} \right] &= \eta_{\nu\rho} M_{\mu\tau} - \eta_{\mu\rho} M_{\nu\tau} + \eta_{\mu\tau} M_{\nu\rho} - \eta_{\nu\tau} M_{\mu\rho}   \\
     \left[ J_0, J_\pm \right] &= \pm 2 i \, J_\pm \qquad
          \left[ J_-, J_+ \right] = 4 i \, J_0 \,.
\end{split}
\end{equation}
The coordinate $X^\mu$ and momentum  $P_\mu$ operators transform contravariantly and covariantly under $SO(d,2)$ subgroup generated by $M_{\mu\nu}$, respectively, and form doublets of the symplectic group
$Sp(2,\mathbb{R}$):
\begin{equation}
     \begin{aligned}
          \left[ J_0, V^\mu \right] &=  - i \, V^\mu \\
          \left[ J_0, U_\mu \right] &=  + i \, U_\mu
     \end{aligned}
         \quad
     \begin{aligned}
         \left[ J_-, V^\mu \right] &= 0 \\
         \left[ J_-, U_\mu \right] &= 2 i \, \eta_{\mu\nu} V^\nu
     \end{aligned}
          \quad
     \begin{aligned}
         \left[ J_+, V^\mu \right] &= -2 i \, \eta^{\mu\nu} U_\nu \\
         \left[ J_+, U_\mu \right] &= 0
     \end{aligned}
\end{equation}

There is a normal ordering ambiguity in defining the quantum operator corresponding to the quartic invariant. We shall choose the quantum quartic invariant \cite{Gunaydin:2005zz}
\begin{equation}
\begin{split}
\mathcal{I}_4 &= \left(X^\mu X^\nu \eta_{\mu\nu}\right) \left(P_\mu P_\nu \eta^{\mu\nu}\right) +
    \left(P_\mu P_\nu \eta^{\mu\nu}\right) \left(X^\mu X^\nu \eta_{\mu\nu}\right) \\
 & \qquad  -  \left(X^\mu P_\mu\right)\left( P_\nu X^\nu \right) -
       \left(P_\mu X^\mu\right) \left( X^\nu P_\nu\right) \,.
       \end{split}
       \end{equation}
Using the quartic invariant, one defines the grade +2 generator as
\begin{equation}
    K_+ = \frac{1}{2} p^2 + \frac{1}{4 \, x^2} \left( \mathcal{I}_4 + \frac{(d+2)^2+3}{2} \right) \,.
\end{equation}
Then the  grade $+1$ generators  are obtained by commutations
\begin{equation}
   \Tilde{V}^\mu = -i \left[ V^\mu, K_+ \right]  \qquad
   \Tilde{U}_\mu = -i \left[ U_\mu, K_+ \right]
\end{equation}
which explicitly read as follows
\begin{equation}
\begin{split}
    \Tilde{V}^\mu &= p \, X^\mu + \frac{1}{2 \, x} \left(  P_\nu X^\lambda X^\rho  +
     X^\lambda X^\rho  P_\nu \right) \eta^{\mu\nu} \eta_{\lambda\rho} \\
       & \quad - \frac{1}{4 \, x} \left[ X^\mu \left(X^\nu P_\nu + P_\nu X^\nu\right) +
      \left(X^\nu P_\nu + P_\nu X^\nu \right) X^\mu \right] \\
    \Tilde{U}_\mu &= p \, P_\mu - \frac{1}{2 \, x}
   \left(  X^\nu P_\lambda P_\rho  +  P_\lambda P_\rho  X^\nu \right) \eta_{\mu\nu} \eta^{\lambda\rho} \\
       & \quad + \frac{1}{4 \, x} \left[ P_\mu \left(X^\nu P_\nu + P_\nu X^\nu\right) +
      \left(X^\nu P_\nu + P_\nu X^\nu\right) P_\mu\right] \,.
\end{split}
\end{equation}
The generators in $\mathfrak{g}^{+1} \oplus
\mathfrak{g}^{+2} $ subspace form an  Heisenberg algebra isomorphic to (\ref{heisenberg})
\begin{equation}
  \left[ \Tilde{V}^\mu , \Tilde{U}_\nu \right] = 2 i \, {\delta^\mu}_\nu K_+ \qquad
 V^\mu = i \left[ \Tilde{V}^\mu, K_- \right] \qquad  U_\mu = i \left[ \Tilde{U}_\mu, K_- \right] .
\end{equation}
Commutators $\left[ \mathfrak{g}^{-1}, \mathfrak{g}^{+1} \right]$
close into grade zero subspace $\mathfrak{g}^0$: 
\begin{equation}
\begin{split}
   \left[ U_\mu, \Tilde{U}_\nu \right] &= i \, \eta_{\mu\nu} J_+ \qquad
   \left[ V^\mu, \Tilde{V}^\nu \right] = i \, \eta^{\mu\nu} J_- \\
   \left[ V^\mu, \Tilde{U}_\nu \right] &= 2 \, \eta^{\mu\rho} M_{\rho\nu} + i \, {\delta^\mu}_\nu \left( J_0 +  \Delta \right)
   \\
    \left[ U_\mu, \Tilde{V}^\nu \right] &= - 2 \, \eta^{\nu\rho} M_{\mu\rho} + i \, {\delta^\nu}_\mu \left( J_0 -  \Delta \right)
\end{split}
\end{equation}
 $\Delta$ is the  generator that determines the 5-grading:
\begin{equation}
\begin{aligned}
\left[ K_-, K_+ \right] &= i \, \Delta
\\
\left[ \Delta, U_\mu \right] &= - i \, U_\mu
\\
\left[ \Delta, \Tilde{U}_\mu \right] &=  i \, \Tilde{U}_\mu
\end{aligned}
\qquad \qquad
\begin{aligned}
\left[ \Delta, K_\pm \right] &= \pm 2 i \, K_\pm
\\
\left[ \Delta, V^\mu \right] &= - i \, V^\mu
\\
\left[ \Delta, \Tilde{V}^\mu \right] &=  i \, \Tilde{V}^\mu 
\end{aligned}
\end{equation}
The quadratic Casimir operators of subalgebras
$\mathfrak{so}\left(d,2\right)$, $\mathfrak{sp}\left(2,
\mathbb{R}\right)_J$  generated by $J_{ab}$ of grade-zero subspace, and
$\mathfrak{sp}\left(2, \mathbb{R}\right)_K$ generated by $K_{\pm}$
and $\Delta$ are given by
\begin{equation}
 \begin{aligned}
   M_{\mu\nu} M^{\mu\nu} &= - \mathcal{I}_4 - 2 \, \left(d+2\right) \cr
   J_- J_+ +  J_+ J_- - 2  \left(J_0\right)^2 &= \mathcal{I}_4 + \frac{1}{2}\left(d+2\right)^2 \cr
    K_- K_+ + K_+ K_- - \frac{1}{2} \Delta^2 &= \frac{1}{4} \mathcal{I}_4 + \frac{1}{8} \left(d+2\right)^2 \,.
 \end{aligned}
\end{equation}
They all reduce to the quartic invariant operator $\mathcal{I}_4$ modulo some additive
constants. Furthermore, the following identity satisfied by the bilinears of grade $\pm 1$ generators 
\begin{equation}
  \left( U_\mu \Tilde{V}^\mu + \Tilde{V}^\mu U_\mu  - V^\mu \Tilde{U}_\mu - \Tilde{U}_\mu V^\mu \right) =
     2 \mathcal{I}_4 + \left(d+2\right)\left(d+6\right)
\end{equation}
prove the existence of  a family of degree 2 polynomials in
the enveloping algebra of $\mathfrak{so}\left(d+2,4\right)$ that
degenerate to a $c$-number for the  minimal unitary  realization, in
accordance with Joseph's theorem \cite{MR0342049}:
\begin{equation}\label{eq:JosephIdeal}
\begin{split}
   &M_{\mu\nu} M^{\mu\nu} + \kappa_1 \left( J_- J_+ +  J_+ J_- - 2  \left(J_0\right)^2 \right)
   +  4 \kappa_2 \left(K_- K_+ + K_+ K_- - \frac{1}{2} \Delta^2\right) \\&-
     \frac{1}{2}\left(\kappa_1+\kappa_2-1\right)
    \left( U_\mu \Tilde{V}^\mu + \Tilde{V}^\mu U_\mu  - V^\mu \Tilde{U}_\mu - \Tilde{U}_\mu V^\mu \right)
  \\ & \qquad \qquad
  = \frac{1}{2}\left(d+2\right)\left( d + 2 - 4\left(\kappa_1+\kappa_2\right) \right)
\end{split}
\end{equation}

The quadratic Casimir of $\mathfrak{so}\left(d+2, 4\right)$
corresponds to the choice $2 \kappa_1 = 2 \kappa_2 = -1$ in \eqref{eq:JosephIdeal}. Hence
the eigenvalue of the quadratic Casimir for the minimal unitary
representation is equal to $\frac{1}{2} \left(d+2\right) \left(d+6\right)$.
This minimal unitary representation is realized on the Hilbert space of square integrable functions in
$(d+3)$ variables.

%%%%%%%%%%%%%%%%%%%%%%%%%%%%%%%%%%%%%%%%%%%%%%%%%%%%%%%%%%%%%%%%%%%%
%%%%%% Section: Minrep of SO(4,2) %%%%%%%%%%%%%%%%%%%%%%%%%%%%%%%%%%
%%%%%%%%%%%%%%%%%%%%%%%%%%%%%%%%%%%%%%%%%%%%%%%%%%%%%%%%%%%%%%%%%%%%

\section{Minimal Unitary Realization of the $4D$ Conformal Group $SO(4,2)$ over  the Hilbert Space of  $L^2$ Functions of  Three Variables}
\label{sec:minrepSO(4,2)}

With applications to AdS/CFT dualities in mind, we shall study the minrep of $SO(4,2)$ in detail.  By setting $d=0$ in the construction of previous section we get the following 5-graded decomposition of $SO(4,2)$ generators in the minrep:

\begin{equation}
\mathfrak{so}(2,4) = K_- \oplus
                     \left[ U_\mu \oplus
                            V^\mu \right] \oplus
                     \left[ \Delta \oplus
                            J_{0 , \pm} \oplus
                            M_{12} \right] \oplus
                     \left[ \tilde{U}_\mu \oplus
                            \tilde{V}^\mu \right] \oplus
                     K_+
\end{equation}
where $\mu , \nu , \dots = 1,2$ and $\eta_{\mu\nu} = \delta_{\mu\nu}$. The  5-grading is determined by the $SO(1,1)$ generator  
\begin{equation*}
\Delta = \frac{1}{2} \left( x p + p x \right) \,.
\end{equation*}

On the other hand, $\mathfrak{so}(4,2)$ has a 3-grading 
\begin{equation}
\mathfrak{so}(2,4) = \mathfrak{N}^- \oplus \mathfrak{N}^0 \oplus \mathfrak{N}^+ 
\end{equation} 
with respect to the noncompact generator 
\begin{equation}
\mathcal{D} = \Delta + J_0 = \frac{1}{2} \left( xp + px + X^\mu P_\mu + P_\mu X^\mu \right) \,.
\end{equation}
Explicitly we have
\begin{equation}
\mathfrak{so}(2,4) = \left[ K_- \oplus
                            J_-  \oplus
                            V^\mu \right] \oplus
                     \left[ \mathcal{D} \oplus \mathcal{B} \oplus
                            M_{\mu \nu} \oplus
                            U_\mu  \oplus
                            \tilde{V}^\mu \right] \oplus
                     \left[ \tilde{U}_\mu \oplus
                            J_+  \oplus
                            K_+ \right]
\end{equation}
where $\mathcal{B} = \Delta - J_0 $ and 
\[ \mathfrak{N}^{0} = \mathfrak{so}(3,1) \oplus \mathfrak{so}(1,1)_{\mathcal{D}} \,. \]
The generators of $\mathfrak{so}(3,1)$ subalgebra are $\mathcal{B}$, 
$M_{\mu \nu}$, $U_\mu $ and  $\tilde{V}^\mu$.
  We shall refer to this as the noncompact 3-grading.    
                         
  Furthermore, the Lie algebra of $\mathfrak{so}(2,4)$ has  a 3-grading with respect to the compact generator 
  \begin{equation}
  H = \frac{1}{2} \left[ \left( K_+ + K_- \right)
                         + \frac{1}{2} ( J_+ + J_- ) \right]
  \end{equation}
   such that
    \begin{equation} 
  \mathfrak{so}(2,4) =    \mathfrak{C}^{-} \oplus \left[ \mathfrak{so}(4) \oplus \mathfrak{so}(2) \right] \oplus 
   \mathfrak{C}^+ \,.
   \end{equation}
    In this decomposition,
\begin{equation}
\begin{split}
\mathfrak{C}^{0}
 &= \mathfrak{so}(4) \oplus \mathfrak{so}(2)
  = \left[ M_{\mu \nu} \oplus
           \left( \left( K_+ + K_- \right)
           - \frac{1}{2} \left( J_+ + J_- \right) \right) \oplus
           \left( U_\mu - \eta_{\mu \nu} \tilde{V}^\nu \right)
    \right.
\\
 & \qquad \qquad \qquad \qquad \qquad \left.
   \oplus
           \left( \tilde{U}_\mu + \eta_{\mu \nu} V^\nu \right) \right] 
   \oplus 
    \frac{1}{2} \left[ \left( K_+ + K_- \right)
    + \frac{1}{2} \left( J_+ + J_- \right)
     \right] \\
\mathfrak{C}^{+}
 &= \left[ \Delta - i \left( K_+ - K_- \right) \right] \oplus
    \left[ J_0 - \frac{i}{2} \left( J_+ - J_- \right) \right] \oplus
    \left[ \frac{1}{2} \left( U_\mu + \eta_{\mu \nu} \tilde{V}^\nu \right)
           - \frac{i}{2} \left( \tilde{U}_\mu - \eta_{\mu \nu} V^\nu \right)
    \right] \\
\mathfrak{C}^{-}
 &= \left[ \Delta + i \left( K_+ - K_- \right) \right] \oplus
    \left[ J_0 + \frac{i}{2} \left( J_+ - J_- \right) \right] \oplus
    \left[ \frac{1}{2} \left( U_\mu + \eta_{\mu \nu} \tilde{V}^\nu \right)
           + \frac{i}{2} \left( \tilde{U}_\mu - \eta_{\mu \nu} V^\nu \right)
    \right] \,.
\end{split}
\end{equation}
We shall refer to this grading as the compact 3-grading. 

 The $\mathfrak{so}
(4)$ generators  in the subspace $\mathfrak{C}^0$ are given by 
\begin{equation}
\begin{split}
\tilde{M}_{4 \mu}
 &:= \frac{1}{2} \left( U_\mu - \eta_{\mu \nu} \tilde{V}^\nu \right)
\qquad \qquad
\tilde{M}_{12}
  := - i M_{12} \\
\tilde{M}_{\mu 3}
 &:= \frac{1}{2} \left( \tilde{U}_\mu + \eta_{\mu \nu} V^\nu \right)
\qquad \qquad
\tilde{M}_{43}
  := \frac{1}{2} \left[ \left( K_+ + K_- \right)
                        - \frac{1}{2} \left( J_+ + J_- \right) \right]
\end{split}
\end{equation}
and satisfy the $\mathfrak{so}(4)$ algebra
\begin{equation}
\commute{\tilde{M}_{AB}}{\tilde{M}_{CD}}
= i \left( \delta_{AC} \tilde{M}_{BD} - \delta_{AD} \tilde{M}_{BC}
           - \delta_{BC} \tilde{M}_{AD} + \delta_{BD} \tilde{M}_{AC} \right)
\end{equation}
where $A,B,\dots=1,2,3,4$.

To analyse the decomposition of the minimal unitary representation of $SO(2,4)$  into K-finite vectors of its maximal compact subgroup, let us 
introduce the oscillators
\begin{equation}
\begin{aligned}
a &:= \frac{1}{\sqrt{2}} \left( x + i \, p \right)
\\
a^\dag &:= \frac{1}{\sqrt{2}} \left( x - i \, p \right)
\end{aligned}
\qquad
\begin{aligned}
b &:= \frac{1}{\sqrt{2}} \left( X^1 + i \, P_1 \right)
\\
b^\dag &:= \frac{1}{\sqrt{2}} \left( X^1 - i \, P_1 \right)
\end{aligned}
\qquad
\begin{aligned}
c &:= \frac{1}{\sqrt{2}} \left( X^2 + i \, P_2 \right)
\\
c^\dag &:= \frac{1}{\sqrt{2}} \left( X^2 - i \, P_2 \right)
\end{aligned}
\end{equation}
so that
\begin{equation}
\begin{aligned}
x &= \frac{1}{\sqrt{2}} \left( a^\dag + a \right)
\\
p &= \frac{i}{\sqrt{2}} \left( a^\dag - a \right)
\end{aligned}
\qquad
\begin{aligned}
X^1 &= \frac{1}{\sqrt{2}} \left( b^\dag + b \right)
\\
P_1 &= \frac{i}{\sqrt{2}} \left( b^\dag - b \right)
\end{aligned}
\qquad
\begin{aligned}
X^2 &= \frac{1}{\sqrt{2}} \left( c^\dag + c \right)
\\
P_2 &= \frac{i}{\sqrt{2}} \left( c^\dag - c \right) \,.
\end{aligned}
\end{equation}
These oscillators satisfy the commutation relations:
\begin{equation}
\begin{aligned}
\commute{a}{a^\dag} &= 1
\\
\commute{x}{p} &= i
\end{aligned}
\qquad
\begin{aligned}
\commute{b}{b^\dag} &= 1
\\
\commute{X^1}{P_1} &= i
\end{aligned}
\qquad
\begin{aligned}
\commute{c}{c^\dag} &= 1
\\
\commute{X^2}{P_2} &= i
\end{aligned}
\end{equation}

The quartic invariant operator  $\mathcal{I}_4$ takes on a simple form in terms of the oscillators $b$, $c$:
\begin{equation}
\mathcal{I}_4
 = - 2 \left( b^\dag c - b c^\dag \right)^2 - 4
\end{equation}
The $\mathfrak{so}(2)$ generator in $\mathfrak{C}^{0}$, which plays the 
role of the AdS energy 
\cite{Gunaydin:1984fk,Gunaydin:1998sw,Gunaydin:1998jc}, is given by:
\begin{equation}
\begin{split}
H &= \frac{1}{2}
     \left[
     \left( K_+ + K_- \right) + \frac{1}{2} \left( J_+ + J_- \right) 
     \right]
\\
  &= \frac{1}{2}
     \left[
     a^\dag a + b^\dag b + c^\dag c
     - \frac{1}{2 \, x^2} \left( b^\dag c - b c^\dag \right)^2
     - \frac{1}{8 \, x^2}
     + \frac{3}{2}
     \right]
\end{split}
\label{SO2generator}
\end{equation}
while 
the    $\mathfrak{so}(4)$ generators  in terms of these oscillators become
\begin{equation}
\begin{split}
\tilde{M}_{12}
 &= - i M_{12}
\\
 &= - i \left( b^\dag c - b c^\dag \right)
\\
\tilde{M}_{43}
 &= \frac{1}{2}
    \left[ \left( K_+ + K_- \right)
           - \frac{1}{2} \left( J_+ + J_- \right) \right]
\\
 &= \frac{1}{2} \left( a^\dag a - b^\dag b - c^\dag c \right)
    - \frac{1}{4 \, x^2} \left( b^\dag c - b c^\dag \right)^2
    - \frac{1}{16 \, x^2}
    - \frac{1}{4}
\\
\tilde{M}_{41}
 &= \frac{1}{2} \left( U_1 - \tilde{V}^1 \right)
\\
 &= \frac{i}{2} \left( a b^\dag - a^\dag b \right)
    - \frac{i}{2 \sqrt{2} \, x}
      \left( b^\dag c - b c^\dag \right) \left( c^\dag + c \right)
    + \frac{i}{4 \sqrt{2} \, x}
      \left( b^\dag + b \right)
\\
\tilde{M}_{42}
 &= \frac{1}{2} \left( U_2 - \tilde{V}^2 \right)
\\
 &= \frac{i}{2} \left( a c^\dag - a^\dag c \right)
    + \frac{i}{2 \sqrt{2} \, x}
      \left( b^\dag c - b c^\dag \right) \left( b^\dag + b \right)
    + \frac{i}{4 \sqrt{2} \, x}
      \left( c^\dag + c \right)
\\
\tilde{M}_{13}
 &= \frac{1}{2} \left( \tilde{U}_1 + V^1 \right)
\\
 &= \frac{1}{2} \left( a b^\dag + a^\dag b \right)
    - \frac{1}{2 \sqrt{2} \, x}
      \left( b^\dag c - b c^\dag \right) \left( c^\dag - c \right)
    + \frac{1}{4 \sqrt{2} \, x}
      \left( b^\dag - b \right)
\\
\tilde{M}_{23}
 &= \frac{1}{2} \left( \tilde{U}_2 + V^2 \right)
\\
 &= \frac{1}{2} \left( a c^\dag + a^\dag c \right)
    + \frac{1}{2 \sqrt{2} \, x}
      \left( b^\dag c - b c^\dag \right) \left( b^\dag - b \right)
    + \frac{1}{4 \sqrt{2} \, x}
      \left( c^\dag - c \right) \,.
\end{split}
\end{equation}

It is useful to list the following commutators between $a$, $a^\dag$ and 
$1/x$, $1/x^2$:
\begin{equation}
\begin{aligned}
\commute{a}{\frac{1}{x}} &= - \frac{1}{\sqrt{2} \, x}
\\
\commute{a}{\frac{1}{x^2}} &= - \frac{\sqrt{2}}{x^3}
\end{aligned}
\qquad \qquad \quad
\begin{aligned}
\commute{a^\dag}{\frac{1}{x}} &= \frac{1}{\sqrt{2} \, x}
\\
\commute{a^\dag}{\frac{1}{x^2}} &= \frac{\sqrt{2}}{x^3}
\end{aligned}
\end{equation}

The generators that belong to the grade $+1$ subspace $\mathfrak{C}^{+}$ have the following form:
\begin{equation}
\begin{split}
\Delta - i \left( K_+ - K_- \right)
 &= i \, a^\dag a^\dag
    + \frac{i}{2 \, x^2} \left( b^\dag c - b c^\dag \right)^2
    + \frac{i}{8 \, x^2}
\\
\frac{1}{2} \left[ \left( U_1 + \tilde{V}^1 \right)
                     - i \left( \tilde{U}_1 - V^1 \right) \right]
 &= i \, a^\dag b^\dag
    + \frac{i}{2 \sqrt{2} \, x}
      \left[ c^\dag \left( b^\dag c - b c^\dag \right)
             + \left( b^\dag c - b c^\dag \right) c^\dag \right]
\\
\frac{1}{2} \left[ \left( U_2 + \tilde{V}^2 \right)
                     - i \left( \tilde{U}_2 - V^2 \right) \right]
 &= i \, a^\dag c^\dag
    - \frac{i}{2 \sqrt{2} \, x}
      \left[ b^\dag \left( b^\dag c - b c^\dag \right)
             + \left( b^\dag c - b c^\dag \right) b^\dag \right]
\\
J_0 - \frac{i}{2} \left( J_+ - J_- \right)
 &= i \, \left( b^\dag b^\dag + c^\dag c^\dag \right)
\\
\end{split}
\end{equation}
and those that belong to grade $-1$ subspaces $\mathfrak{C}^{-}$ are:
\begin{equation}
\begin{split}
\Delta + i \left( K_+ - K_- \right)
 &= - i \, a \, a
    - \frac{i}{2 \, x^2} \left( b^\dag c - b c^\dag \right)^2
    - \frac{i}{8 \, x^2}
\\
\frac{1}{2} \left[ \left( U_1 + \tilde{V}^1 \right)
                     + i \left( \tilde{U}_1 - V^1 \right) \right]
 &= - i \, a \, b
    + \frac{i}{2 \sqrt{2} \, x}
      \left[ c \left( b^\dag c - b c^\dag \right)
             + \left( b^\dag c - b c^\dag \right) c \right]
\\
\frac{1}{2} \left[ \left( U_2 + \tilde{V}^2 \right)
                     + i \left( \tilde{U}_2 - V^2 \right) \right]
 &= - i \, a \, c
    - \frac{i}{2 \sqrt{2} \, x}
      \left[ b \left( b^\dag c - b c^\dag \right)
             + \left( b^\dag c - b c^\dag \right) b \right]
\\
J_0 + \frac{i}{2} \left( J_+ - J_- \right)
 &= - i \, \left( b \, b + c \, c \right)
\\
\end{split}
\end{equation}
They satisfy
\begin{equation}
\commute{H}{\mathfrak{C}^{+}} = + \, \mathfrak{C}^{+}
\qquad \qquad \qquad
\commute{H}{\mathfrak{C}^{-}} = - \, \mathfrak{C}^{-} \,.
\end{equation}

Now the Lie algebra of $SO(4)$ is not simple and can be written as a direct sum 
\begin{equation}
\mathfrak{so}(4) = \mathfrak{su}(2)_L \oplus \mathfrak{su}(2)_R 
\end{equation}
where the 
generators of the two $\mathfrak{su}(2)$ subalgebras  are as follows:
\begin{equation}
\begin{aligned}
L_1 &= \frac{1}{2} \left( \tilde{M}_{23} - \tilde{M}_{41} \right)
\\
R_1 &= \frac{1}{2} \left( \tilde{M}_{23} + \tilde{M}_{41} \right)
\end{aligned}
\qquad
\begin{aligned}
L_2 &= \frac{1}{2} \left( \tilde{M}_{31} - \tilde{M}_{42} \right)
\\
R_2 &= - \frac{1}{2} \left( \tilde{M}_{31} + \tilde{M}_{42} \right)
\end{aligned}
\qquad
\begin{aligned}
L_3 &= \frac{1}{2} \left( \tilde{M}_{12} - \tilde{M}_{43} \right)
\\
R_3 &= - \frac{1}{2} \left( \tilde{M}_{12} + \tilde{M}_{43} \right)
\end{aligned}
\end{equation}

Therefore the raising and lowering generators of the two 
$\mathfrak{su}(2)$'s are given by:
\begin{equation}
\begin{split}
L_+
 &= \frac{1}{\sqrt{2}} \left( L_1 + i L_2 \right)
  = - \frac{i}{2 \sqrt{2}}
    \left[ a
           + \frac{i}{\sqrt{2} \, x} \left( b^\dag c - b c^\dag \right)
           + \frac{1}{2 \sqrt{2} \, x}
    \right]
    \left( b^\dag + i \, c^\dag \right)
\\
L_-
 &= \frac{1}{\sqrt{2}} \left( L_1 - i L_2 \right)
  = \frac{i}{2 \sqrt{2}}
    \left[ a^\dag
           + \frac{i}{\sqrt{2} \, x} \left( b^\dag c - b c^\dag \right)
           - \frac{1}{2 \sqrt{2} \, x}
    \right]
    \left( b - i \, c \right)
\\
R_+
 &= \frac{1}{\sqrt{2}} \left( R_1 + i R_2 \right)
  = \frac{i}{2 \sqrt{2}}
    \left[ a
           - \frac{i}{\sqrt{2} \, x} \left( b^\dag c - b c^\dag \right)
           + \frac{1}{2 \sqrt{2} \, x}
    \right]
    \left( b^\dag - i \, c^\dag \right)
\\
R_-
 &= \frac{1}{\sqrt{2}} \left( R_1 - i R_2 \right)
  = - \frac{i}{2 \sqrt{2}}
    \left[ a^\dag
           - \frac{i}{\sqrt{2} \, x} \left( b^\dag c - b c^\dag \right)
           - \frac{1}{2 \sqrt{2} \, x}
    \right]
    \left( b + i \, c \right)
\end{split}
\end{equation}
while the remaining generators are given by:
\begin{equation}
\begin{split}
L_3 
&= - \frac{1}{4} \left( a^\dag a - b^\dag b - c^\dag c \right)
   + \frac{1}{8 \, x^2} \left( b^\dag c - b c^\dag \right)^2
   - \frac{i}{2} \left( b^\dag c - b c^\dag \right)
   + \frac{1}{32 \, x^2}
   + \frac{1}{8}
\\
R_3
&= - \frac{1}{4} \left( a^\dag a - b^\dag b - c^\dag c \right)
   + \frac{1}{8 \, x^2} \left( b^\dag c - b c^\dag \right)^2
   + \frac{i}{2} \left( b^\dag c - b c^\dag \right)
   + \frac{1}{32 \, x^2}
   + \frac{1}{8}
\end{split}
\end{equation}
They satisfy the commutation relations:
\begin{equation}
\begin{split}
\commute{L_+}{L_-} = L_3
& \qquad \qquad \qquad
\commute{L_3}{L_\pm} = \pm \, L_\pm
\\
\commute{R_+}{R_-} = R_3
& \qquad \qquad \qquad
\commute{R_3}{R_\pm} = \pm \, R_\pm
\end{split}
\end{equation}

Interestingly the two quadratic Casimir operators
\begin{equation}
L^2 = L_+ L_- + L_- L_+ + {L_3}^2
\qquad \qquad \qquad
R^2 = R_+ R_- + R_- R_+ + {R_3}^2
\end{equation}
turn out to be equal and are given by
\begin{equation}
\begin{split}
L^2
 = R^2
&= \frac{1}{16} \left( a^\dag a + b^\dag b + c^\dag c \right)^2
   + \frac{3}{16} \left( a^\dag a + b^\dag b + c^\dag c \right)
   - \frac{1}{64 \, x^2} \left( a^\dag a + b^\dag b + c^\dag c \right)
\\
& \quad
   - \frac{1}{16 \, x^2} \left( a^\dag a + b^\dag b + c^\dag c \right)
                         \left( b^\dag c - b c^\dag \right)^2
   + \frac{\sqrt{2}}{32 \, x^3} \left( a^\dag - a \right)
                                \left( b^\dag c - b c^\dag \right)^2
\\
& \quad
   + \frac{\sqrt{2}}{128 \, x^3} \left( a^\dag - a \right)
   + \frac{1}{64 \, x^4} \left( b^\dag c - b c^\dag \right)^4
   + \frac{13}{128 \, x^4} \left( b^\dag c - b c^\dag \right)^2
\\
& \quad
   - \frac{3}{32 \, x^2} \left( b^\dag c - b c^\dag \right)^2
   + \frac{25}{1024 \, x^4}
   - \frac{3}{128 \, x^2}
   - \frac{7}{64}
\\
&= \frac{1}{16}
    \left[ a^\dag a + b^\dag b + c^\dag c
           - \frac{1}{2 \, x^2} \left( b^\dag c - b c^\dag \right)^2
           - \frac{1}{8 \, x^2}
           + \frac{3}{2}
    \right]^2
   - \frac{1}{4}
\\
&= \frac{1}{4} \left( H^2 - 1 \right)= \mathcal{J} (\mathcal{J}+1 )
\end{split}
\label{L^2R^2inbc}
\end{equation}
with
\eq
\mathcal{J}= \frac{1}{2} \left( H - 1 \right)
\en
where $H$ is the $\mathfrak{so}(2)$ generator, as given in equation 
(\ref{SO2generator}).

%%%%%%%%%%%%%%%%%%%%%%%%%%%%%%%%%%%%%%%%%%%%%%%%%%%%%%%%%%%%%%%%%%%%
%%%%%% Section: Change of basis and minrep of SU(2,2) %%%%%%%%%%%%%%
%%%%%%%%%%%%%%%%%%%%%%%%%%%%%%%%%%%%%%%%%%%%%%%%%%%%%%%%%%%%%%%%%%%%

\section{Change of Basis and the Minimal Unitary Realization of $SU(2,2)$}
\label{sec:changeofbasis}

%%%%%%%%%%%%%%%%%%%%%%%%%%%%%%%%%%%%%%%%%%%%%%%%%%%%%%%%%%%%%%%%%%%%
%%%%%% Subsection: Minrep of SU(2,2) %%%%%%%%%%%%%%%%%%%%%%%%%%%%%%%
%%%%%%%%%%%%%%%%%%%%%%%%%%%%%%%%%%%%%%%%%%%%%%%%%%%%%%%%%%%%%%%%%%%%

\subsection{Minimal unitary realization of $\mathfrak{su}(2,2)$}

Minimal unitary realizations of $SU(n,m)$ obtained from quantization of 
their quasiconformal realizations were given in \cite{Gunaydin:2006vz}, 
which we review here for $SU(2,2)$.
The Lie algebra $\mathfrak{su}(2,2)$ admits a 5-grading with respect 
to its subalgebra $\mathfrak{su}(1,1) \oplus \mathfrak{u}(1)\oplus \mathfrak{so}(1,1) $:
\begin{equation}
\mathfrak{su}(2,2) = \mathbf{1}^{(-2)} \oplus \mathbf{4}^{(-1)} \oplus
                     \left[ \mathfrak{su}(1,1) \oplus
                            \mathfrak{u}(1) \oplus
                            \Delta \right] \oplus
                     \mathbf{4}^{(+1)} \oplus \mathbf{1}^{(+2)}
\end{equation}
where $J_m^n$, $U$ and $\Delta$ are the $SU(1,1)$,  $U(1)$ and $SO(1,1)$ generators, respectively. In \cite{Gunaydin:2006vz} the corresponding generators are labelled as
\begin{equation}
\mathfrak{su}(2,2) = E \oplus ( E^1,E^2,E_1,E_2) \oplus
                     \left[ J_m^n ,
                           U ,
                            \Delta \right] \oplus
                     (F^1,F^2,F_1,F_2)  \oplus F \,.
\end{equation}
The covariant $SU(1,1)$ generators $J_m^n$
are realized as bilinears of 2 pairs of oscillators $d $ and $g$ 
satisfying\footnote{ They are related to the covariant oscillators of \cite{Gunaydin:2006vz}
as $a_1= d$ and $a^2= g$ .}
\begin{equation}
\commute{d}{d^\dag} = \commute{g}{g^\dag} = 1
\end{equation}
as follows
\begin{equation}
J_1^2 = d \, g
\qquad \quad
J_2^1 = - d^\dag g^\dag
\qquad \quad
J^1_1 = -J^2_2 =  \frac{1}{2} ( N_d +N_g +1) 
\end{equation}
 where $N_d = d^\dag d$ and $N_g = g^\dag g$.
Furthermore, these $J^m_n$ can be related to $J_{\pm,0}$ (defined in 
equation (\ref{gradezeroSO(d+2,2)}) as follows:
\begin{equation}
J_+ = J^1_1 + J^1_2 - J^2_1 - J^2_2
\qquad \quad
J_- = J^1_1 - J^1_2 + J^2_1 - J^2_2
\qquad \quad
J_0 = - i \left( J^1_2 + J^2_1 \right)
\end{equation}

The generators of $\mathfrak{su}(2,2)$ 
in the minimal unitary realization take the form:
\begin{equation}
\begin{split}
J^1_1
= \frac{1}{2} \left( N_d + N_g + 1 \right)
\qquad
J^2_2
= - \frac{1}{2} \left( N_d + N_g + 1 \right)
& \qquad
J^1_2
= - d^\dag g^\dag
\qquad
J^2_1
= d \, g
\\
U
= N_d - N_g
\qquad \qquad \qquad &
\Delta
= \frac{1}{2} \left( x p + p x \right)
\\
E
= \frac{1}{2} x^2
&
\\
E^1
= x \, d^\dag
\qquad \qquad
E^2
= x \, g
\qquad \qquad &
E_1
= x \, d
\qquad \qquad
E_2
= - x \, g^\dag
\\
F
= \frac{1}{2} p^2
  + \frac{1}{2 \, x^2} & \left[ \left( N_d - N_g \right)^2
                                - \frac{1}{4} \right]
\\
F^1
= d^\dag 
  \left[ p + \frac{i}{x} \left( N_d - N_g + \frac{1}{2} \right) \right]
\qquad \qquad &
F^2
= g 
  \left[ p + \frac{i}{x} \left( N_d - N_g + \frac{1}{2} \right) \right]
\\
F_1
= d 
  \left[ p - \frac{i}{x} \left( N_d - N_g - \frac{1}{2} \right) \right]
\qquad \qquad &
F_2
= - g^\dag 
  \left[ p - \frac{i}{x} \left( N_d - N_g - \frac{1}{2} \right) \right]
\end{split}
\end{equation}
where $x$ is again the singlet coordinate of the quasiconformal realization and $p$ is its conjugate momentum.

%%%%%%%%%%%%%%%%%%%%%%%%%%%%%%%%%%%%%%%%%%%%%%%%%%%%%%%%%%%%%%%%%%%%
%%%%%% Subsection: From SO(4,2) to SU(2,2) %%%%%%%%%%%%%%%%%%%%%%%%%
%%%%%%%%%%%%%%%%%%%%%%%%%%%%%%%%%%%%%%%%%%%%%%%%%%%%%%%%%%%%%%%%%%%%

\subsection{ From $SO(4,2)$ to $SU(2,2)$ }
The minrep of $SO(4,2)$ to the minrep of $SU(2,2)$ reviewed above  are related very simply by 
 rewriting the oscillators $d$ ($d^\dag$), $g$ ($g^\dag$) in terms of $b$ 
 ($b^\dag$), $c$ ($c^\dag$) as 
\begin{equation}
d = \frac{1}{\sqrt{2}} \left( b - i \, c \right)
\qquad \qquad
g = \frac{1}{\sqrt{2}} \left( b + i \, c \right) \,.
\end{equation}
It is trivial to verify
\begin{equation*}
\commute{d}{d^\dag} = \commute{g}{g^\dag} = 1
\qquad \quad
\commute{d}{g} = \commute{d}{g^\dag} = 0 \,.
\end{equation*}
Then
\begin{equation}
X^1 P_2 - X^2 P_1
    = - i \left( b^\dag c - b c^\dag \right)
    = N_d - N_g
\end{equation}
and
\begin{equation}    
b^\dag b + c^\dag c = N_d + N_g     
\end{equation}
where $N_d = d^\dag d$ and $N_g = g^\dag g$. In terms of these new 
oscillators, the quartic invariant becomes
\begin{equation}
\mathcal{I}_4 = 2 \left( N_d - N_g \right)^2 - 4 \,.
\end{equation}

The $\mathfrak{so}(2)$ generator in $\mathfrak{C}^{0}$, which plays the 
role of the ``energy'' operator (Hamiltonian), becomes
\begin{equation}
\begin{split}
H &= \frac{1}{2}
     \left[
     \left( K_+ + K_- \right) + \frac{1}{2} \left( J_+ + J_- \right)
     \right]
\\
  &= \frac{1}{2}
     \left[
     N_a + N_d + N_g
     + \frac{1}{2 \, x^2} \left( N_d - N_g \right)^2
     - \frac{1}{8 \, x^2}
     + \frac{3}{2}
     \right]
\\
  &= \frac{1}{2}
     \left[ N_d + \frac{1}{2} \right]
     + \frac{1}{2}
       \left[ N_g + \frac{1}{2} \right]
     + \frac{1}{2}
       \left[ N_a + \frac{1}{2} + \frac{G}{x^2} \right] \\
  &= ~~~~~~ H_d ~~~~~~
     + ~~~~~~ H_g ~~~~~~
     + ~~~~~~~~ H_\odot
\end{split}
\label{Hamiltonian}
\end{equation}
where
\begin{equation}
G = \frac{1}{2} \left( N_d - N_g \right)^2 - \frac{1}{8} \,.
\label{SingularPotential}
\end{equation}
Note that $H_d$ and $H_g$ correspond to Hamiltonians of non-singular harmonic oscillators, 
while $H_\odot$ corresponds to a singular harmonic oscillator with a 
potential function $V \left( x \right) = G / x^2$.  $H_\odot$ also arises 
as the Hamiltonian of conformal quantum mechanics \cite{de Alfaro:1976je} with $G$ 
playing the role of coupling constant \cite{Gunaydin:2001bt}. In some 
literature it is also referred to as the isotonic oscillator 
\cite{Casahorran:1995vt,carinena-2007}.
The lowest energy state (vacuum) of the full Hamiltonian is simply the tensor product state of the vacua of $d$ and $g$ type oscillators with the lowest energy state of $H_\odot$.   

Following the literature on singular or isotonic oscillators, we then  introduce the operators 
\begin{equation}
A_{\mathcal{L}} = a - \frac{\mathcal{L}}{\sqrt{2} \, x}
\qquad \qquad \qquad
A_{\mathcal{L}}^\dag = a^\dag - \frac{\mathcal{L}}{\sqrt{2} \, x}
\end{equation}
where 
\begin{equation}
\mathcal{L} = N_d - N_g - \frac{1}{2} 
\end{equation}
which we will refer to as singular (isotonic) oscillators.

These isotonic oscillators satisfy the following commutation relations:
\begin{equation}
\begin{split}
\commute{A_{\mathcal{L}}}{A_{\mathcal{L}^\prime}}
 &= - \frac{\left( \mathcal{L} - \mathcal{L}^\prime \right)}{2 \, x^2} \\
\commute{A_{\mathcal{L}}^\dag}{A_{\mathcal{L}^\prime}^\dag}
 &= + \frac{\left( \mathcal{L} - \mathcal{L}^\prime \right)}{2 \, x^2} \\
\commute{A_{\mathcal{L}}}{A_{\mathcal{L}^\prime}^\dag}
 &= 1 + \frac{\left( \mathcal{L} + \mathcal{L}^\prime \right)}{2 \, x^2}
\end{split}
\end{equation}

In terms of these isotonic oscillators, we can write the singular harmonic 
oscillator of the Hamiltonian as
\begin{equation}
H_\odot = \frac{1}{2}
          \left[
          A_{\mathcal{L}+1}^\dag A_{\mathcal{L}+1}
          + \mathcal{L}
          + \frac{3}{2}
          \right]
        = \frac{1}{2}
          \left[
          A_{\mathcal{L}} A_{\mathcal{L}}^\dag
          + \mathcal{L}
          - \frac{1}{2}
          \right]
\en
and the coupling constant as
\eq
G = \frac{1}{2} \mathcal{L} \left( \mathcal{L} + 1 \right) \,.
\end{equation}

The generators of  $\mathfrak{so}(4)$ in $\mathfrak{C}^{(0)}$  can then be 
expressed in terms of the oscillators $A_{\mathcal{L}}$, $d$ and $g$ as
\begin{equation}
\begin{split}
\tilde{M}_{12}
 &= N_d - N_g
\\
\tilde{M}_{43}
 &= \frac{1}{2} \left( N_a - N_d - N_g \right)
    + \frac{G}{2 \, x^2}
    - \frac{1}{4}
\\
\tilde{M}_{41}
 &= \frac{i}{2 \sqrt{2}} \left( a d^\dag - a^\dag d \right)
    - \frac{i}{4 \, x} \left( N_d - N_g \right) \left( d^\dag - d \right)
    + \frac{i}{8 \, x} \left( d^\dag + d \right)
\\
 & \quad
    + \frac{i}{2 \sqrt{2}} \left( a g^\dag - a^\dag g \right)
    + \frac{i}{4 \, x} \left( N_d - N_g \right) \left( g^\dag - g \right)
    + \frac{i}{8 \, x} \left( g^\dag + g \right)
\\
 &= \frac{i}{2 \sqrt{2}}
    \left[ A_{\mathcal{L}} \, d^\dag
           - A_{\mathcal{L} + 1}^\dag \, d
           + A_{- \left( \mathcal{L} + 1 \right)} \, g^\dag
           - A_{- \mathcal{L}}^\dag \, g
    \right]
\\
\tilde{M}_{42}
 &= \frac{1}{2 \sqrt{2}} \left( a d^\dag + a^\dag d \right)
    - \frac{1}{4 \, x} \left( N_d - N_g \right) \left( d^\dag + d \right)
    + \frac{1}{8 \, x} \left( d^\dag - d \right)
\\
 & \quad
    - \frac{1}{2 \sqrt{2}} \left( a g^\dag + a^\dag g \right)
    - \frac{1}{4 \, x} \left( N_d - N_g \right) \left( g^\dag + g \right)
    - \frac{1}{8 \, x} \left( g^\dag - g \right)
\\
 &= \frac{1}{2 \sqrt{2}}
    \left[ A_{\mathcal{L}} \, d^\dag
           + A_{\mathcal{L} + 1}^\dag \, d
           - A_{- \left( \mathcal{L} + 1 \right)} \, g^\dag
           - A_{- \mathcal{L}}^\dag \, g
    \right]
\\
\tilde{M}_{13}
 &= \frac{1}{2 \sqrt{2}} \left( a d^\dag + a^\dag d \right)
    - \frac{1}{4 \, x} \left( N_d - N_g \right) \left( d^\dag + d \right)
    + \frac{1}{8 \, x} \left( d^\dag - d \right)
\\
 & \quad
    + \frac{1}{2 \sqrt{2}} \left( a g^\dag + a^\dag g \right)
    + \frac{1}{4 \, x} \left( N_d - N_g \right) \left( g^\dag + g \right)
    + \frac{1}{8 \, x} \left( g^\dag - g \right)
\\
 &= \frac{1}{2 \sqrt{2}}
    \left[ A_{\mathcal{L}} \, d^\dag
           + A_{\mathcal{L} + 1}^\dag \, d
           + A_{- \left( \mathcal{L} + 1 \right)} \, g^\dag
           + A_{- \mathcal{L}}^\dag \, g
    \right]
\\
\tilde{M}_{23}
 &= - \frac{i}{2 \sqrt{2}} \left( a d^\dag - a^\dag d \right)
    + \frac{i}{4 \, x} \left( N_d - N_g \right) \left( d^\dag - d \right)
    - \frac{i}{8 \, x} \left( d^\dag + d \right)
\\
 & \quad
    + \frac{i}{2 \sqrt{2}} \left( a g^\dag - a^\dag g \right)
    + \frac{i}{4 \, x} \left( N_d - N_g \right) \left( g^\dag - g \right)
    + \frac{i}{8 \, x} \left( g^\dag + g \right)
\\
 &= - \frac{i}{2 \sqrt{2}}
    \left[ A_{\mathcal{L}} \, d^\dag
           - A_{\mathcal{L} + 1}^\dag \, d
           - A_{- \left( \mathcal{L} + 1 \right)} \, g^\dag
           + A_{- \mathcal{L}}^\dag \, g
    \right] \,.
\end{split}
\label{SO4}
\end{equation}

For reasons that will become evident later we shall work with the following linear combination of generators belonging to the grade $+1$ subspace $\mathfrak{C}^+$ of $\mathfrak{su}(2,2)$:
\begin{equation}
\begin{split}
B^1 &=\Delta - i \left( K_+ - K_- \right)
     = i \left( a^\dag a^\dag - \frac{G}{x^2} \right)
     = i \, A_{- \mathcal{L}}^\dag \, A_{\mathcal{L}}^\dag
     = i \, A_{\mathcal{L} + 1}^\dag \,
            A_{- \left( \mathcal{L} + 1 \right)}^\dag
\\
B^2 &= \frac{1}{2} \left[ \left( U_1 + \tilde{V}^1 \right)
                   - i \left( \tilde{U}_1 - V^1 \right) \right]
       + \frac{i}{2} \left[ \left( U_2 + \tilde{V}^2 \right)
                     - i \left( \tilde{U}_2 - V^2 \right) \right]
\\
    &= \sqrt{2} \, i \left[ a^\dag
                         + \frac{1}{\sqrt{2} \, x} \left( N_d - N_g \right)
                         - \frac{1}{2 \sqrt{2} \, x} \right] d^\dag
  = \sqrt{2} \, i \, A_{- \mathcal{L}}^\dag \, d^\dag
  = \sqrt{2} \, i \, d^\dag \, A_{- \left( \mathcal{L} + 1 \right)}^\dag
\end{split}
\label{g+1generators}
\end{equation}
\begin{equation*}
\begin{split}
B^3 &= \frac{1}{2} \left[ \left( U_1 + \tilde{V}^1 \right)
                   - i \left( \tilde{U}_1 - V^1 \right) \right]
       - \frac{i}{2} \left[ \left( U_2 + \tilde{V}^2 \right)
                     - i \left( \tilde{U}_2 - V^2 \right) \right]
\\
    &= \sqrt{2} \, i \left[ a^\dag
                         - \frac{1}{\sqrt{2} \, x} \left( N_d - N_g \right)
                         - \frac{1}{2 \sqrt{2} \, x} \right] g^\dag
  = \sqrt{2} \, i \, A_{\mathcal{L} + 1}^\dag \, g^\dag
  = \sqrt{2} \, i \, g^\dag \, A_{\mathcal{L}}^\dag
\\
B^4 &= J_0 - \frac{i}{2} \left( J_+ - J_- \right)
     = 2 \, i \, d^\dag \, g^\dag
\end{split}
\end{equation*}

They satisfy the following commutation relations with the energy operator 
$H$ given in equation (\ref{Hamiltonian})
\begin{equation}
\commute{H}{B^i} = B^i  \qquad \qquad i=1,2,3,4.
\en
Furthermore we have the important relation
\begin{equation}
B^3 B^2 = B^4 B^1
\end{equation}
which is valid for the minrep, but is not valid in general. This constraint satisfied by the operators in the minrep will be  important  for its decomposition into  K-finite vectors!

The $\mathfrak{C}^{-}$ generators are given by
\begin{equation}
\begin{split}
B_1 &= \Delta + i \left( K_+ - K_- \right)
     = - i \left( a a - \frac{G}{x^2} \right)
  = - i \, A_{\mathcal{L}} \, A_{- \mathcal{L}}
  = - i \, A_{- \left( \mathcal{L} + 1 \right)} \, A_{\mathcal{L} + 1}
\\
B_2 &= \frac{1}{2} \left[ \left( U_1 + \tilde{V}^1 \right)
                   + i \left( \tilde{U}_1 - V^1 \right) \right]
       - \frac{i}{2} \left[ \left( U_2 + \tilde{V}^2 \right)
                     + i \left( \tilde{U}_2 - V^2 \right) \right]
\\
 &= - \sqrt{2} \, i \, d
      \left[ a
             + \frac{1}{\sqrt{2} \, x} \left( N_d - N_g \right)
             - \frac{1}{2 \sqrt{2} \, x} \right]
  = - \sqrt{2} \, i \, d \, A_{- \mathcal{L}}
  = - \sqrt{2} \, i \, A_{- \left( \mathcal{L} + 1 \right)} \, d
\\
B_3 &= \frac{1}{2} \left[ \left( U_1 + \tilde{V}^1 \right)
                   + i \left( \tilde{U}_1 - V^1 \right) \right]
       + \frac{i}{2} \left[ \left( U_2 + \tilde{V}^2 \right)
                     + i \left( \tilde{U}_2 - V^2 \right) \right]
\\
 &= - \sqrt{2} \, i \, g
      \left[ a
             - \frac{1}{\sqrt{2} \, x} \left( N_d - N_g \right)
             - \frac{1}{2 \sqrt{2} \, x} \right]
  = - \sqrt{2} \, i \, g \, A_{\mathcal{L} + 1}
  = - \sqrt{2} \, i \, A_{\mathcal{L}} \, g
\\
B_4 &= J_0 + \frac{i}{2} \left( J_+ - J_- \right)
     = - 2 \, i \, g \, d \,.
\end{split}
\label{g-1generators}
\end{equation}

The generators of the two $\mathfrak{su}(2)$ subalgebras of 
$\mathfrak{so}(4)$, in terms of the oscillators $A_{\mathcal{L}}$, $d$ and 
$g$ have the following form:
\begin{equation}
\begin{split}
L_+
 &= - \frac{i}{2}
    \left[ a
           - \frac{1}{\sqrt{2} \, x} \left( N_d - N_g \right)
           + \frac{1}{2 \sqrt{2} \, x}
    \right]
    d^\dag
  = - \frac{i}{2} \, A_{\mathcal{L}} \, d^\dag
\\
L_-
 &= \frac{i}{2}
    \left[ a^\dag
           - \frac{1}{\sqrt{2} \, x} \left( N_d - N_g \right)
           - \frac{1}{2 \sqrt{2} \, x}
    \right]
    d
  = \frac{i}{2} \, A_{\mathcal{L} + 1}^\dag \, d
\\
L_3 
&= - \frac{1}{2} \left( H - 1 \right) + N_d
\end{split}
\end{equation}
\begin{equation}
\begin{split}
R_+
 &= \frac{i}{2}
    \left[ a
           + \frac{1}{\sqrt{2} \, x} \left( N_d - N_g \right)
           + \frac{1}{2 \sqrt{2} \, x}
    \right]
    g^\dag
  = \frac{i}{2} \, A_{- ( \mathcal{L} + 1)} \, g^\dag
\\
R_-
 &= - \frac{i}{2}
    \left[ a^\dag
           + \frac{1}{\sqrt{2} \, x} \left( N_d - N_g \right)
           - \frac{1}{2 \sqrt{2} \, x}
    \right]
    g
  = - \frac{i}{2} \, A_{- \mathcal{L}}^\dag \, g \\
R_3
&= - \frac{1}{2} \left( H - 1 \right) + N_g
\end{split}
\end{equation}

Their quadratic Casimir operators take the form 
\begin{equation}
\begin{split}
L^2
 = R^2
&= \frac{1}{16} \left[ \left( N_a + N_d + N_g \right)
                       + \frac{G}{x^2}
                       + \frac{3}{2}
                \right]^2
   - \frac{1}{4}
\\
&= \frac{1}{4} \left( H^2 - 1 \right) \,.
\end{split}
\label{L^2R^2indg}
\end{equation}

%%%%%%%%%%%%%%%%%%%%%%%%%%%%%%%%%%%%%%%%%%%%%%%%%%%%%%%%%%%%%%%%%%%%
%%%%%% Subsection: SU(1,1) subgroup and isotonic oscillators %%%%%%%
%%%%%%%%%%%%%%%%%%%%%%%%%%%%%%%%%%%%%%%%%%%%%%%%%%%%%%%%%%%%%%%%%%%%

\subsection{ $SU(1,1)_L$ subgroup of $SU(2,2)$ generated by the isotonic (singular)  oscillators} 

Consider the singular harmonic oscillator part of the Hamiltonian (\ref{Hamiltonian}), which in coordinate representations has the form:
\begin{equation}
H_\odot
 = \frac{1}{2}
   \left[
   a^\dag a + \frac{1}{2} + \frac{G}{x^2}
   \right]
 = \frac{1}{4} \left( x^2 + p^2 \right) + \frac{G}{2 \, x^2}
 = \frac{1}{4} \left( x^2 - \frac{\partial^2}{\partial x^2} \right) 
   +\frac{G}{2 \, x^2} \,.
\label{SingularHamiltonian}
\end{equation}
Together with the operators $B^1$ and $B_1$ belonging to $\mathfrak{C}^+$ and $\mathfrak{C}^{-}$, respectively,
\begin{equation}
\begin{split}
B^1
 &= i \left( a^\dag \, a^\dag - \frac{G}{x^2} \right)
  = \frac{i}{2} \left( x - i p \right)^2 - i \frac{G}{x^2}
  = \frac{i}{2} \left( x^2 - 2 x \frac{\partial}{\partial x} + \frac{\partial^2}{\partial x^2} - 1 \right) 
    - i \frac{G}{x^2} \\
B_1
 &= - i \left( a \, a - \frac{G}{x^2} \right)
  = - \frac{i}{2} \left( x + i p \right)^2 + i \frac{G}{x^2}
  = - \frac{i}{2} \left( x^2 + 2 x \frac{\partial}{\partial x} + \frac{\partial^2}{\partial x^2}
                         + 1 \right) 
    + i \frac{G}{x^2}
\end{split}
\end{equation}
it
generates a distinguished  $\mathfrak{su}(1,1)_L$ subalgebra\footnote{This is the $SU(1,1)$ subgroup generated by the longest root vector. Hence the  subscript $L$.}
\begin{equation}
\commute{B_1}{B^1} = 8 \, H_\odot
\qquad \qquad
\commute{H_\odot}{B^1} = + \, B^1
\qquad \qquad
\commute{H_\odot}{B_1} = - \, B_1 \,.
\end{equation}
The lowest energy state $\psi_0^{(\alpha)} \left( x \right)$ of this 
singular harmonic oscillator Hamiltonian must satisfy 
\eq
B_1 \, \psi_0^{(\alpha)} \left( x \right) = 0
\en
 whose solution is \cite{MR858831}
\begin{equation}
\psi_0^{(\alpha)} \left( x \right) = C_0 \, x^\alpha e^{-x^2/2}
\en
where 
\eq
\alpha = \frac{1}{2} + \left( 2 \, g + \frac{1}{4} \right)^{\frac{1}{2}}
\end{equation}
and $C_0$ is a normalization constant. Note that $g$ is defined as
\eq
g = \frac{1}{2} \left( n_d - n_g \right)^2 - \frac{1}{8}
\en
where $n_d$ and $n_g$ are the eigenvalues of the number operators $N_d$ 
and $N_g$. Thus we have
\begin{equation}
\alpha = \frac{1}{2} + \left| n_d - n_g \right| \,.
\end{equation}

The normalizability of the state imposes the constraint
\eq
\alpha \geq \frac{1}{2} \,.
\en

Clearly, $\psi_0^{(\alpha)} \left( x \right)$ is an eigenstate of $H_\odot$ with 
eigenvalue $E_{\odot,0}^{(\alpha)}$ given by
\begin{equation}
H_\odot \, \psi_0^{(\alpha)} \left( x \right)
 = E_{\odot,0}^{(\alpha)} \, \psi_0^{(\alpha)} \left( x \right)
\qquad \quad \mbox{where} \quad
E_{\odot,0}^{(\alpha)} = \frac{1}{4} \left( 2 \, \alpha + 1 \right) \,.
\end{equation}
Acting on a tensor product state of $\psi_0^{(\alpha)}$ with the 
eigenstates of the number operators $N_d$ and $N_g$, one may obtain more 
eigenstates of $H_\odot$.

The lowest ``energy'' normalizable eigenstate of $H_\odot$ 
corresponds to the case $n_d = n_g$ (therefore $\alpha = \frac{1}{2}$). 
All the higher ``energy'' eigenstates of $H_\odot$ can be obtained from $\psi_0^{(1/2)} 
\left( x \right)$ by acting on it repeatedly with the raising generator 
$B^1$,
\begin{equation}
\psi_n^{(1/2)} \left( x \right)
 = C_n \, \left( B^1 \right)^n \psi_0^{(1/2)} \left( x \right)
\end{equation}
where $C_n$ are normalization constants, and they have energies 
$E_{\odot,n}$:
\begin{equation}
H_\odot \, \psi_n^{(1/2)} \left( x \right)
 = E_{\odot,n}^{(1/2)} \, \psi_n^{(1/2)} \left( x \right)
 \end{equation}
where 
\eq
E_{\odot,n}^{(1/2)} = E_{\odot,0}^{(1/2)} + n
            = \frac{1}{2} \left| n_d - n_g \right| + n + \frac{1}{2} \,.
\end{equation}

%%%%%%%%%%%%%%%%%%%%%%%%%%%%%%%%%%%%%%%%%%%%%%%%%%%%%%%%%%%%%%%%%%%%
%%%%%% Section: SU(2) x SU(2) x U(1) decomposition %%%%%%%%%%%%%%%%%
%%%%%%%%%%%%%%%%%%%%%%%%%%%%%%%%%%%%%%%%%%%%%%%%%%%%%%%%%%%%%%%%%%%%

\section{$SU(2) \times SU(2) \times U(1)$  Decomposition  of the Minrep of 
$SU(2,2)$ and the Scalar Doubleton}
\label{sec:SU2SU2U1}

Let us label the states that belong to the  Fock spaces of the oscillator 
$d$ as
\eq
\ket{n_d} = \frac{1}{\sqrt{n_d !}} \left( d^\dag \right)^{n_d} \ket{0}
\en
 and similarly the states $\ket{n_g}$ for oscillators $g$.
 As a basis of  the Hilbert space of the minrep we shall consider tensor product states
 \eq
 \ket{\psi_n^{(1/2)} ; n_d , n_g} = \ket{\psi_n^{(1/2)}} \otimes \ket{n_d} \otimes \ket{n_g}
 \en
  where $\ket{\psi_n^{(1/2)}}$ is the state vector corresponding to $\psi_n^{(1/2)}$ defined above. It is an eigenstate of the energy operator $H$ that determines the 3-grading of $SU(2,2)$
  \eq
  H \ket{\psi_n^{(1/2)} ; n_d , n_g} = E \ket{\psi_n^{(1/2)} ; n_d , n_g}
  \en
  where
  \begin{equation}
E = \frac{1}{2} \left( n_d + \frac{1}{2} \right)
    + \frac{1}{2} \left( n_g + \frac{1}{2} \right)
    + \frac{1}{2} \left| n_d - n_g \right|
    + n
    + \frac{1}{2} \,.
\end{equation}

     There  exists a unique lowest energy state, namely 
   \eq
   \ket{\psi_0^{(1/2)} \left( x \right) ; 0 , 0}
   \en
  that is  annihilated by all four operators $B_i$  in $\mathfrak{C}^{-}$ subspace of $\mathfrak{su}(2,2)$ and  
  transforms as a singlet of $SU(2)_L \times SU(2)_R$ with energy
 $E = 1$.  All the other states with higher energies can be obtained 
from $\ket{\psi_0^{(1/2)} \left( x \right) ; 0 , 0}$ by repeatedly acting on it with 
$\mathfrak{C}^{+}$ generators $B^1$, $B^2$, $B^3$ and $B^4$.

The commutation relations between the $\mathfrak{su}(2)_L$  and $\mathfrak{su}(2)_R$ generators and the
operators belonging to $\mathfrak{C}^{\pm}$ are as follows:
\begin{equation}
\begin{aligned}
\commute{L_+}{B^1} &= - \frac{i}{\sqrt{2}} \, B^2
\\
\commute{L_+}{B^2} &= 0
\\
\commute{L_+}{B^3} &= - \frac{i}{\sqrt{2}} \, B^4
\\
\commute{L_+}{B^4} &= 0
\end{aligned}
\qquad \quad
\begin{aligned}
\commute{L_-}{B^1} &= 0
\\
\commute{L_-}{B^2} &= \frac{i}{\sqrt{2}} \, B^1
\\
\commute{L_-}{B^3} &= 0
\\
\commute{L_-}{B^4} &= \frac{i}{\sqrt{2}} \, B^3
\end{aligned}
\qquad \quad
\begin{aligned}
\commute{L_3}{B^1} &= - \frac{1}{2} \, B^1
\\
\commute{L_3}{B^2} &= \frac{1}{2} \, B^2
\\
\commute{L_3}{B^3} &= - \frac{1}{2} \, B^3
\\
\commute{L_3}{B^4} &= \frac{1}{2} \, B^4
\end{aligned}
\end{equation}

\begin{equation}
\begin{aligned}
\commute{R_+}{B^1} &= \frac{i}{\sqrt{2}} \, B^3
\\
\commute{R_+}{B^2} &= \frac{i}{\sqrt{2}} \, B^4
\\
\commute{R_+}{B^3} &= 0
\\
\commute{R_+}{B^4} &= 0
\end{aligned}
\qquad \quad
\begin{aligned}
\commute{R_-}{B^1} &= 0
\\
\commute{R_-}{B^2} &= 0
\\
\commute{R_-}{B^3} &= - \frac{i}{\sqrt{2}} \, B^1
\\
\commute{R_-}{B^4} &= - \frac{i}{\sqrt{2}} \, B^2
\end{aligned}
\qquad \quad
\begin{aligned}
\commute{R_3}{B^1} &= - \frac{1}{2} \, B^1
\\
\commute{R_3}{B^2} &= - \frac{1}{2} \, B^2
\\
\commute{R_3}{B^3} &= \frac{1}{2} \, B^3
\\
\commute{R_3}{B^4} &= \frac{1}{2} \, B^4
\end{aligned}
\end{equation}
showing that  $\left( B^1 , B^2 , B^3 , B^4 \right)$ transform in the $\left( 
\frac{1}{2} , \frac{1}{2} \right)$ representation of $\mathfrak{su}(2)_L 
\oplus \mathfrak{su}(2)_R$. Therefore, we can label them by their 
eigenvalues with respect to $\left( L_3 , R_3 \right)$:
\begin{equation}
\left( B^1 , B^2 , B^3 , B^4 \right)
 = T^{mn}= \left( T^{(-\frac{1}{2},-\frac{1}{2})} , T^{(+\frac{1}{2},-\frac{1}{2})} , T^{(-\frac{1}{2},+\frac{1}{2})} , T^{(+\frac{1}{2},+\frac{1}{2})} \right)
\end{equation}
Similarly, the generators $B_i$ in $\mathfrak{C}^-$ also transform in the 
$\left( 
\frac{1}{2} , \frac{1}{2} \right)$ representation of $\mathfrak{su}(2)_L 
\oplus \mathfrak{su}(2)_R$. 

Operators $B^i$ all  commute with each other, and the $\mathfrak{su}(2)_L \oplus \mathfrak{su}(2)_R$ 
content of the minimal representation of $\mathfrak{so}(4,2)$ is obtained 
by taking the symmetric powers of $\left( \frac{1}{2} , \frac{1}{2} 
\right)$, subject to the constraint $ B^1B^4=B^2B^3$.  For example, this constraint eliminates the $(0,0)$ component in the tensor product $ (\frac{1}{2},\frac{1}{2}) \otimes (\frac{1}{2},\frac{1}{2})$.
\begin{equation}
\begin{split}
\ket{\Omega}
&= \ket{\left( 0 , 0 \right)}
\qquad \qquad \qquad \qquad
\left( E = 2 \right)
\\
T^{mn} \ket{\Omega}
&= \ket{\left( \frac{1}{2} , \frac{1}{2} \right)}
\qquad \qquad \qquad \,\,
\left( E = 4 \right)
\\
\left( T^{mn} \right)^2 \ket{\Omega}
&= \ket{\left( 1 , 1 \right)}
\qquad \qquad \qquad \qquad
\left( E = 6 \right)
\\
\left( T^{mn} \right)^3 \ket{\Omega}
&= \ket{\left( \frac{3}{2} , \frac{3}{2} \right)}
\qquad \qquad \qquad \,\,\,
\left( E = 8 \right)
\\
& \vdots
\\
\left( T^{mn} \right)^P \ket{\Omega}
&= \ket{\left( \frac{P}{2} , \frac{P}{2} \right)}
\qquad \qquad \qquad
\left( E = 2 P + 2 \right)
\\
& \vdots
\end{split}
\label{Tproducts}
\end{equation}
where $m, n = \pm \frac{1}{2}$.

We list all those states that form the relevant irreducible representations 
of $\mathfrak{su}(2)_L \oplus \mathfrak{su}(2)_R$ for the first few energy 
levels and their $l_3 , r_3$ quantum numbers in Table \ref{Table:ScalarDoubleton}.

%%%%%%%%%%%%%%%%%%%%%%%%%%%%%%%%%%%%%%%%%%%%%%%%%%%%%%%%%%%%%%%%%%%%%%%%%%%%
\begin{tiny}
\begin{longtable}[c]{|l|l||c|c|c||c|r|r|}
\kill
%%%%%%%%%%%%%%%%%%%%%%%%%%%%%%%%%%%%%%%%%%%%%%%%%%%%%%%%%%%%%%%%%%%%%%%%%%%%

\caption[The $SU(2)_L \times SU(2)_R \times U(1)$ content of the minimal 
unitary representation of $SO(4,2)$]
{The $SU(2)_L \times SU(2)_R \times U(1)$ content of the minimal unitary
representation of $SO(4,2)$.
\label{Table:ScalarDoubleton}} \\
\hline
& & & & & & & \\
Irrep & State & $E$ & $N_d$ & $N_g$ & $l = r$ & $l_3$ & $r_3$ \\
& & & & & & & \\
\hline
& & & & & & & \\
\endfirsthead
%%%%%%%%%%%%%%%%%%%%%%%%%%%%%%%%%%%%%%%%%%%%%%%%%%%%%%%%%%%%%%%%%%%%%%%%%%%%
\caption[]{(continued)} \\
\hline
& & & & & & & \\
Irrep & State & $E$ & $N_d$ & $N_g$ & $l = r$ & $l_3$ & $r_3$ \\
& & & & & & & \\
\hline
& & & & & & & \\
\endhead
%%%%%%%%%%%%%%%%%%%%%%%%%%%%%%%%%%%%%%%%%%%%%%%%%%%%%%%%%%%%%%%%%%%%%%%%%%%%
& & & & & & & \\
\hline
\endfoot
%%%%%%%%%%%%%%%%%%%%%%%%%%%%%%%%%%%%%%%%%%%%%%%%%%%%%%%%%%%%%%%%%%%%%%%%%%%%
& & & & & & & \\
\hline
\endlastfoot
%%%%%%%%%%%%%%%%%%%%%%%%%%%%%%%%%%%%%%%%%%%%%%%%%%%%%%%%%%%%%%%%%%%%%%%%%%%%

$\ket{\left( 0 , 0 \right)}$ &
$\ket{\psi_0^{(1/2)} ; 0 , 0}$
 & 1 & 0 & 0 & 0 & 0 & 0 \\[8pt]

\hline
 & & & & & & & \\

$\ket{\left( \frac{1}{2} , \frac{1}{2} \right)}$ &
$B^1 \, \ket{\psi_0^{(1/2)} ; 0 , 0}$
 & 2 & 0 & 0 & $\frac{1}{2}$ & $- \frac{1}{2}$ & $- \frac{1}{2}$ \\[8pt]

 &
$B^2 \, \ket{\psi_0^{(1/2)} ; 0 , 0}$
 & 2 & 1 & 0 & $\frac{1}{2}$ & $+ \frac{1}{2}$ & $- \frac{1}{2}$ \\[8pt]

 &
$B^3 \, \ket{\psi_0^{(1/2)} ; 0 , 0}$
 & 2 & 0 & 1 & $\frac{1}{2}$ & $- \frac{1}{2}$ & $+ \frac{1}{2}$ \\[8pt]

 &
$B^4 \, \ket{\psi_0^{(1/2)} ; 0 , 0}$
 & 2 & 1 & 1 & $\frac{1}{2}$ & $+ \frac{1}{2}$ & $+ \frac{1}{2}$ \\[8pt]

\hline
 & & & & & & & \\

$\ket{\left( 1 , 1 \right)}$ & 
$B^1 B^1 \, \ket{\psi_0^{(1/2)} ; 0 , 0}$
 & 3 & 0 & 0 & 1 & $-1$ & $-1$ \\[8pt]

 &
$B^2 B^1 \, \ket{\psi_0^{(1/2)} ; 0 , 0}$
 & 3 & 1 & 0 & 1 & $ 0$ & $-1$ \\[8pt]

 &
$B^2 B^2 \, \ket{\psi_0^{(1/2)} ; 0 , 0}$
 & 3 & 2 & 0 & 1 & $+1$ & $-1$ \\[8pt]

 &
$B^3 B^1 \, \ket{\psi_0^{(1/2)} ; 0 , 0}$
 & 3 & 0 & 1 & 1 & $-1$ & $ 0$ \\[8pt]

 &
$B^3 B^2 \, \ket{\psi_0^{(1/2)} ; 0 , 0}$ =
$B^4 B^1 \, \ket{\psi_0^{(1/2)} ; 0 , 0}$
 & 3 & 1 & 1 & 1 & $ 0$ & $ 0$ \\[8pt]

 &
$B^4 B^2 \, \ket{\psi_0^{(1/2)} ; 0 , 0}$
 & 3 & 2 & 1 & 1 & $+1$ & $ 0$ \\[8pt]

 &
$B^3 B^3 \, \ket{\psi_0^{(1/2)} ; 0 , 0}$
 & 3 & 0 & 2 & 1 & $-1$ & $+1$ \\[8pt]

 &
$B^4 B^3 \, \ket{\psi_0^{(1/2)} ; 0 , 0}$
 & 3 & 1 & 2 & 1 & $ 0$ & $+1$ \\[8pt]

 &
$B^4 B^4 \, \ket{\psi_0^{(1/2)} ; 0 , 0}$
 & 3 & 2 & 2 & 1 & $+1$ & $+1$ \\[16pt]

\hline
 & & & & & & & \\

$\ket{\left( \frac{3}{2} , \frac{3}{2} \right)}$ &
$B^1 B^1 B^1 \, \ket{\psi_0^{(1/2)} ; 0 , 0}$
 & 4 & 0 & 0 & $\frac{3}{2}$ & $- \frac{3}{2}$ & $- \frac{3}{2}$ \\[8pt]

 &
$B^2 B^1 B^1 \, \ket{\psi_0^{(1/2)} ; 0 , 0}$
 & 4 & 1 & 0 & $\frac{3}{2}$ & $- \frac{1}{2}$ & $- \frac{3}{2}$ \\[8pt]

 &
$B^2 B^2 B^1 \, \ket{\psi_0^{(1/2)} ; 0 , 0}$
 & 4 & 2 & 0 & $\frac{3}{2}$ & $+ \frac{1}{2}$ & $- \frac{3}{2}$ \\[8pt]

 &
$B^2 B^2 B^2 \, \ket{\psi_0^{(1/2)} ; 0 , 0}$
 & 4 & 3 & 0 & $\frac{3}{2}$ & $+ \frac{3}{2}$ & $- \frac{3}{2}$ \\[8pt]

 &
$B^3 B^1 B^1 \, \ket{\psi_0^{(1/2)} ; 0 , 0}$
 & 4 & 0 & 1 & $\frac{3}{2}$ & $- \frac{3}{2}$ & $- \frac{1}{2}$ \\[8pt]

 &
$B^3 B^2 B^1 \, \ket{\psi_0^{(1/2)} ; 0 , 0}$ =
$B^4 B^1 B^1 \, \ket{\psi_0^{(1/2)} ; 0 , 0}$
 & 4 & 1 & 1 & $\frac{3}{2}$ & $- \frac{1}{2}$ & $- \frac{1}{2}$ \\[8pt]

 &
$B^3 B^2 B^2 \, \ket{\psi_0^{(1/2)} ; 0 , 0}$ =
$B^4 B^2 B^1 \, \ket{\psi_0^{(1/2)} ; 0 , 0}$
 & 4 & 2 & 1 & $\frac{3}{2}$ & $+ \frac{1}{2}$ & $- \frac{1}{2}$ \\[8pt]

 &
$B^4 B^2 B^2 \, \ket{\psi_0^{(1/2)} ; 0 , 0}$
 & 4 & 3 & 1 & $\frac{3}{2}$ & $+ \frac{3}{2}$ & $- \frac{1}{2}$ \\[8pt]

 &
$B^3 B^3 B^1 \, \ket{\psi_0^{(1/2)} ; 0 , 0}$
 & 4 & 0 & 2 & $\frac{3}{2}$ & $- \frac{3}{2}$ & $+ \frac{1}{2}$ \\[8pt]

 &
$B^3 B^3 B^2 \, \ket{\psi_0^{(1/2)} ; 0 , 0}$ =
$B^4 B^3 B^1 \, \ket{\psi_0^{(1/2)} ; 0 , 0}$
 & 4 & 1 & 2 & $\frac{3}{2}$ & $- \frac{1}{2}$ & $+ \frac{1}{2}$ \\[8pt]

 &
$B^4 B^3 B^2 \, \ket{\psi_0^{(1/2)} ; 0 , 0}$ =
$B^4 B^4 B^1 \, \ket{\psi_0^{(1/2)} ; 0 , 0}$
 & 4 & 2 & 2 & $\frac{3}{2}$ & $+ \frac{1}{2}$ & $+ \frac{1}{2}$ \\[8pt]

 &
$B^4 B^4 B^2 \, \ket{\psi_0^{(1/2)} ; 0 , 0}$
 & 4 & 3 & 2 & $\frac{3}{2}$ & $+ \frac{3}{2}$ & $+ \frac{1}{2}$ \\[8pt]

 &
$B^3 B^3 B^3 \, \ket{\psi_0^{(1/2)} ; 0 , 0}$
 & 4 & 0 & 3 & $\frac{3}{2}$ & $- \frac{3}{2}$ & $+ \frac{3}{2}$ \\[8pt]

 &
$B^4 B^3 B^3 \, \ket{\psi_0^{(1/2)} ; 0 , 0}$
 & 4 & 1 & 3 & $\frac{3}{2}$ & $- \frac{1}{2}$ & $+ \frac{3}{2}$ \\[8pt]

 &
$B^4 B^4 B^3 \, \ket{\psi_0^{(1/2)} ; 0 , 0}$
 & 4 & 2 & 3 & $\frac{3}{2}$ & $+ \frac{1}{2}$ & $+ \frac{3}{2}$ \\[8pt]

 &
$B^4 B^4 B^4 \, \ket{\psi_0^{(1/2)} ; 0 , 0}$
 & 4 & 3 & 3 & $\frac{3}{2}$ & $+ \frac{3}{2}$ & $+ \frac{3}{2}$ \\[8pt]

\hline
 & & & & & & & \\

\vdots &\vdots&\vdots&\vdots&\vdots&\vdots&\vdots&\vdots  \\ [8pt]

 \hline
 & & & & & & & \\

$\ket{\left( \frac{P}{2} , \frac{P}{2} \right)}$ &
$B^{i_1} B^{i_2} \dots B^{i_P} \, \ket{\psi_0^{(1/2)} ; 0 , 0}$
 & $P + 1$ & \vdots & \vdots & $\frac{P}{2}$ & \vdots & \vdots \\ [8pt]

\hline
 & & & & & & & \\

\vdots & \vdots & \vdots & \vdots & \vdots & \vdots & \vdots & \vdots \\ [8pt]

%%%%%%%%%%%%%%%%%%%%%%%%%%%%%%%%%%%%%%%%%%%%%%%%%%%%%%%%%%%%%%%%%%%%%%%%%%%%
\end{longtable}
\end{tiny}
%%%%%%%%%%%%%%%%%%%%%%%%%%%%%%%%%%%%%%%%%%%%%%%%%%%%%%%%%%%%%%%%%%%%%%%%%%%%

  Comparing the $SU(2)\times SU(2)\times U(1)$ decomposition of the minrep of $SU(2,2)$ with that of the scalar doubleton representation of $5D$ AdS group $SU(2,2)$ obtained by the oscillator method \cite{Gunaydin:1984fk,Gunaydin:1998sw,Gunaydin:1998jc}, we see that they coincide exactly.  The quadratic Casimir operator of the $SU(2,2)$  is given by \cite{Gunaydin:2005zz}
  \begin{equation}
\begin{split}
\mathcal{C}_2
&= - \frac{1}{6} J^p_q J^q_p
   + \frac{1}{12} \Delta^2
   - \frac{1}{6} \left( E F + F E \right)
   - \frac{1}{6} U^2
   - \frac{i}{12} \left( E_p F^p + F^p E_p - F_p E^p - E^p F_p \right)
\end{split}
\end{equation}
which  reduces to a c-number 
\eq
\mathcal{C}_2 = \frac{1}{2}
\en
with the higher Casimirs vanishing in the minrep. They agree with the values of the Casimir operators for the scalar doubleton given in \cite{Gunaydin:1998sw,Gunaydin:1998jc}\footnote{ Note that the quadratic Casimir of \cite{Gunaydin:1998sw,Gunaydin:1998jc} differs from that of \cite{Gunaydin:2006vz} by  an overall factor of $-6$, i.e $\mathcal{C}_{2}^{(GMZ)} =  -6 \,\mathcal{C}_{2}^{(GP)} $. }.  
    Hence the 
  minimal unitary representation of the $4D$ conformal group is nothing but the scalar doubleton representation. This representation remains irreducible under restriction to the four dimensional Poincare group and describes a massless and spinless particle \cite{Mack:1969dg,Gunaydin:1984fk,MR848089,Gunaydin:1998sw,Gunaydin:1998jc}. 
  We should also note that the same scalar doubleton representation $SO(4,2)$ was used long time ago to describe the spectrum of the Hydrogen atom \cite{malkin_manko,nambu1,nambu2,barut_kleinert1,Barut:1967zz}. 

%%%%%%%%%%%%%%%%%%%%%%%%%%%%%%%%%%%%%%%%%%%%%%%%%%%%%%%%%%%%%%%%%%%%
%%%%%% Section: Deformations of SU(2,2) %%%%%%%%%%%%%%%%%%%%%%%%%%%%
%%%%%%%%%%%%%%%%%%%%%%%%%%%%%%%%%%%%%%%%%%%%%%%%%%%%%%%%%%%%%%%%%%%%

\section{One Parameter Family of Deformations of the Minrep of $SU(2,2)$ 
and Massless Conformal Fields in Four Dimensions}
\label{sec:deformations}

In the previous section we showed that the minrep of $SU(2,2)$ is simply 
the scalar doubleton representation  that 
describes  a conformal scalar field in four dimensions. The group $SU(2,2)$ admits infinitely many 
doubleton representations corresponding to $4D$ massless conformal fields 
of arbitrary spin 
\cite{ Gunaydin:1984fk,Gunaydin:1998sw,Gunaydin:1998jc,MR848089}. 
The irreducible doubletons  of $SU(2,2)$  remain irreducible under the restriction to the Poincare subgroup and 
describe massless particles of integer and half-integer helicity \cite{Mack:1969dg}. They all 
can be constructed by the oscillator method over the Fock space of two 
pairs of twistorial oscillators transforming in the spinor representation 
of $SU(2,2)$. One important  question is whether the doubleton representations 
corresponding to massless conformal fields of arbitrary spin can all be 
obtained from the quantization of quasiconformal action of $SU(2,2)$.  Remarkably there exists a 
one-parameter ($\zeta$) deformation of the construction given in the previous section such that  
all  doubleton unitary irreducible representation of $SU(2,2)$  can be obtained by choosing the deformation parameter to be an integer. 

In the general formulation of minimal unitary representations of 
noncompact groups obtained by quantizing their quasiconformal 
realizations, the quartic invariant operator $\mathcal{I}_4$ of the grade 
zero subalgebra, modulo the $SO(1,1)$ generator that determines the 
5-grading, enters in the numerator of the singular term in the grade $+2$ 
generator $F$. For the group $SU(n+1,m+1)$ it has the form 
\cite{Gunaydin:2006vz}
   \eq
   F  = \frac{1}{2} p^2
  + \frac{1}{2 \, x^2}  \left[ \mathcal{I}_4 + \frac{\left( m + n \right)^2 -1}{4}  \right] \,.
 \en
 For $SU(2,2)$ the quartic invariant $\mathcal{I}_4$ is related to the Casimir operator of grade zero subalgebra $\mathfrak{su}(1,1)$  as 
   \eq
   \mathcal{I}_4 = 2 J_m^n J_n^m  = \left( N_d - N_g \right)^2 - 1=  U^2 -1
   \,.
   \en
  One parameter deformations of the minimal unitary representation are obtained by replacing the 
  quartic invariant $\mathcal{I}_4$ by
  \eq
  \mathcal{I}_4 \left(\zeta\right) =  \left( N_d - N_g + \zeta \right)^2 - 1
  \,.
  \en
  Then the grade $+2$ generator becomes
     \eq
   F \left(\zeta\right)  = \frac{1}{2} p^2
  + \frac{1}{2 \, x^2}  \left[ \left( N_d - N_g + \zeta \right)^2
                                - \frac{1}{4} \right]    
 \en
while the negative grade generators $E$, $E^m$ and $E_m$ remaining as in 
the undeformed case. The grade $+1$ generators are modified by $\zeta$ 
dependent terms   and are given by 
\eqq
F^{1}(\zeta)
&= & d^\dag 
   \left[ p + \frac{i}{x} 
              \left( N_d - N_g +  \zeta + \frac{1}{2} 
              \right) 
   \right]
\\
F^{2}(\zeta)
&=&  g 
   \left[ p + \frac{i}{x} 
              \left( N_d - N_g + \zeta + \frac{1}{2} 
              \right) 
   \right]
\\
F_{1} (\zeta)
&= & d 
   \left[ p - \frac{i}{x} 
              \left( N_d - N_g + \zeta  - \frac{1}{2} 
              \right) 
   \right]
\\
F_{2}(\zeta)
&=& - g^\dag 
   \left[ p - \frac{i}{x} 
              \left( N_d - N_g +  \zeta  - \frac{1}{2} 
              \right) 
   \right] \,. 
\enn
The only bosonic generator in $\mathfrak{g}^{(0)}$ subspace that changes under this  deformation 
 is $U$ which becomes
\begin{equation}
U(\zeta) = N_d - N_g + \frac{\zeta}{2} \,.
\end{equation}

One can easily verify that all the Jacobi identities are satisfied under this deformation and the quadratic Casimir of $SU(2,2)$ takes on the value
\eq
\mathcal{C}_2 (\zeta) = - \frac{1}{2} \left( \frac{\zeta}{2} - 1 \right) \left( \frac{\zeta}{2} + 1 \right) \,.
\en
The Casimirs of the $SU(2)_L$ and $SU(2)_R$ are no longer equal under this deformation. One finds
\begin{equation}
\begin{split}
 L^2 \left(\zeta\right) 
 &= \left[ \frac{1}{2} \left( H (\zeta) - \frac{\zeta}{2} \right) 
    - \frac{1}{2} \right]
    \left[ \frac{1}{2} \left( H (\zeta) - \frac{\zeta}{2} \right)
    + \frac{1}{2} \right]
 \\
 R^2 \left(\zeta\right) 
 &= \left[ \frac{1}{2} \left( H (\zeta) + \frac{\zeta}{2} \right)
    -\frac{1}{2} \right]
    \left[ \frac{1}{2} \left( H (\zeta) + \frac{\zeta}{2} \right)
    + \frac{1}{2} \right]
\end{split}
\end{equation}
where $H (\zeta)$ is the $\mathfrak{so}(2)$ generator in $\mathfrak{C}^{0}$ 
given in equation \ref{Hamiltonian}, that plays the 
role of the ``AdS energy'' operator (Hamiltonian) and determines the three 
grading in the compact basis, which has now become
\begin{equation}
\begin{split}
H(\zeta) &= \frac{1}{2}
            \left[
            N_a + N_d + N_g
            + \frac{1}{2 \, x^2} \left( N_d - N_g + \zeta \right)^2
            - \frac{1}{8 \, x^2}
            + \frac{3}{2}
            \right]
\\
  &= \frac{1}{2} \left[ N_d + \frac{1}{2} \right]
     + \frac{1}{2} \left[ N_g + \frac{1}{2} \right]
     + \frac{1}{2}
       \left[ N_a + \frac{1}{2} + \frac{G(\zeta)}{x^2} \right] \\
  &= ~~~~~~ H_d ~~~~~~
     + ~~~~~~ H_g ~~~~~~
     + ~~~~~~~~~~ H_\odot(\zeta)
\end{split}
\label{Hamiltonian_deformed}
\end{equation}
where
\begin{equation}
H_\odot (\zeta)  = \frac{1}{2}
                   \left[
                   a^\dag a + \frac{1}{2} + \frac{G(\zeta)}{x^2}
                   \right]
\end{equation}
with
\begin{equation}
G(\zeta) = \frac{1}{2} \left( N_d - N_g +\zeta \right)^2 - \frac{1}{8} \,.
\label{SingularPotential_deformed}
\end{equation}
Together with the operators $B^1(\zeta)$ and $B_1(\zeta)$ belonging to $\mathfrak{C}^+$ and $\mathfrak{C}^{-}$, respectively, 
\eqq
B^1(\zeta)
 &= &i \left( a^\dag \, a^\dag - \frac{G(\zeta)}{x^2} \right) \\
  B_1(\zeta)
 &=& - i \left( a \, a - \frac{G(\zeta)}{x^2} \right)
 \enn
$H_\odot \left( \zeta \right)$ generates the distinguished  $\mathfrak{su}(1,1)_L$ subalgebra:
\begin{equation}
\begin{split}
\commute{B_1(\zeta)}{B^1(\zeta)} &= 8 \, H_\odot (\zeta)
\\
\commute{H_\odot (\zeta)}{B^1(\zeta)} &=  + \, B^1(\zeta)
\\
\commute{H_\odot(\zeta)}{B_1(\zeta)} &= - \, B_1(\zeta)
\end{split}
\end{equation}

We shall denote the eigenfunctions of this deformed singular harmonic 
oscillator Hamiltonian $H_\odot (\zeta)$  as $\psi^\alpha ( x \,;\, \zeta)$
\begin{equation}
H_\odot (\zeta) \, \psi^\alpha\left( x \,;\, \zeta \right)
 = E_{\odot}(\zeta) \, \psi^\alpha \left( x \,;\, \zeta \right)
 \en
 \eq
E_{\odot}(\zeta) = \frac{1}{4} \left[ 2 \,\alpha(\zeta) + 1 \right] \,.
\end{equation}
 For  eigenstates that are lowest weight vectors of a unitary representation of $SU(1,,1)$ we have
 \eq
 B_1(\zeta) \, \psi^\alpha\left( x \,;\, \zeta \right) = 0
 \en
and such states take the form
\begin{equation}
\psi^\alpha \left( x \,;\, \zeta \right) = C \, x^{\alpha(\zeta)} e^{-x^2/2}
\en
where 
\eq
\alpha(\zeta) = \frac{1}{2} + \left( 2 \, g (\zeta) + \frac{1}{4} \right)^{\frac{1}{2}}= \frac{1}{2} + \left| n_d - n_g + \zeta \right| 
\end{equation}
and $C$ is a normalization constant. 
The normalizability condition  requires
\eq
\alpha(\zeta)  \geq \frac{1}{2} \,.
\en
Therefore the fact that the normalizable states corresponding to the lowest energy eigenvalue $E_{\odot}$ of the isotonic oscillator have $\alpha = 1/2$ implies
\eq
n_d - n_g + \zeta = 0 \,.
\label{dgConstraintWithZeta}
\en
This means that  the deformation parameter $\zeta$ is  an integer for such states.  
We shall denote the corresponding eigenfunction as $\psi_0^{(1/2)} \left( x \,;\, \zeta \right)$. There are infinitely many
such states in the tensor product space $\mathcal{H}_\odot \otimes \mathcal{H}_d \otimes \mathcal{H}_g $.

The total energy eigenvalue $E(\zeta)$ of a tensor product state 
\[ \ket{\psi^{(\alpha)}\left(\zeta\right) , n_d , n_d}
\]
is on the other hand given by
\eq
E(\zeta) = E_\odot (\zeta) + E_d + E_g 
= \frac{1}{2} \left| n_d - n_g + \zeta \right| 
  + \frac{1}{2} \left( n_d + n_g \right) 
  + 1 \,.
\en
Therefore, for  a given $\zeta$ we have a unique  lowest energy eigenvalue
\eq
E_0 (\zeta) = \frac{|\zeta|}{2} + 1 \,.
\en
The degeneracy of this energy eigenvalue is $\left|\zeta\right| + 1$. These  degenerate energy eigenstates transform in an irreducible representation of $SU(2)_L \times SU(2)_R \times U(1)$.  For $\zeta = n_r$, where $n_r$ is a positive integer, they transform in the representation $\left( 0 \,,\, \frac{n_r}{2} \right)$ of $SU(2)_L \times SU(2)_R$, and for $\zeta = - n_l$, where $n_l$ is a positive integer, they transform in the representation $\left( \frac{n_l}{2} \,,\, 0 \right)$ of $SU(2)_L\times SU(2)_R$. 
The operators $B_i$ ($i=1,2,3,4$) in grade $-1$ subspace $\mathfrak{C}^-$ annihilate these lowest energy states for a given $\zeta$. Let us label this finite set of  states collectively as $\ket{\Omega}$. Then
\eq
B_i \, \ket{\Omega} = 0 \qquad i=1,2,3,4 \,.
\en
Since the states $\ket{\Omega}$ transform irreducibly under the maximal compact subgroup 
$SU(2) \times SU(2) \times U(1)$, the infinite set of states generated by the repeated action of the operators $B^i \in \mathfrak{C}^+$ on $\ket{\Omega}$
\eq 
\ket{\Omega} \,,\, B^i \ket{\Omega} \,,\, B^i B^j \ket{\Omega} \,,\, \dots
\en       
form the (particle) basis of a positive energy unitary irreducible representation of $SU(2,2)$ \cite{Gunaydin:1981yq,Bars:1982ep,Gunaydin:1984fk,Gunaydin:1998sw}. By going to a noncompact coherent state basis labelled by $4D$ spacetime coordinates one can show that these unitary representations can be identified with  massless conformal fields whose $SL(2,\mathbb{C})$ transformation labels coincide with the $SU(2)\times SU(2)$ labels of their lowest energy states, and the lowest energy eigenvalues $E$ can be identified with their conformal dimensions \cite{Gunaydin:1998sw,Gunaydin:1998jc}. 
Irreducible doubleton  (ladder or most degenerate discrete series) representations remain irreducible under restriction to Poincar\'{e} group and describe massless particles of arbitrary helicity $\lambda$ \cite{Mack:1969dg}, which  is related to our deformation parameter simply as
\eq
\lambda = \frac{\zeta}{2} \,.
\en

%%%%%%%%%%%%%%%%%%%%%%%%%%%%%%%%%%%%%%%%%%%%%%%%%%%%%%%%%%%%%%%%%%%%
%%%%%% Section: Minrep of SU(2,2|p+q) %%%%%%%%%%%%%%%%%%%%%%%%%%%%%%
%%%%%%%%%%%%%%%%%%%%%%%%%%%%%%%%%%%%%%%%%%%%%%%%%%%%%%%%%%%%%%%%%%%%

\section{Minimal Unitary Representations of Supergroups $SU \left( 2 , 2 
\,|\,\mathfrak{p} + \mathfrak{q} \right)$}
\label{sec:minrepSU(2,2|p+q)}

In this section we shall construct the minimal unitary representations of 
the supergroups 
$SU \left( 2 , 2 \,|\, \mathfrak{p} + \mathfrak{q} \right)$ and their 
deformations using the quasiconformal approach. 
Consider a set of $\mathfrak{p}$ pairs of fermionic annihilation and creation 
operators labelled as $\alpha_\mu$ and $\alpha^\mu = \left( \alpha_\mu \right)^\dag$ 
and another set of $\mathfrak{q}$ pairs of fermionic annihilation and creation operators labelled as  
$\beta_y$ and $\beta^y = \left( \beta_y \right)^\dag$, such that they 
satisfy the anti-commutation relations:
\begin{equation}
\begin{aligned}
\anticommute{\alpha_\mu}{\alpha^\nu} &= \delta^\nu_\mu
\\
\anticommute{\beta_y}{\beta^z} &= \delta^y_z
\end{aligned}
\qquad \qquad
\begin{aligned}
\anticommute{\alpha_\mu}{\alpha_\nu} &= 0
\\
\anticommute{\beta_y}{\beta_z} &= 0
\end{aligned}
\qquad \qquad
\begin{aligned}
\anticommute{\alpha^\mu}{\alpha^\nu} &= 0
\\
\anticommute{\beta^y}{\beta^z} &= 0
\end{aligned}
\end{equation}
where $\mu, \nu = 1 , \dots , \mathfrak{p}$ and $y, z = 1 , \dots , 
\mathfrak{q}$. Let $N_\alpha = \alpha^\mu \alpha_\mu$ and $N_\beta = 
\beta^y \beta_y$ be the $\alpha$- and $\beta$-type fermionic number 
operators. Also denote the $\mathfrak{su}(\mathfrak{p})$ and 
$\mathfrak{su}(\mathfrak{q})$ generators inside $\mathfrak{su}(\mathfrak{p} 
+ \mathfrak{q})$ by
\begin{equation}
\mathcal{A}^\nu_\mu 
= \alpha^\nu \alpha_\mu 
  - \frac{1}{\mathfrak{p}} \delta^\nu_\mu \, N_\alpha
\qquad \qquad
\mathcal{B}^z_y 
= \beta^z \beta_y 
  - \frac{1}{\mathfrak{q}} \delta^z_y \, N_\beta
\end{equation}
so that
\begin{equation}
\commute{\mathcal{A}^\nu_\mu}{\mathcal{A}^\rho_\sigma} 
= \delta^\rho_\mu \, \mathcal{A}^\nu_\sigma 
  - \delta^\nu_\sigma \, \mathcal{A}^\rho_\mu
\qquad \qquad
\commute{\mathcal{B}^z_y}{\mathcal{B}^w_x} 
= \delta^w_y \, \mathcal{B}^z_x 
  - \delta^z_x \, \mathcal{B}^w_y \,.
\end{equation}
The remaining generators of $\mathfrak{su}(\mathfrak{p} + \mathfrak{q})$ 
are given by
\begin{equation}
\mathcal{C}_{\mu y} = \alpha_\mu \beta_y
\qquad \qquad
\mathcal{C}^{y \mu} = \left( \mathcal{C}_{\mu y} \right)^\dag
                    = \beta^y \alpha^\mu
\end{equation}
so that
\begin{equation}
\commute{\mathcal{C}_{\mu y}}{\mathcal{C}^{z \nu}}
= - \delta^z_y \, \mathcal{A}^\nu_\mu
  - \delta^\nu_\mu \, \mathcal{B}^z_y
  - \delta^\nu_\mu \delta^z_y \, 
    \left( \frac{1}{\mathfrak{p}} N_\alpha 
           + \frac{1}{\mathfrak{q}} N_\beta 
           - 1 \right)
\end{equation}
where $\frac{1}{\mathfrak{p}} N_\alpha + \frac{1}{\mathfrak{q}} N_\beta - 
1$ is the $\mathfrak{u}(1)$ generator that appears in the decomposition 
$\mathfrak{su}(\mathfrak{p} + \mathfrak{q}) \supseteq 
\mathfrak{su}(\mathfrak{p}) \oplus \mathfrak{su}(\mathfrak{q}) \oplus 
\mathfrak{u}(1)$ and determines a 3-graded decomposition of 
$\mathfrak{su}(\mathfrak{p} + \mathfrak{q})$.

%%%%%%%%%%%%%%%%%%%%%%%%%%%%%%%%%%%%%%%%%%%%%%%%%%%%%%%%%%%%%%%%%%%%
%%%%%% Subsection: 5-grading wrt SU(1,1|p+q) x U(1) x SO(1,1) %%%%%%
%%%%%%%%%%%%%%%%%%%%%%%%%%%%%%%%%%%%%%%%%%%%%%%%%%%%%%%%%%%%%%%%%%%%

\subsection{5-grading of $\mathfrak{su}(2,2 \,|\, \mathfrak{p} + 
\mathfrak{q})$  with respect to the subalgebra $\mathfrak{su} \left( 1 , 1 
\,|\, \mathfrak{p} + \mathfrak{q} \right) \oplus \mathfrak{u} \left( 1 
\right) \oplus \mathfrak{so} \left( 1 , 1 \right)$}

The Lie superalgebra $\mathfrak{su} \left( 2 , 2 \,|\, \mathfrak{p} + 
\mathfrak{q} \right)$ has the following 5-graded decomposition 
with respect to its subalgebra $\mathfrak{su} \left( 1 , 1 \,|\, 
\mathfrak{p} + \mathfrak{q} \right) \oplus \mathfrak{u} \left( 1 \right) \oplus \mathfrak{so} \left( 1 , 1 \right)$:
\begin{equation}
\begin{split}
\mathfrak{su} \left( 2 , 2 \,|\, \mathfrak{p} + \mathfrak{q} \right)
&= 
\mathfrak{g}^{(-2)}
\oplus
\mathfrak{g}^{(-1)}
\oplus
\mathfrak{g}^{(0)}
\oplus
\mathfrak{g}^{(+1)}
\oplus
\mathfrak{g}^{(+2)}
\\
&= 
1^{(-2)}
\, \oplus \,
2 \left( 2 , \mathfrak{p} + \mathfrak{q} \right)^{(-1)}
\, \oplus \,
\left[ \mathfrak{su} \left( 1 , 1 \,|\, \mathfrak{p} + \mathfrak{q} \right) \oplus \mathfrak{u}\left(1\right) \oplus \mathfrak{so} \left( 1 , 1 \right) \right]
\\
& \quad
\oplus \,
2 \left( 2 , \mathfrak{p} + \mathfrak{q} \right)^{(+1)}
\, \oplus \,
1^{(+2)}
\end{split}
\end{equation}

Using these fermionic oscillators, we define the  $2  \left(
\mathfrak{p} + \mathfrak{q} \right)$ supersymmetry generators 
\begin{equation}
S_\mu = \alpha_\mu \, x
\qquad \qquad
S^\mu = \left( S_\mu \right)^\dag
      = \alpha^\mu \, x
\qquad \qquad
S_y = \beta_y \, x
\qquad \qquad
S^y = \left( S_y \right)^\dag
    = \beta^y \, x \,.
\end{equation}
These supersymmetry generators, together with the bosonic generators  
\begin{equation}
E^1
= x \, d^\dag
\qquad \qquad \qquad
E^2
= x \, g
\qquad \qquad  \qquad
E_1
= x \, d
\qquad \qquad \qquad
E_2
= - x \, g^\dag
\end{equation}
span the grade $-1$ subspace
$\mathfrak{g}^{(-1)}$. Clearly, under anticommutation, the 
$\mathfrak{g}^{(-1)}$ supersymmetry generators close into the 
$\mathfrak{g}^{(-2)}$ 
generator $E$:
\begin{equation}
\begin{aligned}
\anticommute{S_\mu}{S^\nu} &= 2 \delta^\mu_\nu \, E
\\
\anticommute{S_y}{S^z} &= 2 \delta^z_y \, E
\end{aligned}
\qquad \qquad
\begin{aligned}
\anticommute{S_\mu}{S_\nu} &= 0
\\
\anticommute{S_y}{S_z} &= 0
\end{aligned}
\qquad \qquad
\begin{aligned}
\anticommute{S^\mu}{S^\nu} &= 0
\\
\anticommute{S^y}{S^z} &= 0
\end{aligned}
\end{equation}

Now based on the results of previous sections and those of \cite{Gunaydin:1998sw,Gunaydin:2006vz} we define  the $\mathfrak{g}^{(+2)}$ generator $F$ with a deformation parameter $\zeta$ as follows
\begin{equation}
F = \frac{1}{2} p^2
   + \frac{1}{2 \, x^2} 
     \left[ \left( N_d - N_g + N_\alpha - N_\beta + \zeta \right)^2
            + \lambda \right]
\end{equation}
where $\zeta$ is the deformation parameter and $\lambda$ is a constant to be determined. 
The $2 \left( \mathfrak{p} + \mathfrak{q} \right)$ 
supersymmetry generators $Q_\mu$, $Q^\mu = \left( Q_\mu \right)^\dag$, 
$Q_y$ and $Q^y = \left( Q_y \right)^\dag$ in $\mathfrak{g}^{(+1)}$ space 
are defined by commutation of grade $-1$ supersymmetry generators with $F$: 
\begin{equation}
\begin{split}
Q_\mu
&= - i \commute{S_\mu}{F}
 = \alpha_\mu 
   \left[ p - \frac{i}{x} 
              \left( N_d - N_g + N_\alpha - N_\beta + \zeta - \frac{1}{2} 
              \right)
   \right]
\\
Q^\mu
&= - i \commute{S^\mu}{F}
 = \alpha^\mu 
   \left[ p + \frac{i}{x} 
              \left( N_d - N_g + N_\alpha - N_\beta + \zeta + \frac{1}{2} 
              \right)
   \right]
\\
Q_y
&= - i \commute{S_y}{F}
 = \beta_y 
   \left[ p + \frac{i}{x} 
              \left( N_d - N_g + N_\alpha - N_\beta + \zeta + \frac{1}{2} 
              \right)
   \right]
\\
Q^y
&= - i \commute{S^y}{F}
 = \beta^y 
   \left[ p - \frac{i}{x} 
              \left( N_d - N_g + N_\alpha - N_\beta + \zeta - \frac{1}{2} 
              \right)
   \right]
\end{split}
\end{equation}
Requiring that
\begin{equation}
\anticommute{Q_\mu}{Q^\nu} = 2 \delta^\nu_\mu \, F
\qquad \qquad
\anticommute{Q_y}{Q^z} = 2 \delta^z_y \, F
\end{equation}
fixes the constant $\lambda = - \frac{1}{4}$. Therefore
\begin{equation}
F = \frac{1}{2} p^2
   + \frac{1}{2 \, x^2} \left[ \left( N_d - N_g + N_\alpha - N_\beta + \zeta 
                               \right)^2
                               - \frac{1}{4} \right] \,.
\end{equation}

The bosonic generators of $SU(2,2)$ in $\mathfrak{g}^{(+1)}$ subspace are modified in the supersymmetric extension as
\begin{equation}
\begin{split}
F^1
&= d^\dag 
   \left[ p + \frac{i}{x} 
              \left( N_d - N_g + N_\alpha - N_\beta + \zeta + \frac{1}{2} 
              \right) 
   \right]
\\
F^2
&= g 
   \left[ p + \frac{i}{x} 
              \left( N_d - N_g + N_\alpha - N_\beta + \zeta + \frac{1}{2} 
              \right) 
   \right]
\\
F_1
&= d 
   \left[ p - \frac{i}{x} 
              \left( N_d - N_g + N_\alpha - N_\beta + \zeta - \frac{1}{2} 
              \right) 
   \right]
\\
F_2
&= - g^\dag 
   \left[ p - \frac{i}{x} 
              \left( N_d - N_g + N_\alpha - N_\beta + \zeta - \frac{1}{2} 
              \right) 
   \right] \,.
\end{split}
\end{equation}
These modifications correspond simply to a shift in the deformation parameter $\zeta$ by $( N_\alpha -N_\beta )$.
The only generator of $SU(2,2)$ in $\mathfrak{g}^{(0)}$ subspace that is modified  
by the supersymmetric extension  is:
\begin{equation}
U = N_d - N_g + \frac{1}{2} \left( N_\alpha - N_\beta + \zeta \right)
\end{equation}
which again represents a shift in the deformation parameter $\zeta$. 
Under anti-commutation, the supersymmetry generators in 
$\mathfrak{g}^{(-1)}$ and $\mathfrak{g}^{(+1)}$ close into the bosonic 
generators in $\mathfrak{g}^{(0)}$:
\begin{equation}
\begin{split}
\anticommute{S_\mu}{Q^\nu}
&= \delta^\nu_\mu \, \Delta
   + 2 i \, \delta^\nu_\mu \, J^1_1
   + 2 i \, \delta^\nu_\mu \, U
   - 2 i \, \mathcal{A}^\nu_\mu
   - 2 i \, \delta^\nu_\mu 
     \left( N_d + \frac{1}{\mathfrak{p}} N_\alpha \right) 
\\
\anticommute{S^\nu}{Q_\mu}
&= \delta^\nu_\mu \, \Delta
   - 2 i \, \delta^\nu_\mu \, J^1_1
   - 2 i \, \delta^\nu_\mu \, U
   + 2 i \, \mathcal{A}^\nu_\mu
   + 2 i \, \delta^\nu_\mu 
     \left( N_d + \frac{1}{\mathfrak{p}} N_\alpha \right)
\\
\anticommute{S_y}{Q^z}
&= \delta^z_y \, \Delta
   - 2 i \, \delta^z_y \, J^2_2
   - 2 i \, \delta^z_y \, U
   - 2 i \, \mathcal{B}^z_y
   - 2 i \, \delta^z_y 
     \left( N_g + \frac{1}{\mathfrak{q}} N_\beta \right) 
\\
\anticommute{S^z}{Q_y}
&= \delta^z_y \, \Delta
   + 2 i \, \delta^z_y \, J^2_2
   + 2 i \, \delta^z_y \, U
   + 2 i \, \mathcal{B}^z_y
   + 2 i \, \delta^z_y 
     \left( N_g + \frac{1}{\mathfrak{q}} N_\beta \right) 
\end{split}
\end{equation}
where $N_d + \frac{1}{\mathfrak{p}} N_\alpha$ and $N_g + 
\frac{1}{\mathfrak{q}} N_\beta$ are the $U(1)$ generators in $SU(1 \,|\, 
\mathfrak{p})$ and $SU(1 \,|\, \mathfrak{q})$, respectively.

The $4 \left( \mathfrak{p} + \mathfrak{q} \right)$ 
supersymmetry generators in the $\mathfrak{g}^{(0)}$ subspace are determined by  the 
commutators between even (odd) generators in $\mathfrak{g}^{(-1)}$ 
and odd (even) generators in $\mathfrak{g}^{(+1)}$:
\begin{equation}
\begin{aligned}
\widetilde{Q}_\mu^1
&= \alpha_\mu d^\dag
 = - \frac{i}{2} \commute{E^1}{Q_\mu}
 = - \frac{i}{2} \commute{S_\mu}{F^1}
\\
\widetilde{S}_\mu^2
&= \alpha_\mu g
 = - \frac{i}{2} \commute{E^2}{Q_\mu}
 = - \frac{i}{2} \commute{S_\mu}{F^2}
\\
\widetilde{Q}^\mu_1
&= \alpha^\mu d
 = - \frac{i}{2} \commute{E_1}{Q^\mu}
 = - \frac{i}{2} \commute{S^\mu}{F_1}
\\
\widetilde{S}^\mu_2
&= \alpha^\mu g^\dag
 = \frac{i}{2} \commute{E_2}{Q^\mu}
 = \frac{i}{2} \commute{S^\mu}{F_2}
\end{aligned}
\qquad
\begin{aligned}
\widetilde{Q}_{1 \, y}
&= \beta_y d
 = - \frac{i}{2} \commute{E_1}{Q_y}
 = - \frac{i}{2} \commute{S_y}{F_1}
\\
\widetilde{S}_{2 \, y}
&= \beta_y g^\dag
 = \frac{i}{2} \commute{E_2}{Q_y}
 = \frac{i}{2} \commute{S_y}{F_2}
\\
\widetilde{Q}^{1 \, y}
&= \beta^y d^\dag
 = - \frac{i}{2} \commute{E^1}{Q^y}
 = - \frac{i}{2} \commute{S^y}{F^1}
\\
\widetilde{S}^{2 \, y}
&= \beta^y g
 = - \frac{i}{2} \commute{E^2}{Q^y}
 = - \frac{i}{2} \commute{S^y}{F^2}
\end{aligned}
\end{equation}

They satisfy the anti-commutation relations:
\begin{equation}
\begin{split}
\anticommute{\widetilde{Q}_\mu^1}{\widetilde{Q}^\nu_1}
= \mathcal{A}^\nu_\mu
  + \delta^\nu_\mu \left( N_d + \frac{1}{\mathfrak{p}} N_\alpha \right)
& \qquad \qquad
\anticommute{\widetilde{Q}_\mu^1}{\widetilde{S}^\nu_2}
= - \delta^\nu_\mu \, J^1_2
\\
\anticommute{\widetilde{S}_\mu^2}{\widetilde{S}^\nu_2}
= - \mathcal{A}^\nu_\mu
  + \delta^\nu_\mu \left( N_g +1 -  \frac{1}{\mathfrak{p}} N_\alpha \right)
& \qquad \qquad
\anticommute{\widetilde{S}_\mu^2}{\widetilde{Q}^\nu_1}
= \delta^\nu_\mu \, J^2_1
\\
\anticommute{\widetilde{Q}_{1 \, y}}{\widetilde{Q}^{1 \, z}}
= - \mathcal{B}^z_y
  + \delta^z_y \left( N_d +1 - \frac{1}{\mathfrak{q}} N_\beta \right)
& \qquad \qquad
\anticommute{\widetilde{Q}_{1 \, y}}{\widetilde{S}^{2 \, z}}
= \delta^z_y \, J^2_1
\\
\anticommute{\widetilde{S}_{2 \, y}}{\widetilde{S}^{2 \, z}}
= \mathcal{B}^z_y
  + \delta^z_y \left( N_g + \frac{1}{\mathfrak{q}} N_\beta \right)
& \qquad \qquad
\anticommute{\widetilde{S}_{2 \, y}}{\widetilde{Q}^{1 \, z}}
= - \delta^z_y \, J^1_2
\end{split}
\end{equation}

\begin{equation}
\begin{aligned}
\anticommute{\widetilde{Q}_\mu^1}{Q^\nu}
&= \delta^\nu_\mu \, F^1
\\
\anticommute{\widetilde{S}_\mu^2}{Q^\nu}
&= \delta^\nu_\mu \, F^2
\\
\anticommute{\widetilde{Q}_{1 \, y}}{Q^z}
&= \delta^z_y \, F_1
\\
\anticommute{\widetilde{S}_{2 \, y}}{Q^z}
&= - \delta^z_y \, F_2
\end{aligned}
\qquad \qquad \qquad
\begin{aligned}
\anticommute{\widetilde{Q}_\mu^1}{S^\nu}
&= \delta^\nu_\mu \, E^1
\\
\anticommute{\widetilde{S}_\mu^2}{S^\nu}
&= \delta^\nu_\mu \, E^2
\\
\anticommute{\widetilde{Q}_{1 \, y}}{S^z}
&= \delta^z_y \, E_1
\\
\anticommute{\widetilde{S}_{2 \, y}}{S^z}
&= - \delta^z_y \, E_2
\end{aligned}
\end{equation}

Thus the decomposition of the generators in the 5-grading of the superalgebra takes the form:
\begin{equation*}
\begin{split}
& E^{(-2)}
  \oplus
  \left[ E^p \,,\, E_p \,,\, S_\mu \,,\, S^\mu \,,\, S_y \,,\, S^y \right]^{(-1)}
\\
& \qquad \qquad
  \oplus
  \left[ J^p_q \,,\, U \,,\, \Delta \,,\, 
         \mathcal{A}^\nu_\mu \,,\, \mathcal{B}^z_y \,,\,
         \mathcal{C}^{y \mu} \,,\, \mathcal{C}_{\mu y} \,,\,
         \widetilde{Q}_\mu^1\,,\, \widetilde{Q}^\mu_1\,,\, 
          \widetilde{S}_\mu^2 \,,\, \widetilde{S}^\mu_2 \,,\,
         \widetilde{Q}_{1 \, y} \,,\, \widetilde{Q}^{1 \, y} \,,\, 
          \widetilde{S}_{2 \, y} \,,\, \widetilde{S}^{2 \, y}
  \right]^{(0)}
\\
& \qquad \qquad \qquad \qquad \qquad \qquad \qquad \qquad \qquad \qquad
  \oplus
  \left[ F^p \,,\, F_p \,,\, Q_\mu \,,\, Q^\mu \,,\, Q_y \,,\, Q^y \right]^{(+1)}
  \oplus
  F^{(+2)}
\end{split}
\end{equation*}

Then the quadratic Casimir of $SU(2,2)$ subgroup of the superalgebra  in the normalization of \cite{Gunaydin:2006vz} becomes
\begin{equation}
\begin{split}
\mathcal{C}_2
&= - \frac{1}{6} J^p_q J^q_p
   + \frac{1}{12} \Delta^2
   - \frac{1}{6} \left( E F + F E \right)
   - \frac{1}{6} U^2
   - \frac{i}{12} \left( E_p F^p + F^p E_p - F_p E^p - E^p F_p \right)
\\
&= \frac{1}{2}
   - \frac{1}{8} \left( N_\alpha - N_\beta + \zeta \right)^2
\end{split}
\end{equation}
where we have used 
\begin{equation}
\begin{split}
J^p_q J^q_p
&= \frac{1}{2} \left( N_d - N_g \right)^2 
  - \frac{1}{2} 
\\
\Delta^2
&= x^2 p^2 - 2 i \, x p - \frac{1}{4}
\\
E F + F E
&= \frac{1}{2} \, x^2 p^2
  - i \, x p
  + \frac{1}{2} \left( N_d - N_g \right)^2
  + \frac{1}{2} \left( N_\alpha - N_\beta + \zeta \right)^2
\\
& \quad
  + \left( N_d - N_g \right) \left( N_\alpha - N_\beta + \zeta \right)
  - \frac{5}{8}
\\
U^2
&= \left( N_d - N_g \right)^2
  + \left( N_d - N_g \right) \left( N_\alpha - N_\beta + \zeta \right) \\
& \quad
  + \frac{1}{4} \left( N_\alpha - N_\beta + c \right)^2
\\
E_p F^p + F^p E_p - F_p E^p - E^p F_p
&= 4 i \left( N_d - N_g \right)^2
  + 4 i \left( N_d - N_g \right) \left( N_\alpha - N_\beta + \zeta \right)
  + 4 i \,.
\end{split}
\end{equation}

%%%%%%%%%%%%%%%%%%%%%%%%%%%%%%%%%%%%%%%%%%%%%%%%%%%%%%%%%%%%%%%%%%%%
%%%%%% Subsection: 3-grading wrt SU(2|p) x SU(2|q) x U(1) %%%%%%%%%%
%%%%%%%%%%%%%%%%%%%%%%%%%%%%%%%%%%%%%%%%%%%%%%%%%%%%%%%%%%%%%%%%%%%%

\subsection{3-grading of $\mathfrak{su}(2,2 \,|\, \mathfrak{p} + 
\mathfrak{q})$  with respect to the compact subalgebra  $\mathfrak{su}(2 
\,|\, \mathfrak{p}) \oplus \mathfrak{su}(2 \,|\, \mathfrak{q}) \oplus 
\mathfrak{u}(1)$}

The Lie superalgebra $\mathfrak{su}(2,2 \,|\, \mathfrak{p} + \mathfrak{q})$ can be given a 3-graded decomposition with respect to its compact subalgebra $\mathfrak{su}(2 \,|\, 
\mathfrak{p}) \oplus \mathfrak{su}(2 \,|\, \mathfrak{q}) \oplus 
\mathfrak{u}(1)$ 
\eq 
\mathfrak{su}(2,2 \,|\, \mathfrak{p} + \mathfrak{q}) = \mathfrak{C}^- \oplus 
\mathfrak{C}^0 \oplus \mathfrak{C}^+ 
\en 
where $\mathfrak{C}^0 =\mathfrak{su}(2 \,|\, 
\mathfrak{p}) \oplus \mathfrak{su}(2 \,|\, \mathfrak{q}) \oplus 
\mathfrak{u}(1)$. The $\mathfrak{u}(1)$ generator in $\mathfrak{C}^{0} $ that defines the 3-grading is given by
\begin{equation}
\begin{split}
\mathcal{H}
&= \frac{1}{2}
   \left[
   \left( F + E \right) + \left( J^1_1 - J^2_2 \right)
   + \frac{1}{2} \left( \alpha^\mu \alpha_\mu 
                        - \alpha_\mu \alpha^\mu \right)
   + \frac{1}{2} \left( \beta^y \beta_y 
                        - \beta_y \beta^y \right)
   \right]
\\
&= \frac{1}{4} \left( x^2 + p^2 \right)
    + \frac{1}{4 \, x^2}
      \left[ \left( N_d - N_g + N_\alpha - N_\beta + \zeta \right)^2 
             - \frac{1}{4} \right]
\\
& \qquad
    + \frac{1}{2} \left( N_d + N_g + N_\alpha + N_\beta \right)
    + \frac{2 - \mathfrak{p} - \mathfrak{q}}{4}
\end{split}
\label{SU2pSU2qU1generator}
\end{equation}
and once again it plays the role of the ``total energy'' operator.

The $\mathfrak{u}(1)$ generator $H$ in $\mathfrak{su}(2,2)$, which is the 
AdS energy, resides within this $\mathcal{H}$.
\begin{equation}
\begin{split}
H
&= \frac{1}{2}
   \left[
   \left( F + E \right) + \left( J^1_1 - J^2_2 \right)
   \right]
\\
&= \frac{1}{4} \left( x^2 + p^2 \right)
   + \frac{1}{4 \, x^2}
     \left[ \left( N_d - N_g + N_\alpha - N_\beta + \zeta \right)^2 
            - \frac{1}{4} \right]
   + \frac{1}{2} \left( N_d + N_g + 1 \right)
\\
&= H_\odot + H_d + H_g
\end{split}
\label{NonSusyH}
\end{equation}
Recall that
\begin{equation}
H_\odot
= \frac{1}{4} \left( x^2 + p^2 \right)
  + \frac{1}{4 \, x^2} 
    \left[ \left( N_d - N_g + N_\alpha - N_\beta \right)^2 - \frac{1}{4} 
    \right]
\end{equation}
is the singular harmonic oscillator part of the AdS Hamiltonian $H$ and
\begin{equation}
H_d = \frac{1}{2} \left( N_d + \frac{1}{2} \right)
\qquad \qquad \qquad
H_g = \frac{1}{2} \left( N_g + \frac{1}{2} \right)
\end{equation}
are its non-singular parts.
 
The generators of  $\mathfrak{su}(2)_L \oplus 
\mathfrak{su}(2)_R$ subalgebra in $\mathfrak{C}^0$ are given by:
\begin{equation}
\begin{split}
L_+
&= - \frac{i}{2 \sqrt{2}} \left( E^1 + i \, F^1 \right)
\\
&= - \frac{i}{2 \sqrt{2}} d^\dag
   \left[ \left( x + i \, p \right)
          - \frac{1}{x} 
            \left( N_d - N_g + N_\alpha - N_\beta + \zeta + \frac{1}{2} \right)
   \right]
\\
L_-
&= \frac{i}{2 \sqrt{2}} \left( E_1 - i \, F_1 \right)
\\
&= \frac{i}{2 \sqrt{2}} d
   \left[ \left( x - i \, p \right)
          - \frac{1}{x} 
            \left( N_d - N_g + N_\alpha - N_\beta + \zeta - \frac{1}{2} \right)
   \right]
   \\
L_3 
&= \frac{1}{2} 
   \left[ U - \frac{1}{2} \left( F + E \right) + J^1_1 \right]
\\
&= N_d 
   - \frac{1}{2}
     \left[
     H - \frac{1}{2} \left( N_\alpha - N_\beta + \zeta \right) - 1 
     \right] 
\end{split}
\end{equation}
\begin{equation}
\begin{split}
R_+
&= - \frac{i}{2 \sqrt{2}} \left( E_2 + i \, F_2 \right)
\\
&= \frac{i}{2 \sqrt{2}} g^\dag
   \left[ \left( x + i \, p \right)
          + \frac{1}{x} 
            \left( N_d - N_g + N_\alpha - N_\beta + \zeta - \frac{1}{2} \right)
   \right]
\\
R_-
&= - \frac{i}{2 \sqrt{2}} \left( E^2 - i \, F^2 \right)
\\
&= - \frac{i}{2 \sqrt{2}} g
   \left[ \left( x - i \, p \right)
          + \frac{1}{x} 
            \left( N_d - N_g + N_\alpha - N_\beta + \zeta + \frac{1}{2} \right)
   \right]
   \\
R_3 
&= - \frac{1}{2} 
   \left[ U + \frac{1}{2} \left( F + E \right) + J^2_2 \right]
\\
&= N_g 
   - \frac{1}{2}
     \left[
     H + \frac{1}{2} \left( N_\alpha - N_\beta + \zeta \right) + 1
     \right]
\end{split}
\end{equation}

The $\mathfrak{su}(2)$ generators satisfy the commutation relations
\begin{equation}
\begin{split}
\commute{L_+}{L_-} = L_3
& \qquad \qquad \qquad
\commute{L_3}{L_\pm} = \pm \, L_\pm
\\
\commute{R_+}{R_-} = R_3
& \qquad \qquad \qquad
\commute{R_3}{R_\pm} = \pm \, R_\pm \,.
\end{split}
\end{equation}

The quadratic Casimir operators of the two $\mathfrak{su}(2)$'s are 
different once again, as opposed to the non-supersymmetric non-deformed 
case (see equations (\ref{L^2R^2inbc}) and (\ref{L^2R^2indg})):
\begin{equation}
L^2
= \frac{1}{4} 
  \left[ \left( H - 
         \frac{1}{2} \left( N_\alpha - N_\beta + \zeta \right) \right)^2 
         - 1
  \right]
\qquad
R^2
= \frac{1}{4} 
  \left[ \left( H + 
         \frac{1}{2} \left( N_\alpha - N_\beta + \zeta \right) \right)^2 
         - 1
  \right]
\label{L^2R^2inSusy}
\end{equation}

Bosonic generators in $\mathfrak{C}^{-}$ are:
\begin{equation}
\begin{split}
B_1
&= \Delta + i \left( F - E \right)
 = - \frac{i}{2} 
     \left\{ \left( x + i \, p \right)^2 
             - \frac{1}{x^2} 
               \left[ \left( N_d - N_g + N_\alpha - N_\beta + \zeta \right)^2 
                      - \frac{1}{4} 
               \right] 
     \right\}
\\
B_2
&= - i \left( E_1 + i \, F_1 \right)
 = - i \, d 
   \left[ \left( x + i \, p \right)
          + \frac{1}{x} 
            \left( N_d - N_g + N_\alpha - N_\beta + \zeta - \frac{1}{2} 
            \right)
   \right]
\\
B_3
&= - i \left( E^2 + i \, F^2 \right)
 = - i \, g 
   \left[ \left( x + i \, p \right)
          - \frac{1}{x} 
            \left( N_d - N_g + N_\alpha - N_\beta + \zeta + \frac{1}{2} 
            \right)
   \right]
\\
B_4
&= - 2 i \, J^2_1
 = - 2 i \, d \, g
\end{split}
\end{equation}
while bosonic generators in $\mathfrak{C}^{+}$ are:
\begin{equation}
\begin{split}
B^1
&= \Delta - i \left( F - E \right)
 = \frac{i}{2} 
   \left\{ \left( x - i \, p \right)^2 
           - \frac{1}{x^2} 
             \left[ \left( N_d - N_g + N_\alpha - N_\beta + \zeta \right)^2 
                    - \frac{1}{4} 
             \right] 
   \right\}
\\
B^2
&= i \left( E^1 - i \, F^1 \right)
 = i \, d^\dag 
   \left[ \left( x - i \, p \right)
          + \frac{1}{x} 
            \left( N_d - N_g + N_\alpha - N_\beta + \zeta + \frac{1}{2} 
            \right)
   \right]
\\
B^3
&= - i \left( E_2 - i \, F_2 \right)
 = i \, g^\dag 
   \left[ \left( x - i \, p \right)
          - \frac{1}{x} 
            \left( N_d - N_g + N_\alpha - N_\beta + \zeta - \frac{1}{2} 
            \right)
   \right]
\\
B^4
&= - 2 i \, J^1_2
 = 2 i \, d^\dag \, g^\dag
\end{split}
\end{equation}

The  satisfy the following commutation relations:
\begin{equation}
\begin{split}
& \commute{\mathcal{H}}{B_i}
= - \, B_i
\qquad \qquad
\commute{\mathcal{H}}{B^i}
= + \, B^i
\qquad \qquad \mbox{where} \qquad
i = 1, 2, 3, 4
\\
& \commute{B_1}{B^1}
= 8 \, H_\odot
\qquad \qquad \qquad \qquad \qquad
\commute{B_2}{B^2}
= 4 \left( H + L_3 - R_3 \right)
\\
& \commute{B_3}{B^3} 
= 4 \left( H - L_3 + R_3 \right)
\qquad \qquad
\commute{B_4}{B^4}
= 8 \left( H_d + H_g \right)
\end{split}
\end{equation}

Once again, we have the important relation
\begin{equation}
B^3 \, B^2 = B^4 \, B^1
\end{equation}
in the deformed minrep.

In this basis, the supersymmetry generators in $\mathfrak{C}^{-}$ are given by:
\begin{equation}
\begin{split}
\mathfrak{S}_\mu 
&= \frac{1}{2} \left( S_\mu + i \, Q_\mu \right)
 = \frac{1}{2} \, \alpha_\mu 
   \left[ \left( x + i \, p \right)
          + \frac{1}{x} 
            \left( N_d - N_g + N_\alpha - N_\beta + \zeta - \frac{1}{2} 
            \right)
   \right]
\\
\mathfrak{Q}_\mu
&= \widetilde{S}_\mu^2
 = \alpha_\mu g
\\
\mathfrak{S}_y 
&= \frac{1}{2} \left( S_y + i \, Q_y \right)
 = \frac{1}{2} \, \beta_y 
   \left[ \left( x + i \, p \right)
          - \frac{1}{x} 
            \left( N_d - N_g + N_\alpha - N_\beta + \zeta + \frac{1}{2} 
            \right)
   \right]
\\
\mathfrak{Q}_y
&= \widetilde{Q}_{1 \, y}
 = \beta_y d
\end{split}
\end{equation}
and the supersymmetry  generators in $\mathfrak{C}^{+}$ are given by:
\begin{equation}
\begin{split}
\mathfrak{S}^\mu 
&= \frac{1}{2} \left( S^\mu - i \, Q^\mu \right)
 = \frac{1}{2} \, \alpha^\mu 
   \left[ \left( x - i \, p \right)
          + \frac{1}{x} 
            \left( N_d - N_g + N_\alpha - N_\beta + \zeta + \frac{1}{2} 
            \right)
   \right]
\\
\mathfrak{Q}^\mu
&= \widetilde{S}^\mu_2
 = \alpha^\mu g^\dag
\\
\mathfrak{S}^y 
&= \frac{1}{2} \left( S^y - i \, Q^y \right)
 = \frac{1}{2} \, \beta^y 
   \left[ \left( x - i \, p \right)
          - \frac{1}{x} 
            \left( N_d - N_g + N_\alpha - N_\beta + \zeta - \frac{1}{2} 
            \right)
   \right]
\\
\mathfrak{Q}^y
&= \widetilde{Q}^{1 \, y}
 = \beta^y d^\dag
\end{split}
\end{equation}

Note that
\begin{equation}
\mathfrak{S}^\mu = \left( \mathfrak{S}_\mu \right)^\dag
\qquad \qquad
\mathfrak{Q}^\mu = \left( \mathfrak{Q}_\mu \right)^\dag
\qquad \qquad
\mathfrak{S}^y = \left( \mathfrak{S}_y \right)^\dag
\qquad \qquad
\mathfrak{Q}^y = \left( \mathfrak{Q}_y \right)^\dag \,.
\end{equation}

These supersymmetry generators in $\mathfrak{C}^{(\pm )}$ satisfy the following (anti-)commutation relations:
\begin{equation}
\begin{split}
\commute{\mathcal{H}}{\mathfrak{S}_\mu} = - \, \mathfrak{S}_\mu
& \qquad
\commute{\mathcal{H}}{\mathfrak{Q}_\mu} = - \, \mathfrak{Q}_\mu
\qquad
\commute{\mathcal{H}}{\mathfrak{S}_y} = - \, \mathfrak{S}_y
\qquad
\commute{\mathcal{H}}{\mathfrak{Q}_y} = - \, \mathfrak{Q}_y
\\
\commute{\mathcal{H}}{\mathfrak{S}^\mu} = + \, \mathfrak{S}^\mu
& \qquad
\commute{\mathcal{H}}{\mathfrak{Q}^\mu} = + \, \mathfrak{Q}^\mu
\qquad
\commute{\mathcal{H}}{\mathfrak{S}^y} = + \, \mathfrak{S}^y
\qquad
\commute{\mathcal{H}}{\mathfrak{Q}^y} = + \, \mathfrak{Q}^y
\end{split}
\end{equation}

\begin{equation}
\begin{split}
\anticommute{\mathfrak{S}_\mu}{\mathfrak{S}^\nu}
&= \delta^\nu_\mu \left( H + L_3 - R_3 \right)
   - \mathcal{A}^\nu_\mu
   - \delta^\nu_\mu \left( N_d + \frac{1}{\mathfrak{p}} N_\alpha \right)
\\
\anticommute{\mathfrak{Q}_\mu}{\mathfrak{Q}^\nu}
&= \delta^\nu_\mu \left( H_d + H_g \right)
   - \mathcal{A}^\nu_\mu
   - \delta^\nu_\mu \left( N_d + \frac{1}{\mathfrak{p}} N_\alpha \right)
\\
\anticommute{\mathfrak{S}_y}{\mathfrak{S}^z}
&= \delta^z_y \left( H - L_3 + R_3 \right)
   - \mathcal{B}^z_y
   - \delta^z_y \left( N_g + \frac{1}{\mathfrak{q}} N_\beta \right)
\\
\anticommute{\mathfrak{Q}_y}{\mathfrak{Q}^z}
&= \delta^\nu_\mu \left( H_d + H_g \right)
   - \mathcal{B}^z_y 
   - \delta^z_y \left( N_g + \frac{1}{\mathfrak{q}} N_\beta \right)
\end{split}
\end{equation}

The supersymmetry  generators in grade 0 space $\mathfrak{C}^{0}$ are 
obtained by taking the commutators between fermionic (bosonic) generators 
in $\mathfrak{C}^{-}$ space and bosonic (fermionic) generators in 
$\mathfrak{C}^{+}$ space. They are as follows:
\begin{equation}
\begin{split}
\widetilde{\mathfrak{S}}_\mu 
&= \frac{1}{2} \left( S_\mu - i \, Q_\mu \right)
 = \frac{1}{2} \, \alpha_\mu 
   \left[ \left( x - i \, p \right)
          - \frac{1}{x} 
            \left( N_d - N_g + N_\alpha - N_\beta + \zeta - \frac{1}{2} \right)
   \right]
\\
\widetilde{\mathfrak{Q}}_\mu
&= \widetilde{Q}_\mu^1
 = \alpha_\mu d^\dag
\\
\widetilde{\mathfrak{S}}^\mu 
&= \frac{1}{2} \left( S^\mu + i \, Q^\mu \right)
 = \frac{1}{2} \, \alpha^\mu 
   \left[ \left( x + i \, p \right)
          - \frac{1}{x} 
            \left( N_d - N_g + N_\alpha - N_\beta + \zeta + \frac{1}{2} \right)
   \right]
\\
\widetilde{\mathfrak{Q}}^\mu
&= \widetilde{Q}^\mu_1
 = \alpha^\mu d
\\
\widetilde{\mathfrak{S}}_y 
&= \frac{1}{2} \left( S_y - i \, Q_y \right)
 = \frac{1}{2} \, \beta_y 
   \left[ \left( x - i \, p \right)
          + \frac{1}{x} 
            \left( N_d - N_g + N_\alpha - N_\beta + \zeta + \frac{1}{2} \right)
   \right]
\\
\widetilde{\mathfrak{Q}}_y
&= \widetilde{S}_{2 \, y}
 = \beta_y g^\dag
\\
\widetilde{\mathfrak{S}}^y 
&= \frac{1}{2} \left( S^y + i \, Q^y \right)
 = \frac{1}{2} \, \beta^y 
   \left[ \left( x + i \, p \right)
          + \frac{1}{x} 
            \left( N_d - N_g + N_\alpha - N_\beta + \zeta - \frac{1}{2} \right)
   \right]
\\
\widetilde{\mathfrak{Q}}^y
&= \widetilde{S}^{2 \, y}
 = \beta^y g
\end{split}
\end{equation}
The anti-commutators between the supersymmetry generators in 
$\mathfrak{C}^-$ and those in $\mathfrak{C}^+$ take the following form:
\begin{equation}
\begin{split}
\anticommute{\widetilde{\mathfrak{S}}_\mu}{\widetilde{\mathfrak{S}}^\nu}
&= - 2 \, \delta^\nu_\mu \, L_3
   + A^\nu_\mu
   + \delta^\nu_\mu \left( N_d + \frac{1}{\mathfrak{p}} N_\alpha \right)
\\
\anticommute{\widetilde{\mathfrak{Q}}_\mu}{\widetilde{\mathfrak{Q}}^\nu}
&= A^\nu_\mu
   + \delta^\nu_\mu \left( N_d + \frac{1}{\mathfrak{p}} N_\alpha \right)
\\
\anticommute{\widetilde{\mathfrak{S}}_y}{\widetilde{\mathfrak{S}}^z}
&= - 2 \, \delta^z_y \, R_3
   + B^z_y
   + \delta^z_y \left( N_g + \frac{1}{\mathfrak{q}} N_\beta \right)
\\
\anticommute{\widetilde{\mathfrak{Q}}_y}{\widetilde{\mathfrak{Q}}^z}
&= B^z_y
   + \delta^z_y \left( N_g + \frac{1}{\mathfrak{q}} N_\beta \right)
\end{split}
\end{equation}

The commutators between bosonic generators in $\mathfrak{C}^-$ and 
supersymmetry generators in $\mathfrak{C}^+$ are as follows:
\begin{equation}
\begin{aligned}
\commute{B_1}{\mathfrak{S}^\mu}
 &= - 2 i \, \widetilde{\mathfrak{S}}^\mu
\\
\commute{B_1}{\mathfrak{Q}^\mu}
 &= 0
\\
\commute{B_1}{\mathfrak{S}^y}
 &= - 2 i \, \widetilde{\mathfrak{S}}^y
\\
\commute{B_1}{\mathfrak{Q}^y}
 &= 0
\end{aligned}
\qquad
\begin{aligned}
\commute{B_2}{\mathfrak{S}^\mu}
 &= - 2 i \, \widetilde{\mathfrak{Q}}^\mu
\\
\commute{B_2}{\mathfrak{Q}^\mu}
 &= 0
\\
\commute{B_2}{\mathfrak{S}^y}
 &= 0
\\
\commute{B_2}{\mathfrak{Q}^y}
 &= - 2 i \, \widetilde{\mathfrak{S}}^y
\end{aligned}
\qquad
\begin{aligned}
\commute{B_3}{\mathfrak{S}^\mu}
 &= 0
\\
\commute{B_3}{\mathfrak{Q}^\mu}
 &= - 2 i \, \widetilde{\mathfrak{S}}^\mu
\\
\commute{B_3}{\mathfrak{S}^y}
 &= - 2 i \, \widetilde{\mathfrak{Q}}^y
\\
\commute{B_3}{\mathfrak{Q}^y}
 &= 0
\end{aligned}
\qquad
\begin{aligned}
\commute{B_4}{\mathfrak{S}^\mu}
 &= 0
\\
\commute{B_4}{\mathfrak{Q}^\mu}
 &= - 2 i \, \widetilde{\mathfrak{Q}}^\mu
\\
\commute{B_4}{\mathfrak{S}^y}
 &= 0
\\
\commute{B_4}{\mathfrak{Q}^y}
 &= - 2 i \, \widetilde{\mathfrak{Q}}^y
\end{aligned}
\end{equation}

The anticommutators of supersymmetry generators in $\mathfrak{C}^0$ and 
those in $\mathfrak{C}^+$ can be written as
\begin{equation}
\begin{aligned}
\anticommute{\widetilde{\mathfrak{S}}^\mu}{\mathfrak{S}^\nu}
 &= 0
\\
\anticommute{\widetilde{\mathfrak{Q}}^\mu}{\mathfrak{S}^\nu}
 &= 0
\\
\anticommute{\widetilde{\mathfrak{S}}^y}{\mathfrak{S}^\nu}
 &= \mathcal{C}^{y \nu}
\\
\anticommute{\widetilde{\mathfrak{Q}}^y}{\mathfrak{S}^\nu}
 &= 0
\end{aligned}
\qquad
\begin{aligned}
\anticommute{\widetilde{\mathfrak{S}}^\mu}{\mathfrak{Q}^\nu}
 &= 0
\\
\anticommute{\widetilde{\mathfrak{Q}}^\mu}{\mathfrak{Q}^\nu}
 &= 0
\\
\anticommute{\widetilde{\mathfrak{S}}^y}{\mathfrak{Q}^\nu}
 &= 0
\\
\anticommute{\widetilde{\mathfrak{Q}}^y}{\mathfrak{Q}^\nu}
 &= \mathcal{C}^{y \nu}
\end{aligned}
\qquad
\begin{aligned}
\anticommute{\widetilde{\mathfrak{S}}^\mu}{\mathfrak{S}^z}
 &= - \mathcal{C}^{z \mu}
\\
\anticommute{\widetilde{\mathfrak{Q}}^\mu}{\mathfrak{S}^z}
 &= 0
\\
\anticommute{\widetilde{\mathfrak{S}}^y}{\mathfrak{S}^z}
 &= 0
\\
\anticommute{\widetilde{\mathfrak{Q}}^y}{\mathfrak{S}^z}
 &= 0
\end{aligned}
\qquad
\begin{aligned}
\anticommute{\widetilde{\mathfrak{S}}^\mu}{\mathfrak{Q}^z}
 &= 0
\\
\anticommute{\widetilde{\mathfrak{Q}}^\mu}{\mathfrak{Q}^z}
 &= - \mathcal{C}^{z \mu}
\\
\anticommute{\widetilde{\mathfrak{S}}^y}{\mathfrak{Q}^z}
 &= 0
\\
\anticommute{\widetilde{\mathfrak{Q}}^y}{\mathfrak{Q}^z}
 &= 0
\end{aligned}
\end{equation}
where $\mathcal{C}^{z \mu} = \beta^z \alpha^\mu$ are the 
$\mathfrak{su}(\mathfrak{p} + \mathfrak{q})$ generators that belong to the 
$\mathfrak{C}^+$ subspace.

%%%%%%%%%%%%%%%%%%%%%%%%%%%%%%%%%%%%%%%%%%%%%%%%%%%%%%%%%%%%%%%%%%%%
%%%%%% Section: Minimal supermultiplets of SU(2,2|4) %%%%%%%%%%%%%
%%%%%%%%%%%%%%%%%%%%%%%%%%%%%%%%%%%%%%%%%%%%%%%%%%%%%%%%%%%%%%%%%%%%

\section{Minimal Unitary Supermultiplet of $\mathfrak{su}\left( 2 , 
2 \,|\, 4 \right)$ and its Deformations}
\label{sec:SU(2,2|4)}

From equations (\ref{L^2R^2inSusy}), it follows that the quadratic Casimir 
operators of  $SU(2)_L$ and $SU(2)_R$ can be written as 
\eq
L^2 = \mathfrak{L} \left( \mathfrak{L} + 1 \right)
\qquad \qquad \qquad
R^2 = \mathfrak{R} \left( \mathfrak{R} + 1 \right)
\en
where 
\begin{equation}
\mathfrak{L}
 = \frac{1}{2}
   \left[ H - \frac{1}{2} \left( N_\alpha - N_\beta + \zeta \right) - 1 
   \right]
\qquad \qquad
\mathfrak{R}
 = \frac{1}{2}
   \left[ H + \frac{1}{2} \left( N_\alpha - N_\beta + \zeta \right) - 1 
   \right] \,.
\end{equation}
As we have shown earlier, $H$ (AdS energy) is given by equation 
(\ref{NonSusyH}).

As  before we shall denote the lowest energy state of the singular 
(isotonic) oscillator with coordinate wave function
\begin{equation}
C_0 \, x^\alpha \, e^{- x^2 / 2} 
\end{equation}
as $\ket{\psi^{(\alpha)}}$ and its tensor product with the vacua of the bosonic 
and fermionic oscillators as $\ket{\alpha ; 0 , 0 ; 0 , 0}$. Note that 
we use the notation $\ket{\alpha ; n_d , n_g ; n_\alpha , n_\beta}$, 
where $n_d$, $n_g$, $n_\alpha$, $n_\beta$ are the eigenvalues of the 
respective bosonic and fermionic number operators.

Clearly,
 \eq
 d \, \ket{\alpha ; 0 , 0 ; 0 , 0} 
 = g \, \ket{\alpha ; 0 , 0 ; 0 , 0} 
 = \alpha_\mu \, \ket{\alpha ; 0 , 0 ; 0 , 0} 
 = \beta_y \, \ket{\alpha ; 0 , 0 ; 0 , 0} 
 = 0 \,.
 \en 

We shall study the case  $\mathfrak{p} = \mathfrak{q} = 
2$. Twistorial oscillator construction of the unitary supermultiplets of $SU(2,2\,|\,4)$ has been studied in  \cite{Gunaydin:1984fk,Gunaydin:1998sw,Gunaydin:1998jc,Claus:1999xr}.

%%%%%%%%%%%%%%%%%%%%%%%%%%%%%%%%%%%%%%%%%%%%%%%%%%%%%%%%%%%%%%%%%%%%
%%%%%% Subsection: Doubleton supermultiplets when \zeta = 0 %%%%%%%%
%%%%%%%%%%%%%%%%%%%%%%%%%%%%%%%%%%%%%%%%%%%%%%%%%%%%%%%%%%%%%%%%%%%%

\subsection{Minimal Unitary Supermultiplet  of $\mathfrak{su}\left( 2 , 2 
\,|\, 4 \right)$ }

Let us first analyze the minimal unitary supermultiplet of  $\mathfrak{su} 
\left( 2 , 2 \,|\, 4 \right)$
for which the deformation parameter is zero
\begin{equation}
\zeta = 0 \,.
\end{equation}

The state $\ket{\frac{1}{2} ; 0 , 0 ; 0 , 0}$ is the unique normalizable 
lowest energy state annihilated by all bosonic generators $B_i$ ($i = 1, 2, 
3, 4$) as well as supersymmetry generators in $\mathfrak{C}^{-}$ subspace. 
It is  a singlet of $SU(2 \,|\, 2) \times SU(2 \,|\, 2)$ subalgebra. 
By acting on it with the grade $+1$ generators in the subspace 
$\mathfrak{C}^+$ one obtains  an infinite set of states which form a basis 
for the minimal unitary irreducible representation of $\mathfrak{su}(2,2 
\,|\, 4)$. This infinite set of states can be decomposed into a finite 
number of irreducible representations of the even subgroup $SU(2,2) 
\times SU(4) $, with each irrep of $SU(2,2)$ corresponding to a massless 
conformal field in four dimensions.

In Table \ref{Table:Supermultiplet_zeta=0}, we present the supermultiplet 
that is obtained by starting from this unique lowest weight vector
\begin{equation}
\ket{\Omega} = \ket{\frac{1}{2} ; 0 , 0 ; 0 , 0}
\end{equation}
and acting on it with the supersymmetry generators of grade $+1$ space 
$\mathfrak{C}^+$.  The  resultant supermultiplet is  the $N=4$ Yang-Mills supermultiplet constructed 
long ago in \cite{Gunaydin:1984fk} which was called the CPT self-conjugate doubleton supermultiplet. In the twistorial oscillator approach the lowest weight vector $\ket{\Omega}$  is the vacuum vector $ \ket{0}$ of all the oscillators in 
the $SU(2\,|\,2)\times SU(2\,|\,2) \times U(1)$ basis \cite{Gunaydin:1984fk,Gunaydin:1998sw,Gunaydin:1998jc}.\footnote{Note that in \cite{Gunaydin:1984fk,Gunaydin:1998sw,Gunaydin:1998jc}  $H$ is the $\mathfrak{u}(1)$ generator corresponding to  $AdS_5$ energy, which is denoted as $E$ .} We should note that  the positive energy 
   unitary representations of $SU(2,2)$ are uniquely determined by the $SU(2)_L \times SU(2)_R \times U(1)$ labels
$ \mathcal{L},\mathcal{R}, H$ of their lowest energy states. Thus the $SU(2,2)\times SU(4)$ decomposition of the unitary supermultiplets of $SU(2,2\,|\,4)$ given in the tables below can be read off from these labels together with the dimensions of irreps of $SU(4)$ that are listed. The eigenvalue of $H$ is simply the conformal dimension of the corresponding massless field in four dimensions.

% Supermultiplet zeta = 0
%%%%%%%%%%%%%%%%%%%%%%%%%%%%%%%%%%%%%%%%%%%%%%%%%%%%%%%%%%%%%%%%%%%%%%%%%%%%
\begin{tiny}
\begin{longtable}[c]{|r|l||c|c||c|r||c|r||c|}
\kill
%%%%%%%%%%%%%%%%%%%%%%%%%%%%%%%%%%%%%%%%%%%%%%%%%%%%%%%%%%%%%%%%%%%%%%%%%%%%
\caption[The doubleton supermultiplet corresponding to $zeta = 0$]
{The minimal unitary supermultiplet of $\mathfrak{su}\left( 2 , 2 
\,|\, 4 \right)$ with the lowest weight vector indicated with an asterisk. $SU(2)_L \times SU(2)_R \times U(1)$  labels of the unitary representations of $SU(2,2)$ are denoted as $\mathcal{L},\mathcal{R}$ and $H$. 
\label{Table:Supermultiplet_zeta=0}} \\
\hline
& & & & & & & & \\
State & $\ket{\alpha ; n_d , n_g ; n_\alpha , n_\beta}$ & 
$H$ & $\mathcal{H}$ & 
$\mathfrak{L}$ & $\mathfrak{L}_3$ & $\mathfrak{R}$ & $\mathfrak{R}_3$ & 
$SU(4)$ \\
& & & & & & & & \\
\hline
& & & & & & & & \\
\endfirsthead
%%%%%%%%%%%%%%%%%%%%%%%%%%%%%%%%%%%%%%%%%%%%%%%%%%%%%%%%%%%%%%%%%%%%%%%%%%%%
\caption[]{(continued)} \\
\hline
& & & & & & & & \\
State & $\ket{\alpha ; n_d , n_g ; n_\alpha , n_\beta}$ & 
$H$ & $\mathcal{H}$ & 
$\mathfrak{L}$ & $\mathfrak{L}_3$ & $\mathfrak{R}$ & $\mathfrak{R}_3$ & 
$SU(4)$ \\
& & & & & & & & \\
\hline
& & & & & & & & \\
\endhead
%%%%%%%%%%%%%%%%%%%%%%%%%%%%%%%%%%%%%%%%%%%%%%%%%%%%%%%%%%%%%%%%%%%%%%%%%%%%
& & & & & & & & \\
\hline
\endfoot
%%%%%%%%%%%%%%%%%%%%%%%%%%%%%%%%%%%%%%%%%%%%%%%%%%%%%%%%%%%%%%%%%%%%%%%%%%%%
& & & & & & & & \\
\hline
\endlastfoot
%%%%%%%%%%%%%%%%%%%%%%%%%%%%%%%%%%%%%%%%%%%%%%%%%%%%%%%%%%%%%%%%%%%%%%%%%%%%

$\ket{\frac{1}{2} ; 0 , 0 ; 0 , 0}^*$ &
$\ket{\frac{1}{2} ; 0 , 0 ; 0 , 0}^*$ &
1 & 0 &
0 & 0 & 0 & 0 &
6 \\[8pt]

\hline
& & & & & & & & \\

$\mathfrak{S}^y \ket{\frac{1}{2} ; 0 , 0 ; 0 , 0}$ &
$\ket{\frac{3}{2} ; 0 , 0 ; 0 , 1}$ &
$\frac{3}{2}$ & 1 & 
$\frac{1}{2}$ & $-\frac{1}{2}$ & 0 & 0 &
$\overline{4}$ \\[8pt]

$\mathfrak{Q}^y \ket{\frac{1}{2} ; 0 , 0 ; 0 , 0}$ &
$\ket{\frac{1}{2} ; 1 , 0 ; 0 , 1}$ &
 & & 
$\frac{1}{2}$ & $+\frac{1}{2}$ & 0 & 0 &
 \\[8pt]

\hline
& & & & & & & & \\

$\mathfrak{S}^\mu \ket{\frac{1}{2} ; 0 , 0 ; 0 , 0}$ &
$\ket{\frac{3}{2} ; 0 , 0 ; 1 , 0}$ &
$\frac{3}{2}$ & 1 & 
0 & 0 & $\frac{1}{2}$ & $-\frac{1}{2}$ &
4 \\[8pt]

$\mathfrak{Q}^\mu \ket{\frac{1}{2} ; 0 , 0 ; 0 , 0}$ &
$\ket{\frac{1}{2} ; 0 , 1 ; 1 , 0}$ &
 & & 
0 & 0 & $\frac{1}{2}$ & $+\frac{1}{2}$ &
 \\[8pt]

\hline
& & & & & & & & \\

$\mathfrak{S}^y \mathfrak{S}^z \ket{\frac{1}{2} ; 0 , 0 ; 0 , 0}$ &
$\ket{\frac{5}{2} ; 0 , 0 ; 0 , 2}$ &
2 & 2 & 
1 & $-1$ & 0 & 0 &
$\overline{1}$ \\[8pt]

$\mathfrak{S}^y \mathfrak{Q}^z \ket{\frac{1}{2} ; 0 , 0 ; 0 , 0}$ &
$\ket{\frac{3}{2} ; 1 , 0 ; 0 , 2}$ &
 & & 
1 & 0 & 0 & 0 &
 \\[8pt]

$\mathfrak{Q}^y \mathfrak{Q}^z \ket{\frac{1}{2} ; 0 , 0 ; 0 , 0}$ &
$\ket{\frac{1}{2} ; 2 , 0 ; 0 , 2}$ &
 & & 
1 & $+1$ & 0 & 0 &
 \\[8pt]

\hline
& & & & & & & & \\

$\mathfrak{S}^\mu \mathfrak{S}^\nu \ket{\frac{1}{2} ; 0 , 0 ; 0 , 0}$ &
$\ket{\frac{5}{2} ; 0 , 0 ; 2 , 0}$ &
2 & 2 & 
0 & 0 & 1 & $-1$ &
1 \\[8pt]

$\mathfrak{S}^\mu \mathfrak{Q}^\nu \ket{\frac{1}{2} ; 0 , 0 ; 0 , 0}$ &
$\ket{\frac{3}{2} ; 0 , 1 ; 2 , 0}$ &
 & & 
0 & 0 & 1 & 0 &
 \\[8pt]

$\mathfrak{Q}^\mu \mathfrak{Q}^\nu \ket{\frac{1}{2} ; 0 , 0 ; 0 , 0}$ &
$\ket{\frac{1}{2} ; 0 , 2 ; 2 , 0}$ &
 & & 
0 & 0 & 1 & $+1$ &
 \\[8pt]

%%%%%%%%%%%%%%%%%%%%%%%%%%%%%%%%%%%%%%%%%%%%%%%%%%%%%%%%%%%%%%%%%%%%%%%%%%%%
\end{longtable}
\end{tiny}
%%%%%%%%%%%%%%%%%%%%%%%%%%%%%%%%%%%%%%%%%%%%%%%%%%%%%%%%%%%%%%%%%%%%%%%%%%%%

%%%%%%%%%%%%%%%%%%%%%%%%%%%%%%%%%%%%%%%%%%%%%%%%%%%%%%%%%%%%%%%%%%%%
%%%%%% Subsection: Doubleton supermultiplets when \zeta \ne 0 %%%%%%
%%%%%%%%%%%%%%%%%%%%%%%%%%%%%%%%%%%%%%%%%%%%%%%%%%%%%%%%%%%%%%%%%%%%

\subsection{Deformed minimal unitary  supermultiplets of $\mathfrak{su}\left( 2 , 2 
\,|\, 4 \right)$ for  $\zeta \ne 0$}

When $\zeta \ne 0$, there is a multiplet of  states that are annihilated by the generators in $\mathfrak{C}^-$  and transform irreducibly under the subalgebra $\mathfrak{C}^0$. 
By an abuse of terminology we shall refer to them  as ``lowest 
weight vectors'' $\ket{\Omega}$ for any given non-zero integer value of 
$\zeta$.

In Table \ref{Table:zetaLWVs}, we list all those states 
$\ket{\Omega}$ that are annihilated by all the generators (bosonic and 
fermionic) in grade $-1$ space $\mathfrak{C}^-$ 
of $SU \left( 2 , 2 \,|\, 4 \right)$ for various values of $\zeta \ne 0$. 
For a given $\zeta$, they form an irreducible representation of 
$SU(2\,|\,2)_L \times SU(2\,|\,2)_R$ whose supertableau is
\begin{equation}
\begin{split}
& | \, \underbrace{\sgenrowbox}_{-\zeta} \,,\, 1 \rangle
\qquad \qquad \mbox{for $\zeta < 0$}
\\
& | 1 \,,\, \underbrace{\sgenrowbox}_{\zeta} \, \rangle
\qquad \qquad \mbox{for $\zeta > 0$} \,.
\end{split}
\label{supertabirreps}
\end{equation}

% Complete Table - LWV zeta
%%%%%%%%%%%%%%%%%%%%%%%%%%%%%%%%%%%%%%%%%%%%%%%%%%%%%%%%%%%%%%%%%%%%%%%%%%%%
\begin{tiny}
\begin{longtable}[c]{|r|l||c|r||c|c||c|c|}
\kill
%%%%%%%%%%%%%%%%%%%%%%%%%%%%%%%%%%%%%%%%%%%%%%%%%%%%%%%%%%%%%%%%%%%%%%%%%%%%
\caption[States $\ket{\Omega}$ that are annihilated by all grade $-1$ 
generators in $\mathfrak{C}^-$ within the minimal unitary representation 
space of $SU\left( 2 , 2 \,|\, 4 \right)$ with a deformation parameter 
$\zeta \ne 0$]
{States $\ket{\Omega}$ that are annihilated by all grade $-1$ generators 
in $\mathfrak{C}^-$ within the minimal unitary representation space of 
$SU \left( 2 , 2 \,|\, 4 \right)$ with a deformation parameter $\zeta \ne 
0$.
\label{Table:zetaLWVs}} \\
\hline
& & & & & & & \\
LWV & Range & $H$ & $\mathcal{H}$ & 
$\mathfrak{L}$ & $\mathfrak{L}_3$ & $\mathfrak{R}$ & $\mathfrak{R}_3$ \\
& & & & & & & \\
\hline
& & & & & & & \\
\endfirsthead
%%%%%%%%%%%%%%%%%%%%%%%%%%%%%%%%%%%%%%%%%%%%%%%%%%%%%%%%%%%%%%%%%%%%%%%%%%%%
\caption[]{(continued)} \\
\hline
& & & & & & & \\
LWV & Range & $H$ & $\mathcal{H}$ & 
$\mathfrak{L}$ & $\mathfrak{L}_3$ & $\mathfrak{R}$ & $\mathfrak{R}_3$ \\
& & & & & & & \\
\hline
& & & & & & & \\
\endhead
%%%%%%%%%%%%%%%%%%%%%%%%%%%%%%%%%%%%%%%%%%%%%%%%%%%%%%%%%%%%%%%%%%%%%%%%%%%%
& & & & & & & \\
\hline
\endfoot
%%%%%%%%%%%%%%%%%%%%%%%%%%%%%%%%%%%%%%%%%%%%%%%%%%%%%%%%%%%%%%%%%%%%%%%%%%%%
& & & & & & & \\
\hline
\endlastfoot
%%%%%%%%%%%%%%%%%%%%%%%%%%%%%%%%%%%%%%%%%%%%%%%%%%%%%%%%%%%%%%%%%%%%%%%%%%%%

$\ket{\frac{1}{2} - \zeta ; 0 , 0 ; 0 , 0}$ &
$\zeta = -1, -2, -3, \dots$ &
$1 - \frac{\zeta}{2}$ & $- \frac{\zeta}{2}$ &
$- \frac{\zeta}{2}$ & $\frac{\zeta}{2}$ & 0 & 0 \\[8pt]

$\ket{\frac{1}{2} + \zeta ; 0 , 0 ; 0 , 0}$ &
$\zeta = 1, 2, 3, \dots$ &
$1 + \frac{\zeta}{2}$ & $\frac{\zeta}{2}$ &
0 & 0 & $\frac{\zeta}{2}$ & $- \frac{\zeta}{2}$ \\[8pt]

$\ket{\frac{1}{2} - n - \zeta ; n , 0 ; 0 , 0}$ &
$\zeta = -1, -2, -3, \dots$ &
$1 - \frac{\zeta}{2}$ & $- \frac{\zeta}{2}$ & 
$- \frac{\zeta}{2}$ & $n + \frac{\zeta}{2}$ & 0 & 0 \\
 &
$n = 1, 2, \dots , -\zeta$ &
 & & 
 & & & \\[8pt]

$\ket{\frac{1}{2} - m + \zeta ; 0 , m ; 0 , 0}$ &
$\zeta = 1, 2, 3, \dots$ &
$1 + \frac{\zeta}{2}$ & $\frac{\zeta}{2}$ &
0 & 0 & $\frac{\zeta}{2}$ & $m - \frac{\zeta}{2}$ \\
 &
$m = 1, 2, \dots , \zeta$ &
 & & 
 & & & \\[8pt]

$\ket{\frac{1}{2} - p - \zeta ; 0 , 0 ; p , 0}$ &
$\zeta = -1, -2, -3, \dots$ &
$1 - \frac{p + \zeta}{2}$ & $- \frac{\zeta}{2}$ &
$- \frac{\zeta + p}{2}$ & $\frac{\zeta + p}{2}$ & 0 & 0 \\
 &
$p = 1 \mbox{ for } \zeta=-1$ &
 & & 
 & & & \\
 &
$p = 1, 2 \mbox{ otherwise}$ &
 & & 
 & & & \\[8pt]

$\ket{\frac{1}{2} - q + \zeta ; 0 , 0 ; 0 , q}$ &
$\zeta = 1, 2, 3, \dots$ &
$1 - \frac{q - \zeta}{2}$ & $\frac{\zeta}{2}$ &
0 & 0 & $\frac{\zeta - q}{2}$ & $- \frac{\zeta - q}{2}$ \\
 &
$q = 1 \mbox{ for } \zeta=1$ &
 & & 
 & & & \\
 &
$q = 1, 2 \mbox{ otherwise}$ &
 & & 
 & & & \\[8pt]

$\ket{\frac{1}{2} - n - p - \zeta ; n , 0 ; p , 0}$ &
$\zeta = -1, -2, -3, \dots$ &
$1 - \frac{p + \zeta}{2}$ & $- \frac{\zeta}{2}$ &
$- \frac{\zeta + p}{2}$ & $n + \frac{\zeta + p}{2}$ & 0 & 0 \\
 &
$p = 1 \mbox{ for } \zeta=-1$ &
 & & 
 & & & \\
 &
$p = 1, 2 \mbox{ otherwise}$ &
 & & 
 & & & \\[8pt]

$\ket{\frac{1}{2} - m - q + \zeta ; 0 , m ; 0 , q}$ &
$\zeta = 1, 2, 3, \dots$ &
$1 - \frac{q - \zeta}{2}$ & $\frac{\zeta}{2}$ &
0 & 0 & $\frac{\zeta - q}{2}$ & $m - \frac{\zeta - q}{2}$ \\
 &
$q = 1 \mbox{ for } \zeta=1$ &
 & & 
 & & & \\
 &
$q = 1, 2 \mbox{ otherwise}$ &
 & & 
 & & & \\[8pt]

%%%%%%%%%%%%%%%%%%%%%%%%%%%%%%%%%%%%%%%%%%%%%%%%%%%%%%%%%%%%%%%%%%%%%%%%%%%%
\end{longtable}
\end{tiny}
%%%%%%%%%%%%%%%%%%%%%%%%%%%%%%%%%%%%%%%%%%%%%%%%%%%%%%%%%%%%%%%%%%%%%%%%%%%%

Now, for each $\zeta \left( \ne 0 \right)$, we can identify separately the 
lowest weight vectors 
$\ket{\Omega}$ that are annihilated by all the generators in 
$\mathfrak{C}^-$. For convenience, below in Tables \ref{Table:LWVs_zeta=-1}, 
\ref{Table:LWVs_zeta=+1}, \ref{Table:LWVs_zeta=-2}, and 
\ref{Table:LWVs_zeta=+2}, we list those states $\ket{\Omega}$ for $\zeta 
= -1, +1, -2, +2$. Clearly, in each case the possible lowest weight vectors 
form an irreducible representation of $SU(2\,|\,2)_L \times SU(2\,|\,2)_R$ 
whose supertableau is given by equation (\ref{supertabirreps}).

% LWV zeta = -1
%%%%%%%%%%%%%%%%%%%%%%%%%%%%%%%%%%%%%%%%%%%%%%%%%%%%%%%%%%%%%%%%%%%%%%%%%%%%
\begin{tiny}
\begin{longtable}[c]{|r||c|c||c|c||c|c|}
\kill
%%%%%%%%%%%%%%%%%%%%%%%%%%%%%%%%%%%%%%%%%%%%%%%%%%%%%%%%%%%%%%%%%%%%%%%%%%%%
\caption[States $\ket{\Omega}$ that are annihilated by all grade $-1$ 
generators within the minimal unitary representation space of $SU\left( 2 
, 2 \,|\, 4 \right)$ when $\zeta = -1$]
{States $\ket{\Omega}$ that are annihilated by all grade $-1$ 
generators within the minimal unitary representation space of $SU\left( 2 
, 2 \,|\, 4 \right)$ when $\zeta = -1$. They transform in the irreducible 
representation $| \, \scalebox{0.6}{\sonebox} \,,\, 1 \rangle$ of 
$SU(2\,|\,2)_L \times SU(2\,|\,2)_R$.
\label{Table:LWVs_zeta=-1}} \\
\hline
& & & & & & \\
LWV & $H$ & $\mathcal{H}$ & 
$\mathfrak{L}$ & $\mathfrak{L}_3$ & $\mathfrak{R}$ & $\mathfrak{R}_3$ \\
& & & & & & \\
\hline
& & & & & & \\
\endfirsthead
%%%%%%%%%%%%%%%%%%%%%%%%%%%%%%%%%%%%%%%%%%%%%%%%%%%%%%%%%%%%%%%%%%%%%%%%%%%%
\caption[]{(continued)} \\
\hline
& & & & & & \\
LWV & $H$ & $\mathcal{H}$ & 
$\mathfrak{L}$ & $\mathfrak{L}_3$ & $\mathfrak{R}$ & $\mathfrak{R}_3$ \\
& & & & & & \\
\hline
& & & & & & \\
\endhead
%%%%%%%%%%%%%%%%%%%%%%%%%%%%%%%%%%%%%%%%%%%%%%%%%%%%%%%%%%%%%%%%%%%%%%%%%%%%
& & & & & & \\
\hline
\endfoot
%%%%%%%%%%%%%%%%%%%%%%%%%%%%%%%%%%%%%%%%%%%%%%%%%%%%%%%%%%%%%%%%%%%%%%%%%%%%
& & & & & & \\
\hline
\endlastfoot
%%%%%%%%%%%%%%%%%%%%%%%%%%%%%%%%%%%%%%%%%%%%%%%%%%%%%%%%%%%%%%%%%%%%%%%%%%%%

$\ket{\frac{3}{2} ; 0 , 0 ; 0 , 0}$ &
$\frac{3}{2}$ & $\frac{1}{2}$ &
$\frac{1}{2}$ & $-\frac{1}{2}$ & 0 & 0 \\[8pt]

$\ket{\frac{1}{2} ; 1 , 0 ; 0 , 0}$ &
$\frac{3}{2}$ & $\frac{1}{2}$ & 
$\frac{1}{2}$ & $\frac{1}{2}$ & 0 & 0 \\[8pt]

$\ket{\frac{1}{2} ; 0 , 0 ; 1 , 0}$ &
1 & $\frac{1}{2}$ &
0 & 0 & 0 & 0 \\[8pt]

%%%%%%%%%%%%%%%%%%%%%%%%%%%%%%%%%%%%%%%%%%%%%%%%%%%%%%%%%%%%%%%%%%%%%%%%%%%%
\end{longtable}
\end{tiny}
%%%%%%%%%%%%%%%%%%%%%%%%%%%%%%%%%%%%%%%%%%%%%%%%%%%%%%%%%%%%%%%%%%%%%%%%%%%%

% LWV zeta = +1
%%%%%%%%%%%%%%%%%%%%%%%%%%%%%%%%%%%%%%%%%%%%%%%%%%%%%%%%%%%%%%%%%%%%%%%%%%%%
\begin{tiny}
\begin{longtable}[c]{|r||c|c||c|c||c|c|}
\kill
%%%%%%%%%%%%%%%%%%%%%%%%%%%%%%%%%%%%%%%%%%%%%%%%%%%%%%%%%%%%%%%%%%%%%%%%%%%%
\caption[States $\ket{\Omega}$ that are annihilated by all grade $-1$ 
generators within the minimal unitary representation space of $SU\left( 2 
, 2 \,|\, 4 \right)$ when $\zeta = +1$]
{States $\ket{\Omega}$ that are annihilated by all grade $-1$ 
generators within the minimal unitary representation space of $SU\left( 2 
, 2 \,|\, 4 \right)$ when $\zeta = +1$. They transform in the irreducible 
representation $| 1 \,,\, \scalebox{0.6}{\sonebox} \rangle$ of 
$SU(2\,|\,2)_L \times SU(2\,|\,2)_R$.
\label{Table:LWVs_zeta=+1}} \\
\hline
& & & & & & \\
LWV & $H$ & $\mathcal{H}$ & 
$\mathfrak{L}$ & $\mathfrak{L}_3$ & $\mathfrak{R}$ & $\mathfrak{R}_3$ \\
& & & & & & \\
\hline
& & & & & & \\
\endfirsthead
%%%%%%%%%%%%%%%%%%%%%%%%%%%%%%%%%%%%%%%%%%%%%%%%%%%%%%%%%%%%%%%%%%%%%%%%%%%%
\caption[]{(continued)} \\
\hline
& & & & & & \\
LWV & $H$ & $\mathcal{H}$ & 
$\mathfrak{L}$ & $\mathfrak{L}_3$ & $\mathfrak{R}$ & $\mathfrak{R}_3$ \\
& & & & & & \\
\hline
& & & & & & \\
\endhead
%%%%%%%%%%%%%%%%%%%%%%%%%%%%%%%%%%%%%%%%%%%%%%%%%%%%%%%%%%%%%%%%%%%%%%%%%%%%
& & & & & & \\
\hline
\endfoot
%%%%%%%%%%%%%%%%%%%%%%%%%%%%%%%%%%%%%%%%%%%%%%%%%%%%%%%%%%%%%%%%%%%%%%%%%%%%
& & & & & & \\
\hline
\endlastfoot
%%%%%%%%%%%%%%%%%%%%%%%%%%%%%%%%%%%%%%%%%%%%%%%%%%%%%%%%%%%%%%%%%%%%%%%%%%%%

$\ket{\frac{3}{2} ; 0 , 0 ; 0 , 0}$ &
$\frac{3}{2}$ & $\frac{1}{2}$ &
0 & 0 & $\frac{1}{2}$ & $-\frac{1}{2}$ \\[8pt]

$\ket{\frac{1}{2} ; 0 , 1 ; 0 , 0}$ &
$\frac{3}{2}$ & $\frac{1}{2}$ & 
0 & 0 & $\frac{1}{2}$ & $\frac{1}{2}$ \\[8pt]

$\ket{\frac{1}{2} ; 0 , 0 ; 0 , 1}$ &
1 & $\frac{1}{2}$ &
0 & 0 & 0 & 0 \\[8pt]

%%%%%%%%%%%%%%%%%%%%%%%%%%%%%%%%%%%%%%%%%%%%%%%%%%%%%%%%%%%%%%%%%%%%%%%%%%%%
\end{longtable}
\end{tiny}
%%%%%%%%%%%%%%%%%%%%%%%%%%%%%%%%%%%%%%%%%%%%%%%%%%%%%%%%%%%%%%%%%%%%%%%%%%%%

% LWV zeta = -2
%%%%%%%%%%%%%%%%%%%%%%%%%%%%%%%%%%%%%%%%%%%%%%%%%%%%%%%%%%%%%%%%%%%%%%%%%%%%
\begin{tiny}
\begin{longtable}[c]{|r||c|c||c|c||c|c|}
\kill
%%%%%%%%%%%%%%%%%%%%%%%%%%%%%%%%%%%%%%%%%%%%%%%%%%%%%%%%%%%%%%%%%%%%%%%%%%%%
\caption[States $\ket{\Omega}$ that are annihilated by all grade $-1$ 
generators within the minimal unitary representation space of $SU\left( 2 
, 2 \,|\, 4 \right)$ when $\zeta = -2$]
{States $\ket{\Omega}$ that are annihilated by all grade $-1$ 
generators within the minimal unitary representation space of $SU\left( 2 
, 2 \,|\, 4 \right)$ when $\zeta = -2$. They transform in the irreducible 
representation $| \, \scalebox{0.6}{\stwobox} \,,\, 1 \rangle$ of 
$SU(2\,|\,2)_L \times SU(2\,|\,2)_R$.
\label{Table:LWVs_zeta=-2}} \\
\hline
& & & & & & \\
LWV & $H$ & $\mathcal{H}$ & 
$\mathfrak{L}$ & $\mathfrak{L}_3$ & $\mathfrak{R}$ & $\mathfrak{R}_3$ \\
& & & & & & \\
\hline
& & & & & & \\
\endfirsthead
%%%%%%%%%%%%%%%%%%%%%%%%%%%%%%%%%%%%%%%%%%%%%%%%%%%%%%%%%%%%%%%%%%%%%%%%%%%%
\caption[]{(continued)} \\
\hline
& & & & & & \\
LWV & $H$ & $\mathcal{H}$ & 
$\mathfrak{L}$ & $\mathfrak{L}_3$ & $\mathfrak{R}$ & $\mathfrak{R}_3$ \\
& & & & & & \\
\hline
& & & & & & \\
\endhead
%%%%%%%%%%%%%%%%%%%%%%%%%%%%%%%%%%%%%%%%%%%%%%%%%%%%%%%%%%%%%%%%%%%%%%%%%%%%
& & & & & & \\
\hline
\endfoot
%%%%%%%%%%%%%%%%%%%%%%%%%%%%%%%%%%%%%%%%%%%%%%%%%%%%%%%%%%%%%%%%%%%%%%%%%%%%
& & & & & & \\
\hline
\endlastfoot
%%%%%%%%%%%%%%%%%%%%%%%%%%%%%%%%%%%%%%%%%%%%%%%%%%%%%%%%%%%%%%%%%%%%%%%%%%%%

$\ket{\frac{5}{2} ; 0 , 0 ; 0 , 0}$ &
2 & 1 &
1 & $-1$ & 0 & 0 \\[8pt]

$\ket{\frac{3}{2} ; 1 , 0 ; 0 , 0}$ &
2 & 1 & 
1 & 0 & 0 & 0 \\[8pt]

$\ket{\frac{1}{2} ; 2 , 0 ; 0 , 0}$ &
2 & 1 & 
1 & 1 & 0 & 0 \\[8pt]

$\ket{\frac{3}{2} ; 0 , 0 ; 1 , 0}$ &
$\frac{3}{2}$ & 1 &
$\frac{1}{2}$ & $- \frac{1}{2}$ & 0 & 0 \\[8pt]

$\ket{\frac{1}{2} ; 0 , 0 ; 2 , 0}$ &
1 & 1 &
0 & 0 & 0 & 0 \\[8pt]

$\ket{\frac{1}{2} ; 1 , 0 ; 1 , 0}$ &
$\frac{3}{2}$ & 1 &
$\frac{1}{2}$ & $\frac{1}{2}$ & 0 & 0 \\[8pt]

%%%%%%%%%%%%%%%%%%%%%%%%%%%%%%%%%%%%%%%%%%%%%%%%%%%%%%%%%%%%%%%%%%%%%%%%%%%%
\end{longtable}
\end{tiny}
%%%%%%%%%%%%%%%%%%%%%%%%%%%%%%%%%%%%%%%%%%%%%%%%%%%%%%%%%%%%%%%%%%%%%%%%%%%%

% LWV zeta = +2
%%%%%%%%%%%%%%%%%%%%%%%%%%%%%%%%%%%%%%%%%%%%%%%%%%%%%%%%%%%%%%%%%%%%%%%%%%%%
\begin{tiny}
\begin{longtable}[c]{|r||c|c||c|c||c|c|}
\kill
%%%%%%%%%%%%%%%%%%%%%%%%%%%%%%%%%%%%%%%%%%%%%%%%%%%%%%%%%%%%%%%%%%%%%%%%%%%%
\caption[States $\ket{\Omega}$ that are annihilated by all grade $-1$ 
generators within the minimal unitary representation space of $SU\left( 2 
, 2 \,|\, 4 \right)$ when $\zeta = +2$]
{States $\ket{\Omega}$ that are annihilated by all grade $-1$ 
generators within the minimal unitary representation space of $SU\left( 2 
, 2 \,|\, 4 \right)$ when $\zeta = +2$. They transform in the irreducible 
representation $| 1 \,,\, \scalebox{0.6}{\stwobox} \rangle$ of 
$SU(2\,|\,2)_L \times SU(2\,|\,2)_R$.
\label{Table:LWVs_zeta=+2}} \\
\hline
& & & & & & \\
LWV & $H$ & $\mathcal{H}$ & 
$\mathfrak{L}$ & $\mathfrak{L}_3$ & $\mathfrak{R}$ & $\mathfrak{R}_3$ \\
& & & & & & \\
\hline
& & & & & & \\
\endfirsthead
%%%%%%%%%%%%%%%%%%%%%%%%%%%%%%%%%%%%%%%%%%%%%%%%%%%%%%%%%%%%%%%%%%%%%%%%%%%%
\caption[]{(continued)} \\
\hline
& & & & & & \\
LWV & $H$ & $\mathcal{H}$ & 
$\mathfrak{L}$ & $\mathfrak{L}_3$ & $\mathfrak{R}$ & $\mathfrak{R}_3$ \\
& & & & & & \\
\hline
& & & & & & \\
\endhead
%%%%%%%%%%%%%%%%%%%%%%%%%%%%%%%%%%%%%%%%%%%%%%%%%%%%%%%%%%%%%%%%%%%%%%%%%%%%
& & & & & & \\
\hline
\endfoot
%%%%%%%%%%%%%%%%%%%%%%%%%%%%%%%%%%%%%%%%%%%%%%%%%%%%%%%%%%%%%%%%%%%%%%%%%%%%
& & & & & & \\
\hline
\endlastfoot
%%%%%%%%%%%%%%%%%%%%%%%%%%%%%%%%%%%%%%%%%%%%%%%%%%%%%%%%%%%%%%%%%%%%%%%%%%%%

$\ket{\frac{5}{2} ; 0 , 0 ; 0 , 0}$ &
2 & 1 &
0 & 0 & 1 & $-1$ \\[8pt]

$\ket{\frac{3}{2} ; 0 , 1 ; 0 , 0}$ &
2 & 1 & 
0 & 0 & 1 & 0 \\[8pt]

$\ket{\frac{1}{2} ; 0 , 2 ; 0 , 0}$ &
2 & 1 & 
0 & 0 & 1 & 1 \\[8pt]

$\ket{\frac{3}{2} ; 0 , 0 ; 0 , 1}$ &
$\frac{3}{2}$ & 1 &
0 & 0 & $\frac{1}{2}$ & $- \frac{1}{2}$ \\[8pt]

$\ket{\frac{1}{2} ; 0 , 0 ; 0 , 2}$ &
1 & 1 &
0 & 0 & 0 & 0 \\[8pt]

$\ket{\frac{1}{2} ; 0 , 1 ; 0 , 1}$ &
$\frac{3}{2}$ & 1 &
0 & 0 & $\frac{1}{2}$ & $\frac{1}{2}$ \\[8pt]

%%%%%%%%%%%%%%%%%%%%%%%%%%%%%%%%%%%%%%%%%%%%%%%%%%%%%%%%%%%%%%%%%%%%%%%%%%%%
\end{longtable}
\end{tiny}
%%%%%%%%%%%%%%%%%%%%%%%%%%%%%%%%%%%%%%%%%%%%%%%%%%%%%%%%%%%%%%%%%%%%%%%%%%%%

Next we construct the supermultiplets that can be obtained for each $\zeta 
\left( \ne 0 \right)$ by starting from the above lowest weight vectors 
$\ket{\Omega}$ and acting on the them with the supersymmetry generators in 
grade $+1$ space $\mathfrak{C}^+$.

The supermultiplet that corresponds to $\zeta = -1$ (given in Table 
\ref{Table:Supermultiplet_zeta=-1}) is exactly the doubleton supermultiplet 
in \cite{Gunaydin:1998sw,Gunaydin:1998jc} obtained by starting from the 
lowest weight vector $\ket{\Omega} = \ket{\, \sonebox \,,\, 1}$.

% Supermultiplet zeta = -1
%%%%%%%%%%%%%%%%%%%%%%%%%%%%%%%%%%%%%%%%%%%%%%%%%%%%%%%%%%%%%%%%%%%%%%%%%%%%
\begin{tiny}
\begin{longtable}[c]{|r|l||c|c||c|r||c|r||c|}
\kill
%%%%%%%%%%%%%%%%%%%%%%%%%%%%%%%%%%%%%%%%%%%%%%%%%%%%%%%%%%%%%%%%%%%%%%%%%%%%
\caption[The doubleton supermultiplet corresponding to $\zeta = -1$]
{The doubleton supermultiplet corresponding to $\zeta = -1$. 
The states that are marked with an asterisk belong to $\ket{\Omega} = 
| \, \scalebox{0.6}{\sonebox} \,,\, 1 \rangle$.
\label{Table:Supermultiplet_zeta=-1}} \\
\hline
& & & & & & & & \\
State & $\ket{\alpha ; n_d , n_g ; n_\alpha , n_\beta}$ & 
$H$ & $\mathcal{H}$ & 
$\mathfrak{L}$ & $\mathfrak{L}_3$ & $\mathfrak{R}$ & $\mathfrak{R}_3$ & 
$SU(4)$ \\
& & & & & & & & \\
\hline
& & & & & & & & \\
\endfirsthead
%%%%%%%%%%%%%%%%%%%%%%%%%%%%%%%%%%%%%%%%%%%%%%%%%%%%%%%%%%%%%%%%%%%%%%%%%%%%
\caption[]{(continued)} \\
\hline
& & & & & & & & \\
State & $\ket{\alpha ; n_d , n_g ; n_\alpha , n_\beta}$ & 
$H$ & $\mathcal{H}$ & 
$\mathfrak{L}$ & $\mathfrak{L}_3$ & $\mathfrak{R}$ & $\mathfrak{R}_3$ & 
$SU(4)$ \\
& & & & & & & & \\
\hline
& & & & & & & & \\
\endhead
%%%%%%%%%%%%%%%%%%%%%%%%%%%%%%%%%%%%%%%%%%%%%%%%%%%%%%%%%%%%%%%%%%%%%%%%%%%%
& & & & & & & & \\
\hline
\endfoot
%%%%%%%%%%%%%%%%%%%%%%%%%%%%%%%%%%%%%%%%%%%%%%%%%%%%%%%%%%%%%%%%%%%%%%%%%%%%
& & & & & & & & \\
\hline
\endlastfoot
%%%%%%%%%%%%%%%%%%%%%%%%%%%%%%%%%%%%%%%%%%%%%%%%%%%%%%%%%%%%%%%%%%%%%%%%%%%%

$\ket{\frac{3}{2} ; 0 , 0 ; 0 , 0}^*$ &
$\ket{\frac{3}{2} ; 0 , 0 ; 0 , 0}^*$ &
$\frac{3}{2}$ & $\frac{1}{2}$ &
$\frac{1}{2}$ & $-\frac{1}{2}$ & 0 & 0 &
6 \\[8pt]

$\ket{\frac{1}{2} ; 1 , 0 ; 0 , 0}^*$ &
$\ket{\frac{1}{2} ; 1 , 0 ; 0 , 0}^*$ &
 & & 
$\frac{1}{2}$ & $+\frac{1}{2}$ & 0 & 0 &
 \\[8pt]

\hline
& & & & & & & & \\

$\mathfrak{S}^y \, \ket{\frac{3}{2} ; 0 , 0 ; 0 , 0}$ &
$\ket{\frac{5}{2} ; 0 , 0 ; 0 , 1}$ &
2 & $\frac{3}{2}$ &
1 & $-1$ & 0 & 0 &
$\overline{4}$ \\[8pt]

$\mathfrak{Q}^y \, \ket{\frac{3}{2} ; 0 , 0 ; 0 , 0}
=
\mathfrak{S}^y \, \ket{\frac{1}{2} ; 1 , 0 ; 0 , 0}$ &
$\ket{\frac{3}{2} ; 1 , 0 , 0 , 1}$ &
 & &
1 & 0 & 0 & 0 &
 \\[8pt]

$\mathfrak{Q}^y \, \ket{\frac{1}{2} ; 1 , 0 ; 0 , 0}$ &
$\ket{\frac{1}{2} ; 2 , 0 , 0 , 1}$ &
 & &
1 & $+1$ & 0 & 0 &
 \\[8pt]

\hline
& & & & & & & & \\

$\mathfrak{S}^y \mathfrak{S}^z \, \ket{\frac{3}{2} ; 0 , 0 ; 0 , 0}$ &
$\ket{\frac{7}{2} ; 0 , 0 , 0 , 2}$ &
$\frac{5}{2}$ & $\frac{5}{2}$ &
$\frac{3}{2}$ & $-\frac{3}{2}$ & 0 & 0 &
$\overline{1}$ \\[8pt]

$\mathfrak{S}^y \mathfrak{Q}^z \, \ket{\frac{3}{2} ; 0 , 0 ; 0 , 0}
=
\mathfrak{S}^y \mathfrak{S}^z \, \ket{\frac{1}{2} ; 1 , 0 ; 0 , 0}$ &
$\ket{\frac{5}{2} ; 1 , 0 , 0 , 2}$ &
 & &
$\frac{3}{2}$ & $-\frac{1}{2}$ & 0 & 0 &
 \\[8pt]

$\mathfrak{Q}^y \mathfrak{Q}^z \, \ket{\frac{3}{2} ; 0 , 0 ; 0 , 0}
=
\mathfrak{S}^y \mathfrak{Q}^z \, \ket{\frac{1}{2} ; 1 , 0 ; 0 , 0}$ &
$\ket{\frac{3}{2} ; 2 , 0 , 0 , 2}$ &
 & &
$\frac{3}{2}$ & $+\frac{1}{2}$ & 0 & 0 &
 \\[8pt]

$\mathfrak{Q}^y \mathfrak{Q}^z \, \ket{\frac{1}{2} ; 1 , 0 ; 0 , 0}$ &
$\ket{\frac{1}{2} ; 3 , 0 , 0 , 2}$ &
 & &
$\frac{3}{2}$ & $+\frac{3}{2}$ & 0 & 0 &
 \\[8pt]

\hline
& & & & & & & & \\

$\ket{\frac{1}{2} ; 0 , 0 ; 1 , 0}^*$ &
$\ket{\frac{1}{2} ; 0 , 0 , 1 , 0}^*$ &
1 & $\frac{1}{2}$ &
0 & 0 & 0 & 0 &
4 \\[8pt]

\hline
& & & & & & & & \\

$\mathfrak{S}^\nu \, \ket{\frac{1}{2} ; 0 , 0 ; 1 , 0}$ &
$\ket{\frac{3}{2} ; 0 , 0 , 2 , 0}$ &
$\frac{3}{2}$ & $\frac{3}{2}$ &
0 & 0 & $\frac{1}{2}$ & $-\frac{1}{2}$ &
1 \\[8pt]

$\mathfrak{Q}^\nu \, \ket{\frac{1}{2} ; 0 , 0 ; 1 , 0}$ &
$\ket{\frac{1}{2} ; 0 , 1 , 2 , 0}$ &
 & &
0 & 0 & $\frac{1}{2}$ & $+\frac{1}{2}$ &
 \\[8pt]

%%%%%%%%%%%%%%%%%%%%%%%%%%%%%%%%%%%%%%%%%%%%%%%%%%%%%%%%%%%%%%%%%%%%%%%%%%%%
\end{longtable}
\end{tiny}
%%%%%%%%%%%%%%%%%%%%%%%%%%%%%%%%%%%%%%%%%%%%%%%%%%%%%%%%%%%%%%%%%%%%%%%%%%%%

Next we give the doubleton supermultiplet that corresponds to $\zeta = +1$ 
in Table \ref{Table:Supermultiplet_zeta=+1}. This supermultiplet was obtained in \cite{Gunaydin:1998sw,Gunaydin:1998jc} by starting from the 
lowest weight vector $\ket{\Omega} = \ket{1 \,,\, \sonebox}$.

% Supermultiplet zeta = +1
%%%%%%%%%%%%%%%%%%%%%%%%%%%%%%%%%%%%%%%%%%%%%%%%%%%%%%%%%%%%%%%%%%%%%%%%%%%%
\begin{tiny}
\begin{longtable}[c]{|r|l||c|c||c|r||c|r||c|}
\kill
%%%%%%%%%%%%%%%%%%%%%%%%%%%%%%%%%%%%%%%%%%%%%%%%%%%%%%%%%%%%%%%%%%%%%%%%%%%%
\caption[The doubleton supermultiplet corresponding to $\zeta = +1$]
{The doubleton supermultiplet corresponding to $\zeta = +1$. 
The states that are marked with an asterisk belong to $\ket{\Omega} = 
| 1 \,,\, \scalebox{0.6}{\sonebox} \rangle$.
\label{Table:Supermultiplet_zeta=+1}} \\
\hline
& & & & & & & & \\
State & $\ket{\alpha ; n_d , n_g ; n_\alpha , n_\beta}$ & 
$H$ & $\mathcal{H}$ & 
$\mathfrak{L}$ & $\mathfrak{L}_3$ & $\mathfrak{R}$ & $\mathfrak{R}_3$ & 
$SU(4)$ \\
& & & & & & & & \\
\hline
& & & & & & & & \\
\endfirsthead
%%%%%%%%%%%%%%%%%%%%%%%%%%%%%%%%%%%%%%%%%%%%%%%%%%%%%%%%%%%%%%%%%%%%%%%%%%%%
\caption[]{(continued)} \\
\hline
& & & & & & & & \\
State & $\ket{\alpha ; n_d , n_g ; n_\alpha , n_\beta}$ & 
$H$ & $\mathcal{H}$ & 
$\mathfrak{L}$ & $\mathfrak{L}_3$ & $\mathfrak{R}$ & $\mathfrak{R}_3$ & 
$SU(4)$ \\
& & & & & & & & \\
\hline
& & & & & & & & \\
\endhead
%%%%%%%%%%%%%%%%%%%%%%%%%%%%%%%%%%%%%%%%%%%%%%%%%%%%%%%%%%%%%%%%%%%%%%%%%%%%
& & & & & & & & \\
\hline
\endfoot
%%%%%%%%%%%%%%%%%%%%%%%%%%%%%%%%%%%%%%%%%%%%%%%%%%%%%%%%%%%%%%%%%%%%%%%%%%%%
& & & & & & & & \\
\hline
\endlastfoot
%%%%%%%%%%%%%%%%%%%%%%%%%%%%%%%%%%%%%%%%%%%%%%%%%%%%%%%%%%%%%%%%%%%%%%%%%%%%

$\ket{\frac{3}{2} ; 0 , 0 ; 0 , 0}^*$ &
$\ket{\frac{3}{2} ; 0 , 0 ; 0 , 0}^*$ &
$\frac{3}{2}$ & $\frac{1}{2}$ &
0 & 0 & $\frac{1}{2}$ & $-\frac{1}{2}$ &
6 \\[8pt]

$\ket{\frac{1}{2} ; 0 , 1 ; 0 , 0}^*$ &
$\ket{\frac{1}{2} ; 0 , 1 ; 0 , 0}^*$ &
 & & 
0 & 0 & $\frac{1}{2}$ & $+\frac{1}{2}$ &
 \\[8pt]

\hline
& & & & & & & & \\

$\mathfrak{S}^\mu \, \ket{\frac{3}{2} ; 0 , 0 ; 0 , 0}$ &
$\ket{\frac{5}{2} ; 0 , 0 ; 1 , 0}$ &
2 & $\frac{3}{2}$ &
0 & 0 & 1 & $-1$ &
4 \\[8pt]

$\mathfrak{Q}^\mu \, \ket{\frac{3}{2} ; 0 , 0 ; 0 , 0}
=
\mathfrak{S}^\mu \, \ket{\frac{1}{2} ; 0 , 1 ; 0 , 0}$ &
$\ket{\frac{3}{2} ; 0 , 1 , 1 , 0}$ &
 & &
0 & 0 & 1 & 0 &
 \\[8pt]

$\mathfrak{Q}^\mu \, \ket{\frac{1}{2} ; 0 , 1 ; 0 , 0}$ &
$\ket{\frac{1}{2} ; 0 , 2 , 1 , 0}$ &
 & &
0 & 0 & 1 & $+1$ &
 \\[8pt]

\hline
& & & & & & & & \\

$\mathfrak{S}^\mu \mathfrak{S}^\nu \, \ket{\frac{3}{2} ; 0 , 0 ; 0 , 0}$ &
$\ket{\frac{7}{2} ; 0 , 0 , 2 , 0}$ &
$\frac{5}{2}$ & $\frac{5}{2}$ &
0 & 0 & $\frac{3}{2}$ & $-\frac{3}{2}$ &
1 \\[8pt]

$\mathfrak{S}^\mu \mathfrak{Q}^\nu \, \ket{\frac{3}{2} ; 0 , 0 ; 0 , 0}
=
\mathfrak{S}^\mu \mathfrak{S}^\nu \, \ket{\frac{1}{2} ; 0 , 1 ; 0 , 0}$ &
$\ket{\frac{5}{2} ; 0 , 1 , 2 , 0}$ &
 & &
0 & 0 & $\frac{3}{2}$ & $-\frac{1}{2}$ &
 \\[8pt]

$\mathfrak{Q}^\mu \mathfrak{Q}^\nu \, \ket{\frac{3}{2} ; 0 , 0 ; 0 , 0}
=
\mathfrak{S}^\mu \mathfrak{Q}^\nu \, \ket{\frac{1}{2} ; 0 , 1 ; 0 , 0}$ &
$\ket{\frac{3}{2} ; 0 , 2 , 2 , 0}$ &
 & &
0 & 0 & $\frac{3}{2}$ & $+\frac{1}{2}$ &
 \\[8pt]

$\mathfrak{Q}^\mu \mathfrak{Q}^\nu \, \ket{\frac{1}{2} ; 0 , 1 ; 0 , 0}$ &
$\ket{\frac{1}{2} ; 0 , 3 , 2 , 0}$ &
 & &
0 & 0 & $\frac{3}{2}$ & $+\frac{3}{2}$ &
 \\[8pt]

\hline
& & & & & & & & \\

$\ket{\frac{1}{2} ; 0 , 0 ; 0 , 1}^*$ &
$\ket{\frac{1}{2} ; 0 , 0 , 0 , 1}^*$ &
1 & $\frac{1}{2}$ &
0 & 0 & 0 & 0 &
$\overline{4}$ \\[8pt]

\hline
& & & & & & & & \\

$\mathfrak{S}^y \, \ket{\frac{1}{2} ; 0 , 0 ; 0 , 1}$ &
$\ket{\frac{3}{2} ; 0 , 0 , 0 , 2}$ &
$\frac{3}{2}$ & $\frac{3}{2}$ &
$\frac{1}{2}$ & $-\frac{1}{2}$ & 0 & 0 &
$\overline{1}$ \\[8pt]

$\mathfrak{Q}^y \, \ket{\frac{1}{2} ; 0 , 0 ; 0 , 1}$ &
$\ket{\frac{1}{2} ; 1 , 0 , 0 , 2}$ &
 & &
$\frac{1}{2}$ & $+\frac{1}{2}$ & 0 & 0 &
 \\[8pt]

%%%%%%%%%%%%%%%%%%%%%%%%%%%%%%%%%%%%%%%%%%%%%%%%%%%%%%%%%%%%%%%%%%%%%%%%%%%%
\end{longtable}
\end{tiny}
%%%%%%%%%%%%%%%%%%%%%%%%%%%%%%%%%%%%%%%%%%%%%%%%%%%%%%%%%%%%%%%%%%%%%%%%%%%%

The supermultiplet we obtain by taking $\zeta = -2$ (given in Table 
\ref{Table:Supermultiplet_zeta=-2}) corresponds to the doubleton 
supermultiplet in \cite{Gunaydin:1998sw,Gunaydin:1998jc} obtained by 
starting from the lowest weight vector $\ket{\Omega} = \ket{\, \stwobox 
\,,\, 1}$.

% Supermultiplet c = -2
%%%%%%%%%%%%%%%%%%%%%%%%%%%%%%%%%%%%%%%%%%%%%%%%%%%%%%%%%%%%%%%%%%%%%%%%%%%%
\begin{tiny}
\begin{longtable}[c]{|r|l||c|c||c|r||c|r||c|}
\kill
%%%%%%%%%%%%%%%%%%%%%%%%%%%%%%%%%%%%%%%%%%%%%%%%%%%%%%%%%%%%%%%%%%%%%%%%%%%%
\caption[The doubleton supermultiplet corresponding to $\zeta = -2$]
{The doubleton supermultiplet corresponding to $\zeta = -2$. The states 
that are marked with an asterisk belong to $\ket{\Omega} = 
| \, \scalebox{0.6}{\stwobox} \,,\, 1 \rangle$.
\label{Table:Supermultiplet_zeta=-2}} \\
\hline
& & & & & & & & \\
State & $\ket{\alpha ; n_d , n_g ; n_\alpha , n_\beta}$ & 
$H$ & $\mathcal{H}$ & 
$\mathfrak{L}$ & $\mathfrak{L}_3$ & $\mathfrak{R}$ & $\mathfrak{R}_3$ & 
$SU(4)$ \\
& & & & & & & & \\
\hline
& & & & & & & & \\
\endfirsthead
%%%%%%%%%%%%%%%%%%%%%%%%%%%%%%%%%%%%%%%%%%%%%%%%%%%%%%%%%%%%%%%%%%%%%%%%%%%%
\caption[]{(continued)} \\
\hline
& & & & & & & & \\
State & $\ket{\alpha ; n_d , n_g ; n_\alpha , n_\beta}$ & 
$H$ & $\mathcal{H}$ & 
$\mathfrak{L}$ & $\mathfrak{L}_3$ & $\mathfrak{R}$ & $\mathfrak{R}_3$ & 
$SU(4)$ \\
& & & & & & & & \\
\hline
& & & & & & & & \\
\endhead
%%%%%%%%%%%%%%%%%%%%%%%%%%%%%%%%%%%%%%%%%%%%%%%%%%%%%%%%%%%%%%%%%%%%%%%%%%%%
& & & & & & & & \\
\hline
\endfoot
%%%%%%%%%%%%%%%%%%%%%%%%%%%%%%%%%%%%%%%%%%%%%%%%%%%%%%%%%%%%%%%%%%%%%%%%%%%%
& & & & & & & & \\
\hline
\endlastfoot
%%%%%%%%%%%%%%%%%%%%%%%%%%%%%%%%%%%%%%%%%%%%%%%%%%%%%%%%%%%%%%%%%%%%%%%%%%%%

$\ket{\frac{5}{2} ; 0 , 0 ; 0 , 0}^*$ &
$\ket{\frac{5}{2} ; 0 , 0 ; 0 , 0}^*$ &
2 & 1 &
1 & $-1$ & 0 & 0 &
6 \\[8pt]

$\ket{\frac{3}{2} ; 1 , 0 ; 0 , 0}^*$ &
$\ket{\frac{3}{2} ; 1 , 0 ; 0 , 0}^*$ &
 & & 
1 & 0 & 0 & 0 &
\\[8pt]

$\ket{\frac{1}{2} ; 2 , 0 ; 0 , 0}^*$ &
$\ket{\frac{1}{2} ; 2 , 0 ; 0 , 0}^*$ &
 & & 
1 & $+1$ & 0 & 0 &
\\[8pt]

\hline
& & & & & & & & \\

$\mathfrak{S}^y \, \ket{\frac{5}{2} ; 0 , 0 ; 0 , 0}$ &
$\ket{\frac{7}{2} ; 0 , 0 ; 0 , 1}$ &
$\frac{5}{2}$ & 2 &
$\frac{3}{2}$ & $-\frac{3}{2}$ & 0 & 0 &
$\overline{4}$ \\[8pt]

$\mathfrak{Q}^y \, \ket{\frac{5}{2} ; 0 , 0 ; 0 , 0}
=
\mathfrak{S}^y \, \ket{\frac{3}{2} ; 1 , 0 ; 0 , 0}$ &
$\ket{\frac{5}{2} ; 1 , 0 ; 0 , 1}$ &
 & &
$\frac{3}{2}$ & $-\frac{1}{2}$ & 0 & 0 &
\\[8pt]

$\mathfrak{Q}^y \, \ket{\frac{3}{2} ; 1 , 0 ; 0 , 0}
=
\mathfrak{S}^y \, \ket{\frac{1}{2} ; 2 , 0 ; 0 , 0}$ &
$\ket{\frac{3}{2} ; 2 , 0 ; 0 , 1}$ &
 & &
$\frac{3}{2}$ & $+\frac{1}{2}$ & 0 & 0 &
\\[8pt]

$\mathfrak{Q}^y \, \ket{\frac{1}{2} ; 2 , 0 ; 0 , 0}$ &
$\ket{\frac{1}{2} ; 3 , 0 ; 0 , 1}$ &
 & &
$\frac{3}{2}$ & $+\frac{3}{2}$ & 0 & 0 &
\\[8pt]

\hline
& & & & & & & & \\

$\ket{\frac{3}{2} ; 0 , 0 ; 1 , 0}^*$ &
$\ket{\frac{3}{2} ; 0 , 0 ; 1 , 0}^*$ &
$\frac{3}{2}$ & 1 &
$\frac{1}{2}$ & $-\frac{1}{2}$ & 0 & 0 &
4 \\[8pt]

$\ket{\frac{1}{2} ; 1 , 0 ; 1 , 0}^*$ &
$\ket{\frac{1}{2} ; 1 , 0 ; 1 , 0}^*$ &
 & &
$\frac{1}{2}$ & $+\frac{1}{2}$ & 0 & 0 &
\\[8pt]

\hline
& & & & & & & & \\

$\mathfrak{S}^y \mathfrak{S}^z \, \ket{\frac{5}{2} ; 0 , 0 ; 0 , 0}$ &
$\ket{\frac{9}{2} ; 0 , 0 ; 0 , 2}$ &
3 & 3 &
2 & $-2$ & 0 & 0 &
$\overline{1}$ \\[8pt]

$\mathfrak{S}^y \mathfrak{Q}^z \, \ket{\frac{5}{2} ; 0 , 0 ; 0 , 0}
=
\mathfrak{S}^y \mathfrak{S}^z \, \ket{\frac{3}{2} ; 0 , 0 ; 1 , 0}$ &
$\ket{\frac{7}{2} ; 1 , 0 ; 0 , 2}$ &
 & &
2 & $-1$ & 0 & 0 &
\\[8pt]

$\mathfrak{Q}^y \mathfrak{Q}^z \ket{\frac{5}{2} ; 0 , 0 ; 0 , 0}
=
\mathfrak{S}^y \mathfrak{Q}^z \ket{\frac{3}{2} ; 1 , 0 ; 0 , 0}$ &
$\ket{\frac{5}{2} ; 2 , 0 ; 0 , 2}$ &
 & &
2 & 0 & 0 & 0 &
\\

$= \mathfrak{S}^y \mathfrak{S}^z \ket{\frac{1}{2} ; 2 , 0 ; 0 , 0}$ &
 &
 & &
 & & & &
\\[8pt]

$\mathfrak{Q}^y \mathfrak{Q}^z \, \ket{\frac{3}{2} ; 1 , 0 ; 0 , 0}
=
\mathfrak{S}^y \mathfrak{Q}^z \, \ket{\frac{1}{2} ; 2 , 0 ; 0 , 0}$ &
$\ket{\frac{3}{2} ; 3 , 0 ; 0 , 2}$ &
 & &
2 & $+1$ & 0 & 0 &
\\[8pt]

$\mathfrak{Q}^y \mathfrak{Q}^z \, \ket{\frac{1}{2} ; 2 , 0 ; 0 , 0}$ &
$\ket{\frac{1}{2} ; 4 , 0 , 0 , 2}$ &
 & &
2 & $+2$ & 0 & 0 &
\\[8pt]

\hline
& & & & & & & & \\

$\ket{\frac{1}{2} ; 0 , 0 , 2 , 0}^*$ &
$\ket{\frac{1}{2} ; 0 , 0 , 2 , 0}^*$ &
1 & 1 &
0 & 0 & 0 & 0 &
1 \\[8pt]

%%%%%%%%%%%%%%%%%%%%%%%%%%%%%%%%%%%%%%%%%%%%%%%%%%%%%%%%%%%%%%%%%%%%%%%%%%%%
\end{longtable}
\end{tiny}
%%%%%%%%%%%%%%%%%%%%%%%%%%%%%%%%%%%%%%%%%%%%%%%%%%%%%%%%%%%%%%%%%%%%%%%%%%%%

Finally, we give the supermultiplet obtain by taking $\zeta = +2$ in 
Table \ref{Table:Supermultiplet_zeta=+2}. This supermultiplet corresponds 
to the doubleton supermultiplet in \cite{Gunaydin:1998sw,Gunaydin:1998jc} 
obtained by starting from the lowest weight vector $\ket{\Omega} = 
\ket{1 \,,\, \stwobox}$.

% Supermultiplet c = +2
%%%%%%%%%%%%%%%%%%%%%%%%%%%%%%%%%%%%%%%%%%%%%%%%%%%%%%%%%%%%%%%%%%%%%%%%%%%%
\begin{tiny}
\begin{longtable}[c]{|r|l||c|c||c|r||c|r||c|}
\kill
%%%%%%%%%%%%%%%%%%%%%%%%%%%%%%%%%%%%%%%%%%%%%%%%%%%%%%%%%%%%%%%%%%%%%%%%%%%%
\caption[The doubleton supermultiplet corresponding to $\zeta = +2$]
{The doubleton supermultiplet corresponding to $\zeta = +2$. The states 
that are marked with an asterisk belong to $\ket{\Omega} = 
| 1 \,,\, \scalebox{0.6}{\stwobox} \rangle$.
\label{Table:Supermultiplet_zeta=+2}} \\
\hline
& & & & & & & & \\
State & $\ket{\alpha ; n_d , n_g ; n_\alpha , n_\beta}$ & 
$H$ & $\mathcal{H}$ & 
$\mathfrak{L}$ & $\mathfrak{L}_3$ & $\mathfrak{R}$ & $\mathfrak{R}_3$ & 
$SU(4)$ \\
& & & & & & & & \\
\hline
& & & & & & & & \\
\endfirsthead
%%%%%%%%%%%%%%%%%%%%%%%%%%%%%%%%%%%%%%%%%%%%%%%%%%%%%%%%%%%%%%%%%%%%%%%%%%%%
\caption[]{(continued)} \\
\hline
& & & & & & & & \\
State & $\ket{\alpha ; n_d , n_g ; n_\alpha , n_\beta}$ & 
$H$ & $\mathcal{H}$ & 
$\mathfrak{L}$ & $\mathfrak{L}_3$ & $\mathfrak{R}$ & $\mathfrak{R}_3$ & 
$SU(4)$ \\
& & & & & & & & \\
\hline
& & & & & & & & \\
\endhead
%%%%%%%%%%%%%%%%%%%%%%%%%%%%%%%%%%%%%%%%%%%%%%%%%%%%%%%%%%%%%%%%%%%%%%%%%%%%
& & & & & & & & \\
\hline
\endfoot
%%%%%%%%%%%%%%%%%%%%%%%%%%%%%%%%%%%%%%%%%%%%%%%%%%%%%%%%%%%%%%%%%%%%%%%%%%%%
& & & & & & & & \\
\hline
\endlastfoot
%%%%%%%%%%%%%%%%%%%%%%%%%%%%%%%%%%%%%%%%%%%%%%%%%%%%%%%%%%%%%%%%%%%%%%%%%%%%

$\ket{\frac{5}{2} ; 0 , 0 ; 0 , 0}^*$ &
$\ket{\frac{5}{2} ; 0 , 0 ; 0 , 0}^*$ &
2 & 1 &
0 & 0 & 1 & $-1$ &
6 \\[8pt]

$\ket{\frac{3}{2} ; 0 , 1 ; 0 , 0}^*$ &
$\ket{\frac{3}{2} ; 0 , 1 ; 0 , 0}^*$ &
 & & 
0 & 0 & 1 & 0 &
\\[8pt]

$\ket{\frac{1}{2} ; 0 , 2 ; 0 , 0}^*$ &
$\ket{\frac{1}{2} ; 0 , 2 ; 0 , 0}^*$ &
 & & 
0 & 0 & 1 & $+1$ &
\\[8pt]

\hline
& & & & & & & & \\

$\mathfrak{S}^\mu \, \ket{\frac{5}{2} ; 0 , 0 ; 0 , 0}$ &
$\ket{\frac{7}{2} ; 0 , 0 ; 1 , 0}$ &
$\frac{5}{2}$ & 2 &
0 & 0 & $\frac{3}{2}$ & $-\frac{3}{2}$ &
4 \\[8pt]

$\mathfrak{Q}^\mu \, \ket{\frac{5}{2} ; 0 , 0 ; 0 , 0}
=
\mathfrak{S}^\mu \, \ket{\frac{3}{2} ; 0 , 1 ; 0 , 0}$ &
$\ket{\frac{5}{2} ; 0 , 1 ; 1 , 0}$ &
 & &
0 & 0 & $\frac{3}{2}$ & $-\frac{1}{2}$ &
\\[8pt]

$\mathfrak{Q}^\mu \, \ket{\frac{3}{2} ; 0 , 1 ; 0 , 0}
=
\mathfrak{S}^\mu \, \ket{\frac{1}{2} ; 0 , 2 ; 0 , 0}$ &
$\ket{\frac{3}{2} ; 0 , 2 ; 1 , 0}$ &
 & &
0 & 0 & $\frac{3}{2}$ & $+\frac{1}{2}$ &
\\[8pt]

$\mathfrak{Q}^\mu \, \ket{\frac{1}{2} ; 0 , 2 ; 0 , 0}$ &
$\ket{\frac{1}{2} ; 0 , 3 ; 1 , 0}$ &
 & &
0 & 0 & $\frac{3}{2}$ & $+\frac{3}{2}$ &
\\[8pt]

\hline
& & & & & & & & \\

$\ket{\frac{3}{2} ; 0 , 0 ; 0 , 1}^*$ &
$\ket{\frac{3}{2} ; 0 , 0 ; 0 , 1}^*$ &
$\frac{3}{2}$ & 1 &
0 & 0 & $\frac{1}{2}$ & $-\frac{1}{2}$ &
$\overline{4}$ \\[8pt]

$\ket{\frac{1}{2} ; 0 , 1 ; 0 , 1}^*$ &
$\ket{\frac{1}{2} ; 0 , 1 ; 0 , 1}^*$ &
 & &
0 & 0 & $\frac{1}{2}$ & $+\frac{1}{2}$ &
\\[8pt]

\hline
& & & & & & & & \\

$\mathfrak{S}^\mu \mathfrak{S}^\nu \, \ket{\frac{5}{2} ; 0 , 0 ; 0 , 0}$ &
$\ket{\frac{9}{2} ; 0 , 0 ; 2 , 0}$ &
3 & 3 &
0 & 0 & 2 & $-2$ &
1 \\[8pt]

$\mathfrak{S}^\mu \mathfrak{Q}^\nu \, \ket{\frac{5}{2} ; 0 , 0 ; 0 , 0}
=
\mathfrak{S}^\mu \mathfrak{S}^\nu \, \ket{\frac{3}{2} ; 0 , 0 ; 0 , 1}$ &
$\ket{\frac{7}{2} ; 0 , 1 ; 2 , 0}$ &
 & &
0 & 0 & 2 & $-1$ &
\\[8pt]

$\mathfrak{Q}^\mu \mathfrak{Q}^\nu \ket{\frac{5}{2} ; 0 , 0 ; 0 , 0}
=
\mathfrak{S}^\mu \mathfrak{Q}^\nu \ket{\frac{3}{2} ; 0 , 1 ; 0 , 0}$ &
$\ket{\frac{5}{2} ; 0 , 2 ; 2 , 0}$ &
 & &
0 & 0 & 2 & 0 &
\\

$= \mathfrak{S}^\mu \mathfrak{S}^\nu \ket{\frac{1}{2} ; 0 , 2 ; 0 , 0}$ &
 &
 & &
 & & & &
\\[8pt]

$\mathfrak{Q}^\mu \mathfrak{Q}^\nu \, \ket{\frac{3}{2} ; 0 , 1 ; 0 , 0}
=
\mathfrak{S}^\mu \mathfrak{Q}^\nu \, \ket{\frac{1}{2} ; 0 , 2 ; 0 , 0}$ &
$\ket{\frac{3}{2} ; 0 , 3 ; 2 , 0}$ &
 & &
0 & 0 & 2 & $+1$ &
\\[8pt]

$\mathfrak{Q}^\mu \mathfrak{Q}^\nu \, \ket{\frac{1}{2} ; 0 , 2 ; 0 , 0}$ &
$\ket{\frac{1}{2} ; 0 , 4 , 2 , 0}$ &
 & &
0 & 0 & 2 & $+2$ &
\\[8pt]

\hline
& & & & & & & & \\

$\ket{\frac{1}{2} ; 0 , 0 , 0 , 2}^*$ &
$\ket{\frac{1}{2} ; 0 , 0 , 0 , 2}^*$ &
1 & 1 &
0 & 0 & 0 & 0 &
$\overline{1}$ \\[8pt]

%%%%%%%%%%%%%%%%%%%%%%%%%%%%%%%%%%%%%%%%%%%%%%%%%%%%%%%%%%%%%%%%%%%%%%%%%%%%
\end{longtable}
\end{tiny}
%%%%%%%%%%%%%%%%%%%%%%%%%%%%%%%%%%%%%%%%%%%%%%%%%%%%%%%%%%%%%%%%%%%%%%%%%%%%

Following this method, one can obtain all the other higher spin doubleton 
supermultiplets by choosing a deformation parameter $\left| \zeta \right| 
> 2$.

%%%%%%%%%%%%%%%%%%%%%%%%%%%%%%%%%%%%%%%%%%%%%%%%%%%%%%%%%%%%%%%%%%%%
%%%%%% Section: Minimal supermultiplets of SU(2,2|p+q) %%%%%%%%%%%%%
%%%%%%%%%%%%%%%%%%%%%%%%%%%%%%%%%%%%%%%%%%%%%%%%%%%%%%%%%%%%%%%%%%%%

\section{Minimal Unitary Supermultiplet of $\mathfrak{su}\left( 2 , 
2 \,|\, \mathfrak{p} + \mathfrak{q} \right)$ and its Deformations}
\label{sec:SU(2,2|p+q)}

It is clear that one can easily generalize the above construction of 
minimal unitary supermultiplet  of $\mathfrak{su}\left( 2 , 2 \,|\, 4 
\right)$ and its deformations to those of  $SU(2,2|\mathfrak{p}$ and $\mathfrak{q})$.

Once again, when $\zeta = 0$, the state $\ket{\frac{1}{2} ; 0 , 0 ; 0 , 0}$ 
is the unique 
normalizable lowest energy state annihilated by all bosonic and fermionic 
generators in grade $-1$ space $\mathfrak{C}^{-}$. Thus it forms a singlet 
of $SU(2 \,|\, \mathfrak{p}) \times SU(2 \,|\, \mathfrak{q}) $ subalgebra. 
By acting on it with grade $+1$ generators in $\mathfrak{C}^+$, one can 
obtain an infinite set of states that form a basis for the minimal unitary 
irreducible representation of $\mathfrak{su}(2,2 \,|\, \mathfrak{p} + 
\mathfrak{q})$. These infinitely many states decompose into a finite number 
of irreps of the even subgroup $SU(2,2) \times SU(\mathfrak{p} + 
\mathfrak{q})$, with each irrep of $SU(2,2)$ corresponding to  a massless conformal field in four dimensions.

When $\zeta \ne 0$, there are multiple states, for any given $\zeta$, that 
are annihilated by all bosonic and fermionic generators in grade $-1$ space 
$\mathfrak{C}^-$ of $SU \left( 2 , 2 \,|\, \mathfrak{p} + \mathfrak{q} 
\right)$. They form an irreducible representation of 
$SU(2\,|\,\mathfrak{p})_L \times SU(2\,|\,\mathfrak{q})_R$ whose 
supertableau is
\begin{equation}
\begin{split}
& | \, \underbrace{\sgenrowbox}_{-\zeta} \,,\, 1 \rangle
\qquad \qquad \mbox{for $\zeta < 0$}
\\
& | 1 \,,\, \underbrace{\sgenrowbox}_{\zeta} \, \rangle
\qquad \qquad \mbox{for $\zeta > 0$} \,.
\end{split}
\end{equation}

In Table \ref{Table:generalLWVs}, we list all such states $\ket{\Omega}$ 
for different values of the deformation parameter $\zeta \ne 0$.

%%%%%%%%%%%%%%%%%%%%%%%%%%%%%%%%%%%%%%%%%%%%%%%%%%%%%%%%%%%%%%%%%%%%%%%%%%%%

\begin{tiny}
\begin{longtable}[c]{|r|l||c|c||c|c||c|c|}
\kill
%%%%%%%%%%%%%%%%%%%%%%%%%%%%%%%%%%%%%%%%%%%%%%%%%%%%%%%%%%%%%%%%%%%%%%%%%%%%
\caption[States $\ket{\Omega}$ that are annihilated by all grade $-1$ 
generators in $\mathfrak{C}^-$ within the minimal unitary representation 
space of $SU\left( 2 , 2 \,|\, \mathfrak{p} + \mathfrak{q} \right)$ with a 
deformation parameter $\zeta \ne 0$]
{States $\ket{\Omega}$ that are annihilated by all grade $-1$ 
generators in $\mathfrak{C}^-$ within the minimal unitary representation 
space of $SU\left( 2 , 2 \,|\, \mathfrak{p} + \mathfrak{q} \right)$ with a 
deformation parameter $\zeta \ne 0$.
\label{Table:generalLWVs}} \\
\hline
& & & & & & & \\
LWV & Range & $H$ & $H_s$ & $l$ & $l_3$ & $r$ & $r_3$ \\
& & & & & & & \\
\hline
& & & & & & & \\
\endfirsthead
%%%%%%%%%%%%%%%%%%%%%%%%%%%%%%%%%%%%%%%%%%%%%%%%%%%%%%%%%%%%%%%%%%%%%%%%%%%%
\caption[]{(continued)} \\
\hline
& & & & & & & \\
LWV & Range & $H$ & $H_s$ & $l$ & $l_3$ & $r$ & $r_3$ \\
& & & & & & & \\
\hline
& & & & & & & \\
\endhead
%%%%%%%%%%%%%%%%%%%%%%%%%%%%%%%%%%%%%%%%%%%%%%%%%%%%%%%%%%%%%%%%%%%%%%%%%%%%
& & & & & & & \\
\hline
\endfoot
%%%%%%%%%%%%%%%%%%%%%%%%%%%%%%%%%%%%%%%%%%%%%%%%%%%%%%%%%%%%%%%%%%%%%%%%%%%%
& & & & & & & \\
\hline
\endlastfoot
%%%%%%%%%%%%%%%%%%%%%%%%%%%%%%%%%%%%%%%%%%%%%%%%%%%%%%%%%%%%%%%%%%%%%%%%%%%%

$\ket{\frac{1}{2} - \zeta ; 0 , 0 ; 0 , 0}$ &
$\zeta = -1, -2, -3, \dots$ &
$2 - \zeta$ & $- \zeta$ &
$- \frac{\zeta}{2}$ & $\frac{\zeta}{2}$ & 0 & 0 \\[8pt]

$\ket{\frac{1}{2} + \zeta ; 0 , 0 ; 0 , 0}$ &
$\zeta = 1, 2, 3, \dots$ &
$2 + \zeta$ & $\zeta$ &
0 & 0 & $\frac{\zeta}{2}$ & $- \frac{\zeta}{2}$ \\[8pt]

$\ket{\frac{1}{2} - n - \zeta ; n , 0 ; 0 , 0}$ &
$\zeta = -1, -2, -3, \dots$ &
$2 - \zeta$ & $- \zeta$ & 
$- \frac{\zeta}{2}$ & $n + \frac{\zeta}{2}$ & 0 & 0 \\
 &
$n = 1, 2, \dots , -\zeta$ &
 & & 
 & & & \\[8pt]

$\ket{\frac{1}{2} - m + \zeta ; 0 , m ; 0 , 0}$ &
$\zeta = 1, 2, 3, \dots$ &
$2 + \zeta$ & $\zeta$ &
0 & 0 & $\frac{\zeta}{2}$ & $m - \frac{\zeta}{2}$ \\
 &
$m = 1, 2, \dots , \zeta$ &
 & & 
 & & & \\[8pt]

$\ket{\frac{1}{2} - p - \zeta ; 0 , 0 ; p , 0}$ &
$\zeta = -1, -2, -3, \dots$ &
$2 - p - \zeta$ & $- \zeta$ &
$- \frac{\zeta + p}{2}$ & $\frac{\zeta + p}{2}$ & 0 & 0 \\
& $p = 1, 2 , \dots , -\zeta ,\mbox{ if }  -\zeta < \mathfrak{p} $ &
 & & 
 & & & \\
 & $p = 1, 2 , \dots , \mathfrak{p} \mbox{ otherwise}$ &
 & & 
 & & & \\[8pt]

$\ket{\frac{1}{2} - q + \zeta ; 0 , 0 ; 0 , q}$ &
$\zeta = 1, 2, 3, \dots$ &
$2 - q + \zeta$ & $\zeta$ &
0 & 0 & $\frac{\zeta - q}{2}$ & $- \frac{\zeta - q}{2}$ \\
 &
$q = 1,2,3, \dots   \zeta , \mbox{ if } \zeta < \mathfrak{q} $ &
 & & 
 & & & \\
 &
$q = 1, 2,  \dots \mathfrak{q}  \mbox{ otherwise}$ &
 & & 
 & & & \\[8pt]

$\ket{\frac{1}{2} - n - p - \zeta ; n , 0 ; p , 0}$ &
$\zeta = -1, -2, -3, \dots$ &
$2 - p - \zeta$ & $- \zeta$ &
$- \frac{\zeta + p}{2}$ & $n + \frac{\zeta + p}{2}$ & 0 & 0 \\
 &
$p, n = 1,2,3 \dots $ &
 & & 
 & & & \\
 &
$p +n  \leq  - \zeta $ &
 & & 
 & & & \\[8pt]

$\ket{\frac{1}{2} - m - q + \zeta ; 0 , m ; 0 , q}$ &
$\zeta = 1, 2, 3, \dots$ &
$2 - q + \zeta$ & $\zeta$ &
0 & 0 & $\frac{\zeta - q}{2}$ & $m - \frac{\zeta - q}{2}$ \\
 &
$q, m = 1,2,3, \dots  $ &
 & & 
 & & & \\
 &
$q + m \leq  \zeta$ &
 & & 
 & & & \\[8pt]

%%%%%%%%%%%%%%%%%%%%%%%%%%%%%%%%%%%%%%%%%%%%%%%%%%%%%%%%%%%%%%%%%%%%%%%%%%%%
\end{longtable}
\end{tiny}
%%%%%%%%%%%%%%%%%%%%%%%%%%%%%%%%%%%%%%%%%%%%%%%%%%%%%%%%%%%%%%%%%%%%%%%%%%%%

Then it is a straightforward exercise to construct the corresponding deformed minimal 
unitary supermultiplets of $SU(2,2| \mathfrak{p} + 
\mathfrak{q})$ for each value of $\zeta$ by acting on the lowest 
weight vectors $\ket{\Omega}$ with the supersymmetry generators in grade 
$+1$ space $\mathfrak{C}^+$ repeatedly.

\section{Conclusions}
\label{sec:conclusions}

In this paper we first studied the minrep of the four dimensional conformal 
group $SU(2,2)$  and  its deformations obtained by quantizing  the quasiconformal 
realization of  $SU(2,2)$.  The resulting representations correspond to massless conformal fields in four spacetime dimensions.We then extended these results to construct the minimal unitary supermultiplet of $SU(2,2|4)$ and its deformations. 
 The minimal unitary supermultiplet of $SU(2,2|4)$ is simply the $N=4$ Yang-Mills supermultiplet. For each integer value of the deformation parameter we obtained a unique supermultiplet  of $SU(2,2|4)$.  These supermultiplets are simply the doubleton supermultiplets studied earlier in \cite{Gunaydin:1984fk,Gunaydin:1998sw,Gunaydin:1998jc}. 

Decomposition of tensoring of minreps into irreducible unitary representations is, in general, a  difficult problem.
Since the decomposition of tensor products of doubleton representations into its irreducible components is relatively  easier in the twistorial oscillator approach we hope to be able to use  our results to solve the tensoring  problem for the minreps of $SU(2,2)$ and $SU(2,2\,|\,\mathfrak{p}+\mathfrak{q})$ as well as their deformations within the quasiconformal approach \cite{workinprogress_sfmg}. We hope that these results will then enable one to tackle the much harder problem of decomposition of tensor products of minreps of noncompact groups that are not of hermitian symmetric type, such as $E_{8(8)}$ or $E_{8(-24)}$. 

The extension of the above results to other noncompact groups which admit  
positive energy unitary representations and their supersymmetric  
extensions, as well as their applications to AdS/CFT dualities will be 
the subjects of separate studies.

%%%%%%%%%%%%%%%%%%%%%%%%%%%%%%%%%%%%%%%%%%%%%%%%%%%%%%%%%%%%%%%%%%%%
%%%%%% Acknowledgements %%%%%%%%%%%%%%%%%%%%%%%%%%%%%%%%%%%%%%%%%%%%
%%%%%%%%%%%%%%%%%%%%%%%%%%%%%%%%%%%%%%%%%%%%%%%%%%%%%%%%%%%%%%%%%%%%

{\bf Acknowledgements:} M.G. would like to thank  Sergei Gukov  and David Vogan for stimulating and  informative discussions on the minimal unitary representations of noncompact groups. He would also like to acknowledge  stimulating  discussions with Juan Maldacena, Shiraz Minwalla, Mukund Rangamani, Augusto Sagnotti and Misha Vasiliev 
on the minrep of the $4D$ conformal group and thank the organizers of the `Fundamental Aspects of Superstring Theory 2009' Workshop at KITP, UC Santa Barbara and of the 
  `New Perspectives in String Theory 2009' Workshop at GGI, Florence where part of this work was carried out. \\
  S.F. would like to thank the Center for Fundamental Theory at the 
Institute for Gravitation and the Cosmos at Pennsylvania State University, where part of this work was completed, for their warm hospitality. \\
  This work was supported in part by the National
Science Foundation under grants numbered PHY-0555605 and PHY-0855356. Any opinions,
findings and conclusions or recommendations expressed in this
material are those of the authors and do not necessarily reflect the
views of the National Science Foundation.

\providecommand{\href}[2]{#2}\begingroup\raggedright\endgroup

%\bibliographystyle{utphys}
%\bibliography{combined}
\end{document}